\documentclass[preprint,12pt]{elsarticle}

\usepackage{hyperref}
\usepackage{amsfonts}
\usepackage{amsthm}
\usepackage{amsmath}
\usepackage{bm}
\usepackage{breqn}

\usepackage[ruled]{algorithm2e}
\usepackage{algpseudocode}
\SetKwInput{KwInit}{Init}

\usepackage{tabularx}

\usepackage{animate}
\usepackage{tikz}

\usepackage{comment}
\usepackage[normalem]{ulem}
\usepackage{graphicx}
\usepackage{subcaption}
\usepackage{xcolor}
\newcommand{\red}[1]{\textcolor{red}{#1}}

\definecolor{vividviolet}{rgb}{0.62, 0.0, 1.0}
\definecolor{coralred}{rgb}{1.0, 0.25, 0.25}

\definecolor{ndblue}{RGB}{12,35,64}
\definecolor{ndgold}{RGB}{201,151,0}
\definecolor{ao}{rgb}{0.0, 0.5, 0.0}
\definecolor{forestgreen}{rgb}{0.13, 0.55, 0.13}

\newcommand\encircle[1]{%
  \tikz[baseline=(X.base)] 
    \node (X) [draw, shape=circle, inner sep=0] {\strut #1};}

\newtheorem{defn}{Definition}[section]

\newdefinition{remark}{Remark}

\def\eqref#1{equation~\ref{#1}}

\def\1{\bm{1}}

\newcommand{\bmf}{\mathbf{f}}
\newcommand{\bmg}{\mathbf{g}}
\newcommand{\bmh}{\mathbf{h}}

\newcommand{\bmn}{\mathbf{n}}

\newcommand{\bmr}{\mathbf{r}}
\newcommand{\bms}{\mathbf{s}}
\newcommand{\bmt}{\mathbf{t}}
\newcommand{\bmu}{\mathbf{u}}

\newcommand{\bmx}{\mathbf{x}}
\newcommand{\bmy}{\mathbf{y}}
\newcommand{\bmz}{\mathbf{z}}

\newcommand{\bmI}{\mathbf{I}}

\newcommand{\bmepsilon}{\bm{\epsilon}}

\newcommand{\bmmu}{\bm{\mu}}

\newcommand{\bmphi}{\bm{\phi}}

\newcommand{\bmsigma}{\bm{\sigma}}
\newcommand{\bmtau}{\bm{\tau}}
\newcommand{\bmtheta}{\bm{\theta}}

\newcommand{\bmxi}{\bm{\xi}}

\newcommand{\bmzero}{\bm{0}}

\newcommand{\E}{\mathbb{E}}

\newcommand{\R}{\mathbb{R}}

\newcommand{\KL}{D_{\mathrm{KL}}}

\newcommand{\mc}{\mathcal}

\DeclareMathOperator*{\argmin}{arg\,min}

\newcommand{\norm}[1]{\left\lVert #1 \right\rVert}

\newcommand{\si}{{(i)}}
\newcommand{\sj}{{(j)}}

\journal{Journal of Computational Physics}

\begin{document}
\abovedisplayskip=6.0pt
\belowdisplayskip=6.0pt
\begin{frontmatter}

\title{Physics-Constrained Deep Learning for High-dimensional Surrogate Modeling and Uncertainty Quantification without Labeled Data} 
 
\author[1]{Yinhao Zhu}
\ead{yzhu10@nd.edu}
\author[1]{Nicholas Zabaras\corref{cor1}}
\ead{nzabaras@gmail.com}
\cortext[cor1]{Corresponding author: 
  Tel.: +1-574-631-2429;  }
\ead[url]{https://www.cics.nd.edu/}
\author[2]{Phaedon-Stelios Koutsourelakis}
\ead[url]{http://www.contmech.mw.tum.de/index.php?id=5}
\ead{p.s.koutsourelakis@tum.de}
\author[3]{Paris Perdikaris}
\ead{pgp@seas.upenn.edu}
\ead[url]{https://www.seas.upenn.edu/directory/profile.php?ID=237}

\address[1]{Center for Informatics and Computational Science, \\311I Cushing Hall, University of Notre Dame, Notre Dame, IN 46556, U.S.A.}
\address[2]{Continuum Mechanics Group,
Technical University of Munich, Boltzmannstra{\ss}e 15, 85748 Garching, Germany}
\address[3]{Department of Mechanical Engineering and Applied Mechanics, 220 South 33rd Street, 229 Towne Building, University of Pennsylvania, Philadelphia, PA  19104-6315, U.S.A.}

\begin{abstract}
Surrogate modeling and uncertainty quantification tasks for PDE systems are most often considered as supervised learning problems where input and output data pairs are used for training. The construction of such emulators is by definition a small data problem which poses challenges to deep learning approaches that have been developed to operate in the big data regime. Even in cases where such models have been shown to have good predictive capability in high dimensions, they fail to address constraints in the data implied by the PDE model. This paper provides a methodology that incorporates the governing equations of the physical model in the loss/likelihood functions. The resulting physics-constrained, deep learning models are trained without any labeled data (e.g. employing only input data) and provide comparable predictive responses with data-driven models while obeying the constraints of the problem at hand. This work employs a convolutional encoder-decoder neural network approach as well as a conditional flow-based generative model for the solution of PDEs, surrogate model construction, and uncertainty quantification tasks. The methodology is posed as a minimization problem of the reverse Kullback-Leibler (KL) divergence between the model predictive density and the reference conditional density, where the later is defined as the Boltzmann-Gibbs distribution at a given inverse temperature with the underlying potential relating to the PDE system of interest. The generalization capability of these models to out-of-distribution input is considered. Quantification and interpretation of the predictive uncertainty is provided for a number of problems.
\end{abstract}

\begin{keyword}
	physics-constrained, energy-based models, label-free, variational inference, surrogate modeling, uncertainty quantification, high-dimensional, porous media flow, encoder-decoder, conditional generative model, normalizing flow
\end{keyword}

\end{frontmatter}

\section{Introduction}

Surrogate modeling is computationally attractive for problems that require repetitive yet expensive simulations, such as determinsitsic design, uncertainty propagation, optimization under uncertainty or inverse modeling. Data-efficiency,  uncertainty quantification and generalization are the main challenges facing surrogate modeling, especially for problems with high-dimensional stochastic input, such as material properties~\cite{bilionis2013multi}, background potentials~\cite{charalampidis2018computing}, etc.

Training surrogate models is commonly posed as a supervised learning problem, which requires simulation data as the target. 
Gaussian process (GP) models are widely used as emulators for physical systems~\cite{kennedy2000predicting} with built-in uncertainty quantification. The recent advances to scale GPs to high-dimensional input include Kronecker product decomposition that exploits the spatial structure~\cite{bilionis2013multi, wilson2015kernel, atkinson2018structured}, convolutional kernels~\cite{NIPS2017_6877} and other algorithmic and software developments~\cite{gardner2018gpytorch}. However, GPs are still struggling to effectively model high-dimensional input-output maps.
Deep neural networks (DNNs) are becoming the most popular surrogate models nowadays across engineering and scientific fields. As universal function approximators, DNNs excel at settings where both the input and output are high-dimensional. Applications in flow simulations include pressure projections in solving Navier-Stokes equations~\cite{doi:10.1002/cav.1695}, fluid flow through random heterogeneous media~\cite{zhu2018bayesian, tripathy2018deep, mo2018deep}, Reynolds-Averaged Navier-Stokes simulations~\cite{ling_kurzawski_templeton_2016, thuerey2018well,Geneva2019Quantifying} and others.
Uncertainty quantification for DNNs is often studied under the re-emerging framework of Bayesian deep learning\footnote{http://bayesiandeeplearning.org/}~\cite{doi:10.1162/neco.1992.4.3.448}, mostly using variational inference for approximate posterior of model parameters, e.g. variational dropout~\cite{NIPS2015_5666, gal2016dropout}, Stein variational gradient descent~\cite{NIPS2016_6338, zhu2018bayesian}, although other methods exist, e.g. ensemble methods~\cite{lakshminarayanan2017simple}.
Another perspective to high-dimensional problems is offered by latent variable models~\cite{2017arXiv171102475G}, where the latent variables encode the information bottleneck between the input and output.

Sufficient amount of training data is usually required for the surrogates to achieve accurate predictions even under restricted settings, e.g. fixed boundary conditions. For physically-grounded domains, baking in the prior knowledge can potentially overcome the challenges of data-efficiency and generalization, etc. The \textit{inductive bias} can be built into the network architecture, e.g. spherical convolutional neural networks (CNNs) for the physical fields on unstructured grid~\cite{jiang2018spherical}, graph networks for object- and relation-centric representations of complex, dynamical systems~\cite{gn_icml18}, learning linear embeddings of nonlinear dynamics based on Koopman operator theory~\cite{lusch2018deep}. Another approach is to \textit{embed physical laws} into the learning systems, such as approximating differential operators with convolutions~\cite{pmlr-v80-long18a}, enforcing hard constraint of mass conservation by learning the stream function~\cite{2018arXiv180602071K} whose curl is guaranteed to be divergence-free. 

A more general way to incorporate physical knowledge is through \textit{constraint learning}~\cite{2016arXiv160905566S}, i.e. learning the models by minimizing the violation of the physical constraints, symmetries, e.g. cycle consistency in domain translation~\cite{CycleGAN2017}, temporal coherence of consecutive frames in fluid simulation~\cite{xie2018tempogan} and video translation~\cite{wang2018vid2vid}. One typical example in computational physics is \textit{learning solutions of deterministic PDEs with neural networks} in space/time, which dates back at least to the early $1990$s, e.g.~\cite{psichogios1992hybrid, meade1994numerical, lagaris1998artificial}. The main idea is to train neural networks to approximate the solution by minimizing the violation of the governing PDEs (e.g. the residual of the PDEs) and also of the  initial and boundary conditions.
In~\cite{lagaris1998artificial}, a one-hidden-layer fully-connected neural network (FC-NN) with spatial coordinates as input is trained to minimize the residual norm evaluated on a  fixed grid. 
The success of deep neural networks brings several new developments: (1) most of the works parameterize the solution with FC-NNs, thus the solution is analytical and meshfree~\cite{raissi2017physics, berg2018unified}; (2) the loss function can be derived from the variational form~\cite{weinan2018deep, nabian2018deep}; (3) stochastic gradient descent is used to train the network by randomly sampling mini-batches of inputs (spatial locations and/or time instances)~\cite{sirignano2018dgm, weinan2018deep}; (4) deeper networks are used to break the curse of dimensionality~\cite{grohs2018proof} allowing for several high-dimensional PDEs to be solved with high accuracy and speed~\cite{han2018solving, beck2017machine, sirignano2018dgm, raissi2018forward}; (5) multiscale numerical solvers are enhanced by replacing the linear basis with learned ones with DNNs~\cite{wang2018deep, fan2018multiscale}; (6) surrogate modeling for PDEs~\cite{CNNFluid2016, khoo2017solving, nabian2018deep}.

Our work focuses on physics-constrained surrogate modeling for stochastic PDEs with high-dimensional spatially-varying coefficients \textit{without} simulation data. 
We first show that when solving deterministic PDEs, the CNN-based parameterizations are more computationally efficient in  capturing multiscale features of the solution fields than the FC-NN ones.  
Furthermore, we demonstrate  that in comparison with image-to-image regression  approaches that employ Deep NNs~\cite{zhu2018bayesian}, the proposed method  achieves comparable predictive performance, despite the fact that it does not make use of any output simulation data. In addition, it produces  better predictions under extrapolative conditions as when   out-of-distribution test input datasets are used.
Finally, a flow-based conditional generative model is proposed to capture the predictive distribution with calibrated uncertainty, without compromising the predictive accuracy.

The paper is organized as follows.
Section~\ref{sec:Definition} provides the definition of the problems of interest including the solution of PDEs, surrogate modeling and uncertainty quantification. Section~\ref{sec:Methodology} provides the parametrization of the solutions with FC-NNs and CNNs, the physics-constrained learning of a deterministic surrogate and the variational learning of a probabilistic surrogate. Section~\ref{sec:numerical_experiments} investigates the performance of the developed techniques with a variety of tests for various PDE systems. We conclude in Section~\ref{sec:Conclusions} with a summary of this work and extensions to address limitations that have been identified.

\section{Problem Definition}
\label{sec:Definition}
Consider  modeling of a physical system governed by PDEs:
\begin{equation}
    \label{eq:PDE}
    \begin{aligned}
    \mc N (u(s); K(s)) &= f(s), \qquad s \in \mc S, \\
    \mc B (u(s))       &= b(s), \qquad s \in \Gamma,
    \end{aligned}
\end{equation}
where $\mc N$ is a general differential operator, $u(s)$ are the field variables of interest, $f(s)$ is the source field, and $K(s)$ denotes an input property field defining the system's constitutive behavior. $\mc B$ is the operator for boundary conditions defined on the boundary $\Gamma$ of the domain $\mc S$.
In particular, we consider the following Darcy flow problem as a motivating example throughout this paper:
\begin{equation}
\label{eq:darcy}
    - \nabla \cdot (K(s) \nabla u(s))) = f(s), \qquad s \in \mc S,
\end{equation}
with boundary conditions
\begin{equation}\label{eq:BCs}
    \begin{aligned}
    u(s) &= u_D(s), \qquad s \in  \Gamma_D, \\
    \nabla u(s) \cdot n &= g(s), \qquad s \in \Gamma_N,
    \end{aligned}
\end{equation}
where $n$ is the unit normal vector to the Neumann boundary $\Gamma_N$, $\Gamma_D$ is the Dirichlet boundary. 

Of particular interest are PDEs for which the field variables can be computed by appropriate minimization of a field energy functional (potential) $V(u; K)$, i.e.  
\begin{equation}
\argmin_u V(u; K).     
\end{equation}
Such potentials are common in many linear and nonlinear problems in  physics and engineering and serve as the basis of the finite element  method.  For problems where such potentials cannot be found~\cite{FilSavSho92}, one can consider $V$ as the square of the residual norm of the PDE evaluated at different trial solutions, e.g.
\begin{equation}
    V(u; K)=R^2\left(u; K\right).
\end{equation}
In this paper, we are interested in the solution of parametric PDEs for a given set of boundary conditions.

\begin{defn}[Solution of a deterministic PDE system]
\label{defn:solution_pde}
Given the potential $V(u; K)$, and the boundary conditions in Eq.~(\ref{eq:BCs}), compute the solution $u(s)$ of the PDE for a given input field $K(s)$.
\end{defn}

The input field $K(s)$ is often modeled as a random field $K(s, \omega)$ in the context of uncertainty quantification, where $\omega$ denotes a random event in the sample space $\Omega$. In practice, discretized versions of this field are employed in computations which is denoted as the random vector $\bmx$, i.e. $\bmx = [K(s_1), \cdots, K(s_{n_s})]$. We note that when fine-scale fluctuations  of the input field $K$ are present, the dimension $n_s$ of $\bmx$ can become very high. Let $p(\bmx)$ be the associated density postulated by mathematical considerations or learned from data, e.g.  CT scans of microstructures, measurement of permeability fields, etc.
Suppose $\bmy$ denotes a discretized version of the PDE solution, i.e.
$$\bmy = [u(s_1), \cdots, u(s_{n_s})].$$ 

Note that all the discretized field variable(s) are denoted as bold, while the continuous field variable(s) are non-bold.

We are interested in developing a surrogate model that allows fast calculation of the system response $\bmy$
for any input realization $\bmx \in p(\bmx)$, and potentially for various boundary conditions.
This leads to the following definition:

\begin{defn}[Deterministic Surrogate Model]
\label{defn:det_surrogate}
 Given the potential $V(u; K)$, the boundary conditions in Eq.~(\ref{eq:BCs}), and a set of training input data $\mc D_{\text{input}} = \{\bmx^\si\}_{i=1}^N, \bmx^\si \sim p(\bmx)$, learn a deterministic surrogate 
$
    \bmy = \hat{\bmy}_\bmtheta(\bmx),
$
for predicting the solution $\bmy$ for any input $\bmx \in p(\bmx)$, where $\bmtheta$ denotes the parameters of the surrogate model.
\end{defn}

Note that often the density $p(\bmx)$ is not known and needs to be approximated from the given data $\{\bmx^\si\}_{i=1}^N$. When the density $p(\bmx)$ is given, the surrogate model can be defined without referring to the particular training data set. In this case, as part of the training process, one can select any dataset of size $N$, $\{\bmx^\si\}_{i=1}^N, \bmx^\si \sim p(\bmx)$, including the most informative one for the surrogate task.   

We note  that the aforementioned problem refers to  a new  type of machine learning  task that falls between unsupervised learning due to the absence of labeled data (i.e. the $\bmy^\si$ corresponding to each $\bmx^\si$ is {\em not} provided) and (semi-)supervised learning because the objective  involves discovering the map from the input $\bmx$ to the output $\bmy$. 
Given the finite training data employed in practice and the inadequacies of the model postulated, $\hat{\bmy}_\bmtheta(\bmx)$, it is often advantageous to obtain 
 a distribution over the possible solutions via a probabilistic surrogate, rather than  a mere point estimate for the solution.

\begin{defn}[Probabilistic Surrogate Model]
Given the potential $V(u; K)$, the boundary conditions in Eq.~(\ref{eq:BCs}), and a set of training input data $\mc D_{\text{input}} = \{\bmx^\si\}_{i=1}^N, \bmx^\si \sim p(\bmx)$, a probabilistic surrogate model specifies a conditional density $p_\bmtheta(\bmy | \bmx)$, where $\bmtheta$ denotes the model parameters.
\end{defn}

Finally, since the input $\bmx$ arises from an underlying probability density, one may be interested to compute the statistics of the output $\bmy$ leading to the following forward uncertainty propagation problem. 

\begin{defn}[Forward Uncertainty Propagation]
\label{defn:uncertainty_propagation} 
Given the potential $V(u; K)$, the boundary conditions in Eq.~(\ref{eq:BCs}), and a set of training input data $\mc {\mathcal D}_{\text{input}} = \{\bmx^\si\}_{i=1}^N, \bmx^\si \sim p(\bmx)$, estimate moments of the response, $\E[\bmy], \text{Var}[\bmy], \ldots$ or more generally any aspect of the  probability density of $\bmy$.
\end{defn}

\section{Methodology}
\label{sec:Methodology}
\subsection{Differentiable Parameterizations of Solutions}\label{sec:solve_pde}
We only consider the parameterizations of solutions using neural networks, primarily FC-NNs and CNNs. Given one input $\bmx = [K(s_1), \cdots, K(s_{n_s})]$, most previous works~\cite{lagaris1998artificial, han2018solving, raissi2017physics, sirignano2018dgm} use FC-NNs to represent the solution as
\begin{equation}\label{eq:solve_fc_nn}
    u(s) = \hat{u}_\bmphi (s),
\end{equation}
where the input to the network is coordinate $s$, the output is the predicted solution at $s$, and $\hat{u}_\bmphi$ denotes a FC-NN with parameters $\bmphi$. The spatial gradients can be evaluated exactly by automatic differentiation. This approach yields a smooth representation of the solution that can be evaluated at any input location. Even though the outputs in this model at two different locations are correlated (as they both depend on the shared parameters $\bmphi$ of the NN), FC-NNs do not have the inductive bias as in CNNs, e.g. translation invariance, parameter sharing, etc.
Despite promising results in a series of canonical problems~\cite{raissi2018physics},  the trainability and predictive performance of FC-NNs deteriorates as the complexity of the underlying solution increases. This drawback is confirmed by our numerical studies presented in Section~\ref{sec:solve_det_pde} involving solution fields with multiscale features.

An alternative parametrization of the solution is through a convolutional decoder network
\begin{equation}\label{eq:solve_cnn}
    \bmy = \hat{\bmy}_\bmtheta (\bmz),
\end{equation}
where $\bmy = [u(s_1), \cdots, u(s_{n_s})]$ denotes the solution on pre-defined fixed grids $s_1, \cdots, s_{n_s}$  that is generated by one pass of the latent variable $\bmz$ through the decoder, similarly as in~\cite{2017arXiv171110925U}. Note that $\bmz$ is usually much lower-dimensional than $n_s$ and initialized arbitrarily. The spatial gradients can be approximated efficiently with Sobel filter\footnote{\url{https://www.researchgate.net/publication/239398674_An_Isotropic_3x3_Image_Gradient_Operator}}, which amounts to one convolution layer with fixed kernel, see~\ref{appendix:sobel_filter} for details.
In contrast to FC-NNs, convolutional architectures can directly capture complex spatial correlations and return a multi-resolution representation of the underlying solution field.

\begin{remark}
The dimensionality  $n_s$ of the input $\bmx$ is not required to be the same as that of  the output $\bmy$. Since our CNN approach would involve  operations between images including pixel-wise multiplication of input and output images (see Section~\ref{sec:LossFnDarcyFlow}), we select herein the same dimensionality for both inputs and outputs. Upsampling/downsampling can always be used to accommodate different dimensionalities $n_{sx}$ and $n_{sy}$ of the input and output images, respectively.
\end{remark}

To solve the deterministic PDE for a given input, we can train the   FC-NN solution as in Eq.~(\ref{eq:solve_fc_nn}) by minimizing the residual loss where the exact derivatives are calculated with automatic differentiation~\cite{lagaris1998artificial, han2018solving, raissi2017physics, sirignano2018dgm}. For the CNN representation, we will detail the loss functions and numerical derivatives in the next section.

\subsection{Physics-constrained Learning of Deterministic Surrogates without Labeled Data}
\label{sec:deterministic_surrogate}

We are particularly interested in surrogate modeling with high-dimensional input and output, i.e.   $\text{dim}(\bmx), \text{dim}(\bmy) \gg 1$. Surrogate modeling is an extension of the solution networks in the previous section by adding the realizations of stochastic input $\bmx$ as the input, e.g. $u(s, \bmx) = \hat{u}_\bmphi(s, \bmx)$ in the FC-NN case~\cite{nabian2018deep}, or $\bmy = \hat{\bmy}_\bmtheta(\bmx)$ in the CNN case~\cite{khoo2017solving}.

Here, we adopt the \textit{image-to-image regression} approach~\cite{zhu2018bayesian} to deal with the problem arising in practice where the realizations of the random input field are image-like data instead of being computed from an analytical formula. More specifically, the surrogate model $\bmy = \hat{\bmy}_\bmtheta (\bmx)$ is an extension of the decoder network in Eq.~(\ref{eq:solve_cnn}) by prepending an encoder network to transform the high-dimensional input $\bmx$ to the latent variable $\bmz$, i.e. $\bmy = \text{decoder} \circ \text{encoder} (\bmx)$.

In contrast to existing convolutional encoder-decoder network structures~\cite{zhu2018bayesian}, the surrogate model studied here is trained \textit{without} labeled data i.e. without computing the solution of the PDE. Instead, it is trained by learning to solve the PDE with given boundary conditions, using the following loss function
\begin{equation}\label{eq:loss_det_surr}
    L(\bmtheta; \{\bmx^\si\}_{i=1}^N) =  \frac{1}{N} \sum_{i=1}^N \Big[V(\hat{\bmy}_{\bmtheta}(\bmx^\si), \bmx^\si) + \lambda B(\hat{\bmy}_{\bmtheta}(\bmx^\si)) \Big],
\end{equation}
where $\hat{\bmy}^\si = \hat{\bmy}_{\bmtheta}(\bmx^\si)$ is the prediction of the surrogate for $\bmx^\si \in \mc D_{\text{input}}$, $V(\hat{\bmy}^\si, \bmx^\si)$ is the equation loss, either in the form of the \textit{residual norm}~\cite{lagaris1998artificial} or the \textit{variational functional}~\cite{weinan2018deep} of the PDE, $B(\hat{\bmy}^\si)$ is the boundary loss of the prediction $\hat{\bmy}^\si$, and $\lambda$ is the weight (Lagrange multiplier) to softly enforce the boundary conditions. Both $V(\hat{\bmy}^\si, \bmx^\si)$ and $B(\hat{\bmy}^\si)$ may involve integration and differentiation with respect to the spatial coordinates, which are approximated with highly efficient discrete operations, detailed below for the Darcy flow problem. The surrogate trained with the loss function in Eq.~(\ref{eq:loss_det_surr}) is called \textit{physics-constrained surrogate} (PCS).

In contrast to the physically motivated loss function advocated above, a typical  data-driven surrogate  employs a loss function of the form
\begin{equation}\label{eq:loss_det_surr_data}
    L_{\text{MLE}}(\bmtheta; \{(\bmx^\si, \bmy^\si)\}_{i=1}^N) = \frac{1}{N} \sum_{i=1}^N  \norm{\bmy^\si - \hat{\bmy}_\bmtheta(\bmx^\si)}_2^2,
\end{equation}
where $\bmy^\si$ is the output data for the input $\bmx^\si$ which must be computed in advance. We refer to the surrogate trained with loss function in Eq.~(\ref{eq:loss_det_surr_data}) as the \textit{data-driven surrogate} (DDS).

\subsubsection{Loss Function for Darcy Flow}
\label{sec:LossFnDarcyFlow}
There are at least four variations of loss functions for a second-order elliptic PDE problem, depending on whether the field variables refer to the primal variable (pressure) or to mixed variables (pressure and fluxes), and whether the loss is expressed in strong form or variational form. Specifically, for the Darcy flow problem defined in Eq.~(\ref{eq:darcy}), we can consider:

\paragraph{Primal residual loss} The residual norm for the primal variable is
\begin{equation}
    V(u; K) = \int_{\mc S} \Big(\nabla \cdot (K \nabla u) + f \Big)^2 ds.
\end{equation}
\paragraph{Primal variational loss}
The energy functional is
\begin{equation}
    \label{eq:energy_functional}
    V(u; K) = \int_{\mc S} \Big( \frac{1}{2} K \nabla u\cdot \nabla u - f u \Big) ds - \int_{\Gamma_N} g u ~ds.
    \end{equation}

Mixed formulation introduces an additional (vector) variable, namely flux $\tau$, which turns Eq.~(\ref{eq:darcy}) into a systems of equations
\begin{equation}
\label{eq:darcy_mixed}
    \begin{aligned}
    \tau &= - K \nabla u, \qquad \text{in } \mc S, \\
    \nabla \cdot \tau &= f, \qquad \text{in } \mc S,
    \end{aligned}
\end{equation}
with the same boundary conditions as in Eq.~(\ref{eq:BCs}). $\tau(s) = [\tau_1(s), \tau_2(s)]$ are the flux field components along the horizontal and vertical directions, respectively.
 
\paragraph{Mixed variational loss} Following the Hellinger-Reissner principle~\cite{arnold1990mixed}, the mixed variational principle states that the solution $(\tau^*, u^*)$ of the Darcy flow problem is the unique critical point of the functional
\begin{equation}
\label{eq:mixed_functional}
    V(\tau, u; K) = \int_\Omega \Big(\frac{1}{2} K^{-1} \tau \cdot \tau + u \nabla \cdot \tau + f u \Big) ds - \int_{\Gamma_D} u_D \tau \cdot \bmn ds, 
\end{equation}
over the space of vector fields $\tau \in \mc H(\text{div})$ satisfying the Neumann boundary condition and all the fields $u \in \mc L^2$.
It should be highlighted that the solution $(\tau^*, u^*)$ is not an extreme point of the functional in Eq.~(\ref{eq:mixed_functional}), but a \textit{saddle point}, i.e.
\begin{equation*}
    V(\tau^*, u) \leq V(\tau^*, u^*) \leq V(\tau, u^*).
\end{equation*}

\paragraph{Mixed residual loss} The residual norm for the mixed variables is
\begin{equation}\label{eq:mixed_residual}
    V(u; K) = \int_{\mc S} \Big[ \Big(\tau + K \nabla u \Big)^2 + \Big(\nabla \cdot \tau - f\Big)^2 \Big] ds.
\end{equation}

Both the variational and mixed formulations have the advantage of lowering the order of differentiation which is approximated numerically in our implementation by a Sobel filter, as detailed in~\ref{appendix:sobel_filter}. For example by employing the discretized representation $\bmx$ for $K$ where the domain is $\mc S=[0, 1] \times [0, 1]$,  the mixed residual loss  is evaluated as
\begin{equation}\label{eq:mixed_residual_impl}
    V(\bmtau, \bmu; \bmx) \approx \frac{1}{n_s} \Big(\norm{\bmtau + \bmx \odot \nabla \bmu}^2_2 + \norm{\nabla \cdot \bmtau - \bmf}^2_2 \Big),
\end{equation}
where $n_s$ is the number of uniform grids, $\nabla \bmu = [\bmu_h, \bmu_v]$, $\bmu_h, \bmu_v$ are two gradient images along the horizontal and vertical directions estimated by Sobel filter, similarly for $\nabla \cdot \bmtau = (\bmtau_1)_h + (\bmtau_2)_v$, and $\odot$ denotes the  element-wise product.

\subsection{Probabilistic Surrogates with Reverse KL Formulation}

While a deterministic surrogate provides fast predictions to new input realizations, it does not model the predictive uncertainty which is important in practice especially when the surrogate is tested on unseen (during training) inputs.
Moreover, many PDEs in physics have multiple solutions~\cite{farrell2015deflation} which cannot be captured with a deterministic model. 
Thus building probabilistic surrogates that can model the distribution over possible solutions given the input is of great importance.

A probabilistic surrogate models the conditional density of the predicted solution given the input, i.e. $p_\bmtheta(\bmy | \bmx)$. 
Instead of learning this conditional density with labeled data~\cite{sohn2015learning, mirza2014conditional, van2016conditional}, we distill it from a reference density $p_\beta(\bmy | \bmx)$. The reference density is a Boltzmann distribution
\begin{equation}
\label{eqn:target_conditional}
    p_{\beta}(\bmy | \bmx) = \frac{\exp{(-\beta L(\bmy, \bmx) )}}{Z_\beta(\bmx)},
\end{equation}
where $L(\bmy, \bmx) = V(\bmy, \bmx) + \lambda B(\bmy)$ is the loss function (Eq.~\ref{eq:loss_det_surr}) for the deterministic surrogate that penalizes the violation of the PDE and boundary conditions, and $\beta$ is an inverse temperature parameter that controls the overall variance of the reference density. This energy-based model is obtained solely from the PDE and boundary conditions, without having  access to labeled output data~\cite{lecun2006tutorial}. However, this PDE-constrained model provides similar information as the labeled data allowing us to learn a probabilistic surrogate.

Since sampling from the probabilistic surrogate $p_\bmtheta(\bmy | \bmx)$ is usually fast and evaluating the (unnormalized) reference density $p_\beta(\bmy | \bmx)$ is often cheap, we choose to minimize the following reverse KL divergence:
\begin{equation}\label{eq:rev_kl}
\begin{split}
     \KL (p(\bmx)~p_{\bmtheta}(\bmy | \bmx) \parallel p(\bmx)~ p_{\beta}( \bmy | \bmx )) 
     & = \E_{p(\bmx)} \Big[ - \E_{p_{\bm\theta}(\bmy | \bmx)}[\log p_{\beta} (\bmy | \bmx)] + \E_{p_{\bm\theta}(\bmy | \bmx)} [\log p_{\bm\theta}(\bmy | \bmx)] \Big]\\
     & = \beta \E_{p(\bmx)p_{\bm\theta}(\bmy | \bmx)} [L (\bmy, \bmx)] - \mathbb{H}_\bmtheta (\bmy | \bmx) + \E_{p(\bmx)}[\log Z_{\beta}(\bmx)].
\end{split}
\end{equation}
The first term is the expectation of the loss function $L(\bmy,\bmx)$  w.r.t. the joint density $p(\bmx) p_{\bmtheta}(\bmy | \bmx)$, which enforces the satisfaction of PDEs and boundary conditions.
The second term is the negative conditional entropy of $p_{\bmtheta}(\bmy | \bmx)$ which promotes the diversity of model predictions. It also helps to stabilize the training of flow-based conditional generative model introduced in Section~\ref{sec:cflow}. 
The third term is the variational free energy, which is constant when optimizing $\bmtheta$. For the models with intractable log-likelihood $\log p_{\bm\theta}(\bmy | \bmx)$, one can derive a lower bound for the conditional entropy $\mathbb{H}_{\bmtheta}(\bmy | \bmx)$ that helps to regularize training and avoid mode collapse as in~\cite{yang2018adversarial}. In this work, the log-likelihood can be exactly evaluated for the model introduced in Section~\ref{sec:cflow}.

This idea is similar to probability density distillation~\cite{oord2017parallel} to learn generative models for real-time speech synthesis,  neural renormalization group~\cite{li2018neural} to accelerate sampling for Ising models, and Boltzmann generators~\cite{noe2018boltzmann}
 to efficiently sample equilibrium states of many-body systems.

The reverse KL divergence itself is not enough to guarantee that the predictive uncertainty is well-calibrated. Even if this divergence is optimized to zero, i.e. $p_\bmtheta(\bmy|\bmx) = p_\beta(\bmy|\bmx)$, the predictive uncertainty is still controlled by $\beta$. Thus we add an uncertainty calibration constraint to the optimization problem, i.e. 
\begin{equation}
\begin{aligned}
    \min_{\beta, \bmtheta} \quad & \KL (p(\bmx) p_{\bmtheta}(\bmy | \bmx)
    \parallel p(\bmx) p_{\beta}( \bmy | \bmx )), \\
    s.t. \quad & p_{\bmtheta}(\bmy | \bmx) \text{ is calibrated on validation data.}
\end{aligned}
\end{equation}
Here, the predictive uncertainty is calibrated using the reliability diagram~\cite{degroot1983comparison}. 
The naive approach to select $\beta$ is through grid search, i.e. train the probabilistic surrogate with different values of $\beta$, and select the one under which the trained surrogate is well-calibrated w.r.t. validation data, which includes input-output data pairs.

\remark{Instead of tuning $\beta$ with grid search, we can also re-calibrate the trained model \textit{post-hoc}~\cite{guo2017calibration, kuleshov2018accurate} by learning an auxiliary regression model. For a small amount of miscalibration, sampling latent variables with different temperature (Section~6 in~\cite{kingma2018glow}) can also change the variance of the output with a slight drop of predictive accuracy.}

\begin{remark}
Similar to our approach, Probabilistic Numerical Methods (PNMs)~\cite{hennig2015probabilistic, cockayne2016probabilistic, cockayne2017bayesian} take a statistical point of view of classical numerical methods (e.g. a finite element solver) that treat the output  as a point estimate of the true solution. Given finite information (e.g. finite number of evaluations of the PDE operator and boundary conditions) and prior belief about the solution, PNMs output the posterior distribution of the solution. 
PNM focuses on inference of the solution for one input, instead of amortized inference as what the probabilistic surrogate does.
\end{remark}

\subsubsection{Conditional Flow-based Generative Models}\label{sec:cflow}
This section presents flow-based generative models~\cite{dinh2016density} as our probabilistic surrogates. This family of models offers several advantages over other generative models~\cite{2013arXiv1312.6114K, NIPS2014_5423}, such as exact inference and exact log-likelihood evaluation that is particularly attractive for learning the conditional distribution with the reverse KL divergence as in Eq.~(\ref{eq:rev_kl}). The generative model $\bmy = \bmg_\bmtheta(\bmz)$ consists of a sequence of \textit{invertible} layers (also called normalizing flows~\cite{rezende2015variational}) that transforms a simple distribution $p(\bmz)$ to a target distribution $p(\bmy)$, i.e. 
$$\bmy := \bmh_0 \overset{\bmg_\bmtheta^1}{\longleftrightarrow} \bmh_1 \overset{\bmg_\bmtheta^2}{\longleftrightarrow} \bmh_2 \cdots \overset{\bmg_\bmtheta^L}{\longleftrightarrow} \bmh_L := \bmz,$$
where $\bmg_\bmtheta = \bmg_\bmtheta^1 \circ \bmg_\bmtheta^2 \circ \cdots \circ \bmg_\bmtheta^L$.
By the change of variables formula, the log-likelihood of the model given $\bmy$ can be calculated as
\begin{align*}
    \log p_\bmtheta(\bmy) = \log p_\bmtheta(\bmz) + \sum_{l=1}^L \log |\det(d\bmh_l / d\bmh_{l-1}))|,
\end{align*}
where the log-determinant of the absolute value of the Jacobian term $\log |\det(d\bmh_l / d\bmh_{l-1}))|$ for each transform $(\bmg_\bmtheta^l)^{-1}$ can be easily computed for certain design of invertible layers~\cite{rezende2015variational, dinh2016density} similar to the Feistel cipher. Given training data of $\bmy$, the model can be optimized stably with maximum likelihood estimation.

\begin{figure}
    \centering
    \begin{subfigure}[b]{0.65\textwidth}
        \hspace{-1em}
        \includegraphics[width=1.1\textwidth]{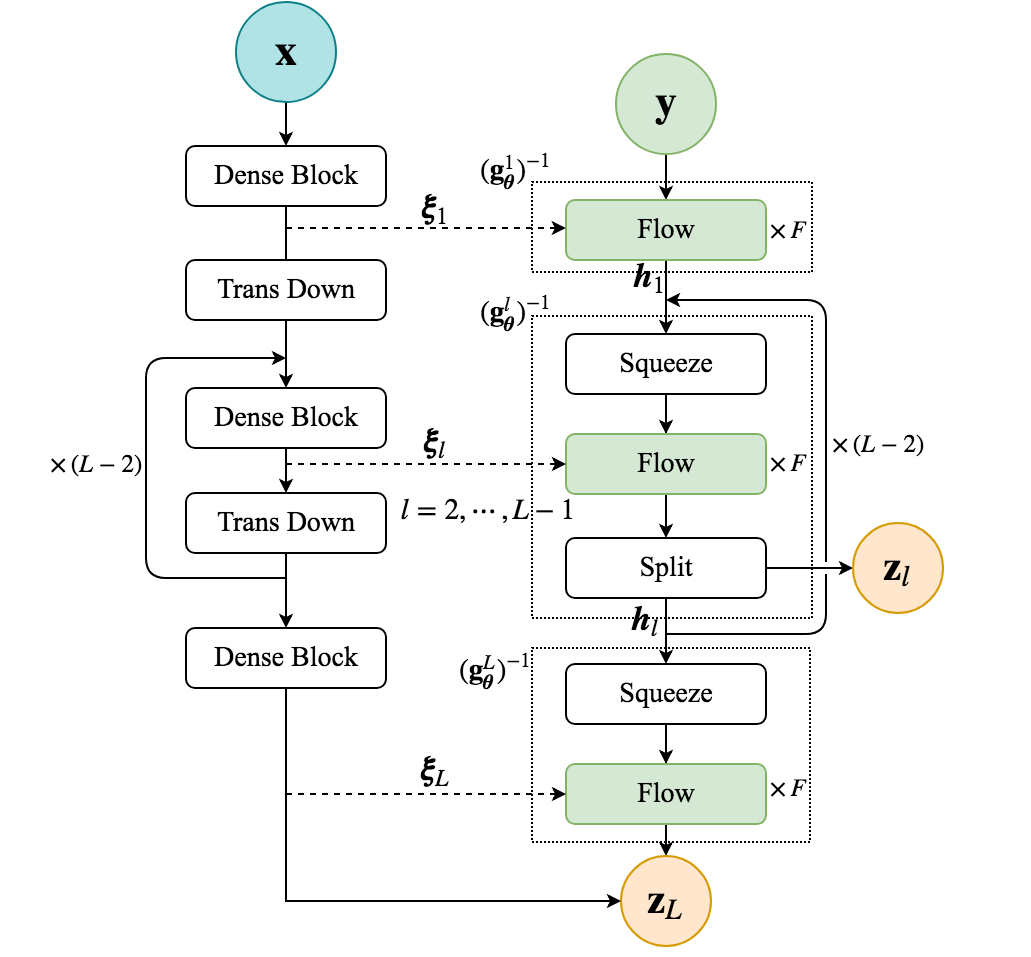}
        \caption{Mutliscale conditioning.}
        \label{fig:glow_msc_sub}
    \end{subfigure}
    \hfill
    \begin{minipage}[b]{0.33\textwidth}
        \begin{subfigure}[b]{\textwidth}
            \centering
            \includegraphics[width=.95\textwidth]{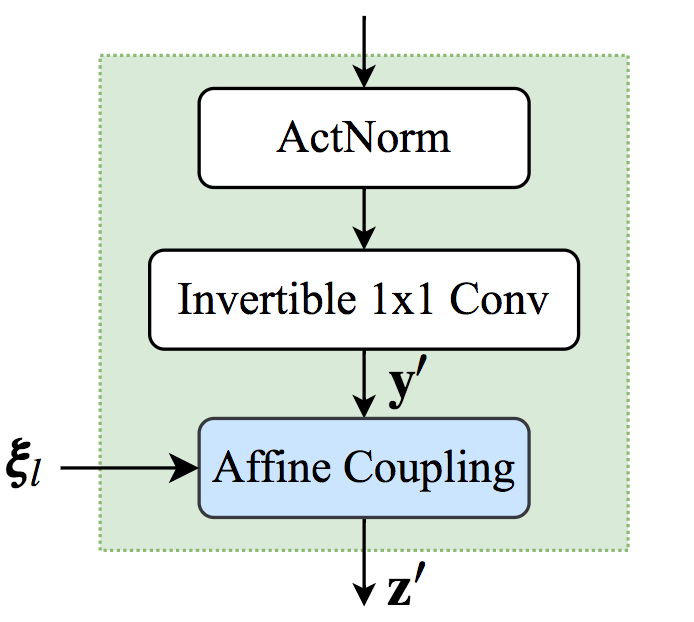}
            \caption{One step of flow.}
            \label{fig:flow_step}
        \end{subfigure}
        \\[\baselineskip]
        \begin{subfigure}[b]{\textwidth}
            \centering
            \includegraphics[width=0.98\textwidth]{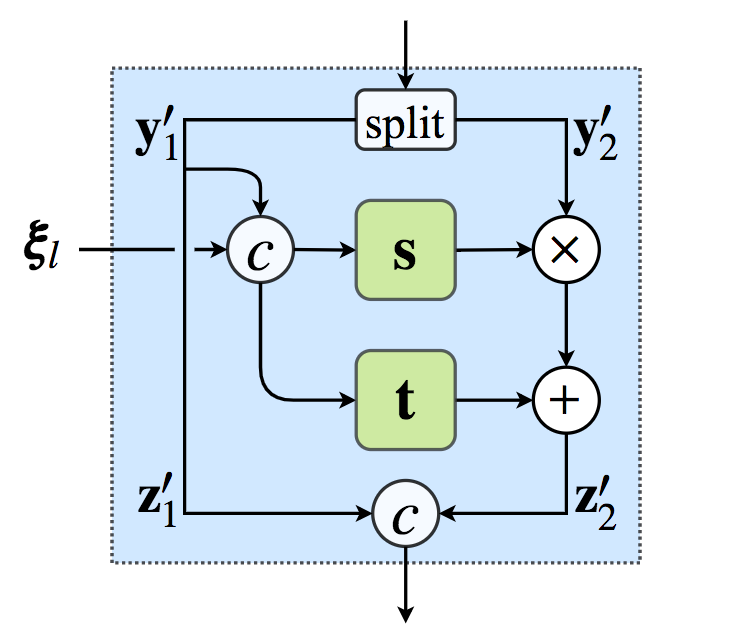}
            \caption{Affine coupling layer.}
            \label{fig:affine_coupling}
        \end{subfigure}
    \end{minipage}
    \caption{Multiscale conditional Glow. (a) Multiscale features extracted with the encoder network (left) are used as conditions to generate output with the Glow model (right). $\times F$, $\times (L-2)$ denotes repeating for $F$ times and $L-2$ times respectively. (b) One step of flow, i.e. \texttt{Flow} block in (a), and (c) Affine coupling layer following the structure of Glow (Fig.~$2$ in~\cite{kingma2018glow}) except conditioning on the input features. The figure shows the forward path from $\{\bmy; \bmx \}$ to $\bmz = \{\bmz_2, \cdots, \bmz_L\}$. The reverse (sampling) path from $\{\bmz; \bmx \}$ to $\bmy$ is used during training, where $\bmz$ are sampled from diagonal Gaussians, see Algorithm~\ref{algo:cglow}. See~\ref{appendix:cglow} for the details of all modules in the model.}
    \label{fig:glow_msc}
\end{figure}
A recently developed generative flow model called Glow~\cite{kingma2018glow} proposed to learn invertible $1\times1$ convolution to replace the fixed permutation and synthesize large photo-realistic images using  the log-likelihood objective.
We extend Glow to condition on high-dimensional input $\bmx$, e.g. images, as shown in Fig.~\ref{fig:glow_msc}. The conditional model consists of two components (Fig.~\ref{fig:glow_msc_sub}): an encoder network which extracts multiscale features $\{\bmxi_l\}_{l=1}^L$ from the input $\bmx$ through a cascade of alternating dense blocks and downsampling layers, and a Glow model (with multiscale structure) which transforms the latent variables $\bmz=\{\bmz_2, \cdots, \bmz_L\}$ distributed at different scales to the output $\bmy$ conditioned on $\{\bmxi_l\}_{l=1}^L$ through skip connections (dashed lines in Fig.~\ref{fig:glow_msc_sub}, as in Unet~\cite{ronneberger2015u}) between the encoder and the Glow.

More specifically, the input features $\bmxi_l$ enter the Glow model as the condition for the affine coupling layers at the same scale, as shown in Fig.~\ref{fig:flow_step}, whose input and output are denoted as $\bmy'$ and $\bmz'$ in the forward path. As shown in Fig.~\ref{fig:affine_coupling}, the input features $\bmxi_l$ are concatenated \encircle{c} with half of the flow features $\bmy_1'$ before passing to scale \fbox{$\bms$} and shift \fbox{$\bmt$} networks which specify arbitrarily nonlinear transforms that need not to be invertible. Given $\bmz' = [\bmz_1', \bmz_2']$ and $\bmxi_l$, $\bmy'=[\bmy_1', \bmy_2']$ can be recovered exactly by reversing the shift and scaling operations, as detailed in Table~\ref{tab:affine_coupling}. Note that $\bmxi_l$ is the condition for all $F$ steps of flow at scale $l=1, \cdots, L$, where $L$ denotes the number of scales (or levels). More details of the model including dense blocks, transition down layers, split, squeeze, and affine coupling layers are given in~\ref{appendix:cglow}.

In a data-driven scenario, the conditional Glow is trained by passing data $\bmy$ through the model to compute the latent $\bmz$ and maximizing the evaluated log-likelihood of data given $\bmx$. But to train with the loss in Eq.~(\ref{eq:rev_kl}), we need to sample the output $\hat{\bmy}$ from the conditional density $p_{\bmtheta}(\bmy | \bmx)$ given $\bmx$, which goes in the opposite direction of the data-driven case. Algorithm~\ref{algo:cglow} shows the details of training conditional Glow. 
The sampling/generation process is shown within the for-loop before computing the loss. Note that for one input sample only one output sample is used to approximate the expectation over $p_{\bmtheta}(\bmy | \bmx)$ during training. To obtain multiple output samples for an input e.g. to compute the predictive mean and variance during prediction, we only need to sample the noise variables $\{\bmepsilon_l\}_{l=2}^L$ multiple times, and pass them through the reverse path of the Glow.
The conditional log-likelihood $p_{\bmtheta}(\hat{\bmy} | \bmx)$ can be exactly evaluated as the following:
\begin{equation}
    \label{eq:flow_likelihood}
    \log p_{\bmtheta}(\hat{\bmy} | \bmx) = \log p_{\bmtheta}(\bmz) + \log |\det (d\bmz / d\hat{\bmy})|,
\end{equation}
where both the latent $\bmz$ and $\log |\det (d\bmz / d\hat{\bmy})|$ depend on $\bmx$ and realizations of the noise $\{\bmepsilon_l\}_{l=2}^L$. The density of the latent $p_{\bmtheta}(\bmz)$ is usually a simple distribution, e.g. diagonal Gaussian, which is computed with the second (for $\bmz_L$) and third (for $\{\bmz_l\}_{l=2}^{L-1}$) terms within the bracket of the reverse KL divergence loss in Algorithm~\ref{algo:cglow}. Also $\log |\det (d\bmz / d\bmy)|$ is computed with the fourth term. Notably, the log-determinant of the Jacobian for the affine coupling layer is just $\texttt{sum}(\log |\bms|)$, where $\bms$ is the output of the scaling network.
Thus the conditional density $p_\bmtheta(\hat{\bmy} | \bmx)$ can be evaluated exactly and efficiently, enabling us to directly approximate the entropy term in  Eq.~(\ref{eq:rev_kl}), e.g. via Monte Carlo approximation.

\remark{
The training process does not require output data. However, validation data with input-output pairs are necessary to calibrate the predictive uncertainty of the trained model. Careful initialization of the model is important to stabilize the training process. In this work,  we initialize the \texttt{ActNorm} to be the identity transform, the weight matrix of \texttt{Invertible $1\times1$ Convolution} to be a random rotation matrix, and the \texttt{Affine Coupling} layer to be close to the identity transform ($\hat{\bms}=\bmzero$ and $\bmt=\bmzero$ in Table~\ref{tab:affine_coupling}). 
We can also use data-dependent initialization to speed up the training process. More specifically, one mini-batch $\mc D_{\text{init}} = \{(\bmx^\sj, \bmr^\sj)\}_{j=1}^{M'}$ (e.g. $M'=32$) of input-output data pairs can be passed forward from $\{\bmy; \bmx\}$ to $\bmz$ to initialize the parameters of \texttt{ActNorm} such that the post-\texttt{ActNorm} activations per-channel have zero mean and unit variance given $\mc D_{\text{init}}$~\cite{kingma2018glow}. The reference output $\bmr$ can be the solution from standard deterministic PDE solvers or more appropriately here from the methods presented in Sections~\ref{sec:solve_pde} and~\ref{sec:solve_det_pde}.
}

\begin{algorithm}
\SetAlgoLined
\KwIn{Inverse temperature $\beta$, input samples $\{\bmx^\si\}_{i=1}^N$. Mini-batch size M, number of steps $F$ of \texttt{Flow} in each scale, number of scales $L$.}
\KwOut{Model parameters $\bmtheta$}
\For{number of training iterations}{
    Sample a mini-batch of input $\{\bmx^\si\}_{i=1}^M$, pass it through the encoder to compute the multiscale input features $\{\bmxi_l^\si\}_{l=1, i=1}^{L, M}$; 
    
    Sample the latent $\bmz^\si_L = \bmmu_\bmtheta^L(\bmxi^\si_L) + \bmsigma_\bmtheta^L(\bmxi^\si_L) \odot \bmepsilon_L^\si, \bmepsilon_L^\si \sim \mc N(\bmzero, \bmI)$; 
    
    Compute flow feature $\bmh_{L-1}^\si = \bmg_{\bmtheta}^L (\bmh_{L}^\si; \bmxi_L^\si)$; \algorithmiccomment{$\bmh_L = \bmz_L$, $\bmg_\bmtheta^L$ includes the reverse path of \texttt{Sequeeze} and $F$ steps of \texttt{Flow}} 
    
    \For{$l=L-1:2$}{
        Sample the split latent variable at level $l$ $\bmz_l^\si = \bmmu_\bmtheta^l(\bmh^\si_l) + \bmsigma_\bmtheta^l(\bmh^\si_l) \odot \bmepsilon_l^\si, \bmepsilon_l^\si \sim \mc N(\bmzero, \bmI), i=1, \cdots, M$; 
        
        Compute flow feature $\bmh_{l-1}^\si = \bmg_{\bmtheta}^l (\bmh_l^\si, \bmz_l^\si; \bmxi_l^\si)$;  \algorithmiccomment{$\bmg_\bmtheta^l$ includes the reverse path of \texttt{Sequeeze}, $F$ steps of \texttt{Flow} and \texttt{Split}}
        }
    
    Compute output $\hat{\bmy}^\si = \bmg_\bmtheta^1(\bmh_1^\si; \bmxi_1^\si)$;
    \algorithmiccomment{$\bmg_\bmtheta^1$ includes the reverse path of $F$ steps of \texttt{Flow}}
    
    Minimize the \textit{reverse KL} divergence in Eq.~(\ref{eq:rev_kl}) with Adam optimizer w.r.t. $\bmtheta$ 
        $ \frac{1}{M} \sum_{i=1}^M \Big[ \beta L(\hat{\bmy}^\si, \bmx^\si) 
        + \sum_{l=2}^{L-1} \log \mc N(\bmz_l^\si | \bmmu_\bmtheta^l(\bmh_l^\si), (\bmsigma_\bmtheta^l(\bmh_l^\si))^2) 
         + \log \mc N(\bmz_L^\si | \bmmu_\bmtheta^L(\bmxi_L^\si), (\bmsigma_\bmtheta^L(\bmxi_L^\si))^2) 
        + \sum_{l=1}^L \log |\det(d\bmh_l^\si / d\bmh_{l-1}^\si)| \Big].$

    \algorithmiccomment{$\bmh_0=\hat{\bmy}$, see Table 1 in~\cite{kingma2018glow} for formula to compute $\log |\det(d\bmh_l^\si / d\bmh_{l-1}^\si)|$, i.e. log-determinant of Jacobian for \texttt{ActNorm}, \texttt{Invertible $1\times 1$ Conv} and \texttt{Affine Coupling} layer.}
}
\caption{Training conditional Glow.}
\label{algo:cglow}
\end{algorithm}

\section{Numerical Experiments}\label{sec:numerical_experiments}

\paragraph{Model problem} Steady-state flow in random heterogeneous media is studied as the model problem throughout the experiments, as in Eqs.~(\ref{eq:darcy}),~(\ref{eq:darcy_mixed}),~(\ref{eq:BCs}). We consider the domain $\mc S = [0, 1] \times [0, 1]$, the left and right boundaries are Dirichlet, with pressure values $1$ and $0$, respectively. The upper and lower boundaries are Neumann, with zero flux. The source field is zero.

\paragraph{Dataset} Only input samples are needed to train the physics-constrained surrogates (PCSs). Additional simulated output data for training data-driven surrogates (DDSs) and evaluating surrogate performance are obtained with FEniCS~\cite{AlnaesBlechta2015a}. Here, we mainly introduce three types of input datasets, which are Gaussian random field (GRF), warped GRF, and channelized field.
\begin{figure}
    \centering
    \includegraphics[width=0.9\textwidth]{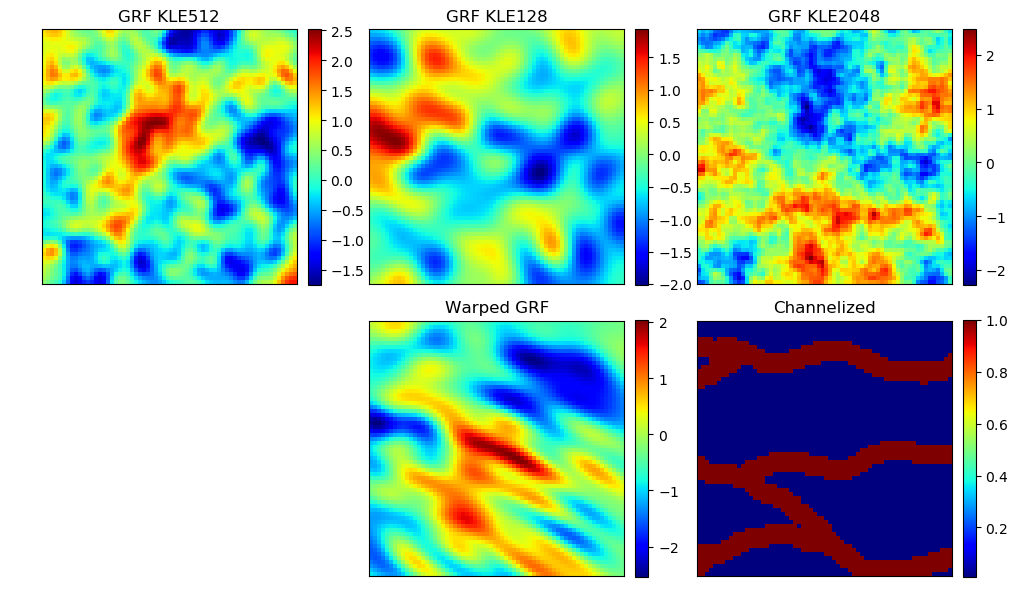}
    \caption{Samples from $5$ test input distributions over a $64\times 64$ uniform grid, i.e. GRF KLE$512$, GRF KLE$128$, GRF KLE$2048$, warped GRF, channelized field. Log   permeability samples are shown except the last channelized field that is defined with binary values $0.01$ and $1.0$.}
    \label{fig:generalization_test_input_samples}
\end{figure}

The first input dataset is the exponential of a GRF, i.e. $K(s) = \exp(G(s))$, $G(\cdot) \sim \mathcal{GP}(0, k(\cdot, \cdot))$, where $k(s, s') = \exp(-\norm{s - s'}_2 / l)$, $l$ is the length scale. The field realization is generated with Karhunen-Lo\`eve expansion (KLE) with the leading $N$ terms, paired with Latin hypercube sampling. See Section 4.1 in~\cite{zhu2018bayesian} for more details. This type of dataset is called \texttt{GRF KLE$N$}. For the deterministic surrogate experiments in Section~\ref{sec:deterministic_surrogate}, the training input GRF KLE$512$ is generated with length scale $l=0.25$, $N=512$ leading terms, discretized over a $64 \times 64$ uniform grid,  which accumulates $95.04\%$ energy. For the probabilistic surrogate in Section~\ref{sec:prob_surrogate}, the parameters for the training input GRF KLE$100$ are $N=100$, $l=0.2$, over $32 \times 32$ uniform grid.
The test set may have other KLE truncations, but with the same length scale in each case, i.e. $l=0.25$ for $64 \times 64$, and $l=0.2$ for $32\times 32$. The dataset for uncertainty propagation consists of 10,000 input-output data pairs unseen during training.

A slightly different test input is \texttt{warped GRF}, where there are two Gaussian fields and the output of the first GRF is the input to the second GRF. The kernel for both GRFs is squared exponential kernel, the length scale and KLE terms are $2, 16$ for the first GRF and $0.1, 128$ for the second GRF.

The last type of input field considered is a  \texttt{channelized} field. Samples are obtained by cropping $64\times 64$ patches from one large training image~\cite{laloy2018training} of size $2500\times 2500$, or $32\times 32$ patches from the resized $1250\times 1250$ image (resized with nearest neighborhood). Typical samples of the input datasets considered are shown in Fig.~\ref{fig:generalization_test_input_samples}.

We begin our experiments by solving deterministic PDEs with spatially-varying coefficient (input) with convolutional decoder networks, and compare with FC-NNs. Then we show experiments for surrogate modeling for solving random PDEs, and compare with the data-driven approach. The last part is on experiments of using the conditional Glow as our probabilistic surrogate for uncertainty quantification tasks. The code and datasets for this work will become available at \url{https://github.com/cics-nd/pde-surrogate} upon publication.

\subsection{Solving Deterministic PDEs}
\label{sec:solve_det_pde}
In this section, we explore the relative merit of using CNNs and FC-NNs to parameterize the solutions of deterministic PDEs with image-like input field, including both linear and nonlinear PDEs. Since our focus is on surrogate modeling, the results below are mostly qualitative. The network architectures and training details are described in~\ref{appendix:solve}.

\begin{figure}[h!]
    \centering
    \begin{subfigure}[b]{0.48\textwidth}
        \centering
        \includegraphics[width=\textwidth]{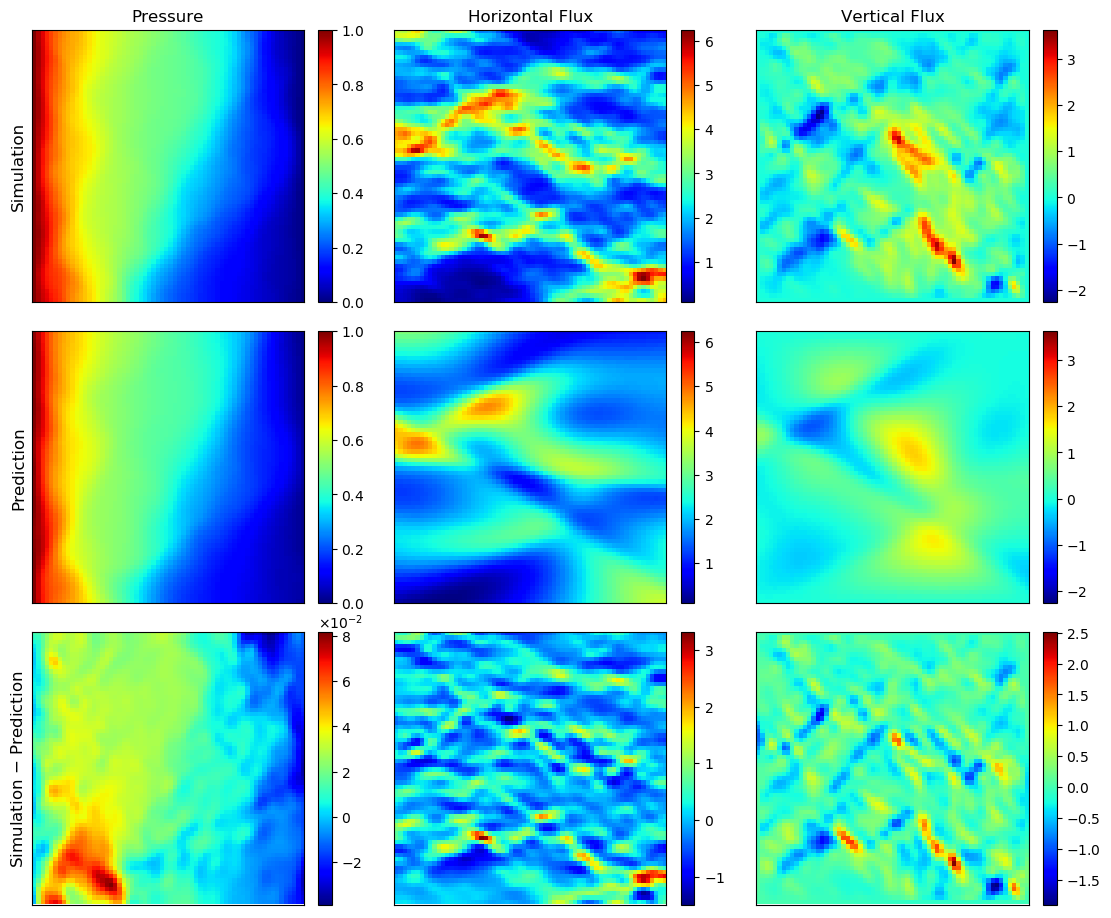}
        \caption{FC-NN, iteration $500$.}
    \end{subfigure}
    ~
    \begin{subfigure}[b]{0.48\textwidth}
        \centering
        \includegraphics[width=\textwidth]{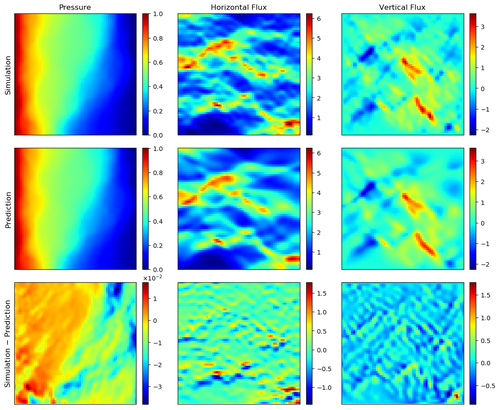}
        \caption{FC-NN, iteration $2000$.}
    \end{subfigure}
    
    \begin{subfigure}[b]{0.48\textwidth}
        \centering
        \includegraphics[width=\textwidth]{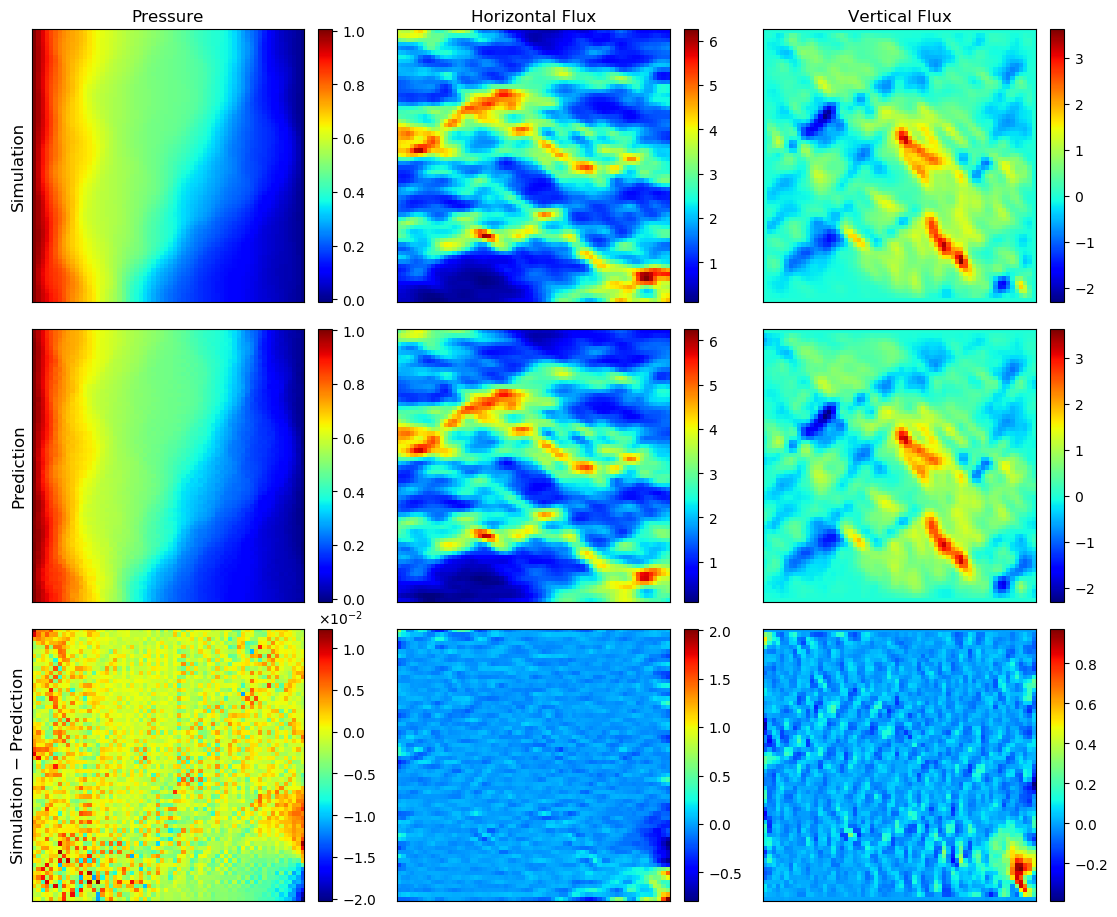}
        \caption{CNN, iteration $250$.}
    \end{subfigure}
    ~
    \begin{subfigure}[b]{0.48\textwidth}
        \centering
        \includegraphics[width=\textwidth]{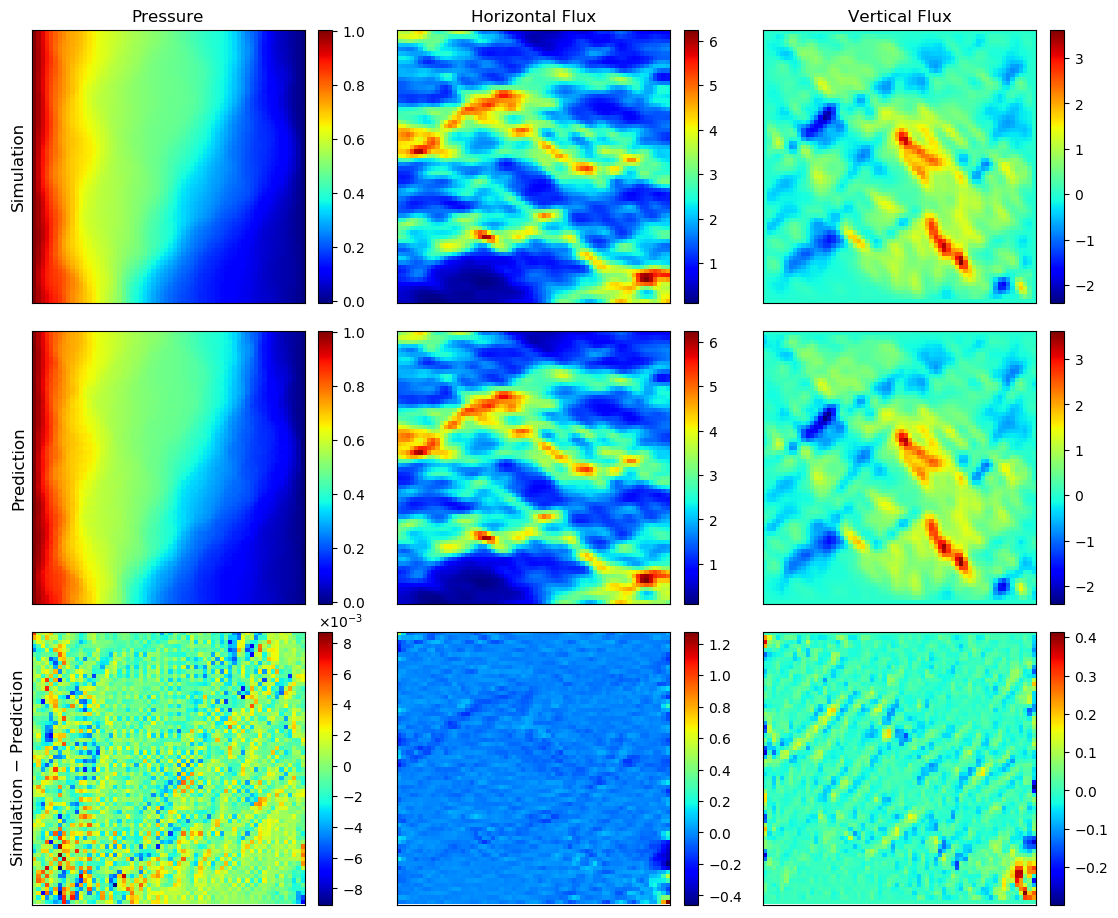}
        \caption{CNN, iteration $500$.}
    \end{subfigure}

    \caption{Solving Darcy flow for one sample from GRF KLE1024 under mixed residual loss. FC-NN takes much longer to resolve the fine details of flux fields, while the pressure field does not improve much. CNN obtains more accurate solution within shorter time.}
    \label{fig:solve_k1024}
\end{figure}
\begin{figure}[h!]
    \centering
    \begin{subfigure}[b]{0.48\textwidth}
        \centering
        \includegraphics[width=\textwidth]{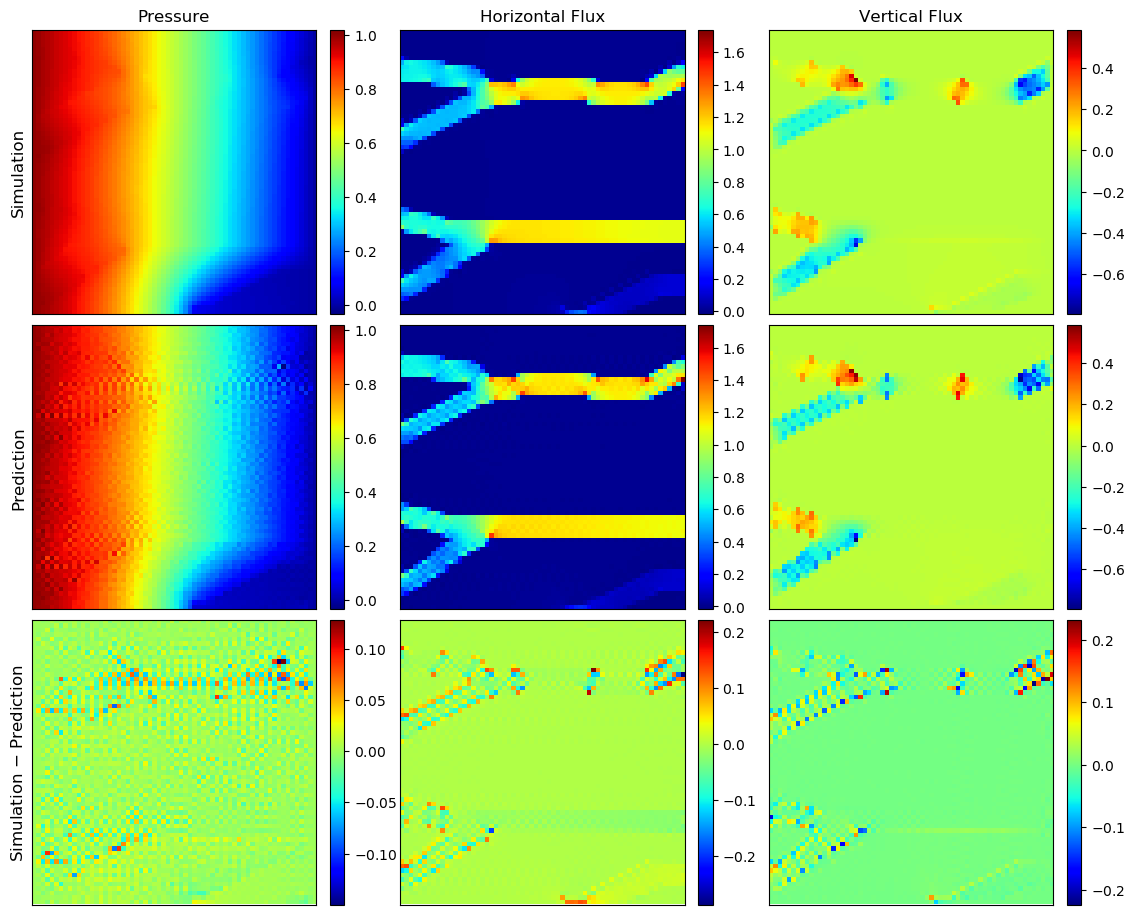}
        \caption{CNN, iteration $500$.}
    \end{subfigure}
    ~
    \begin{subfigure}[b]{0.48\textwidth}
        \centering
        \includegraphics[width=\textwidth]{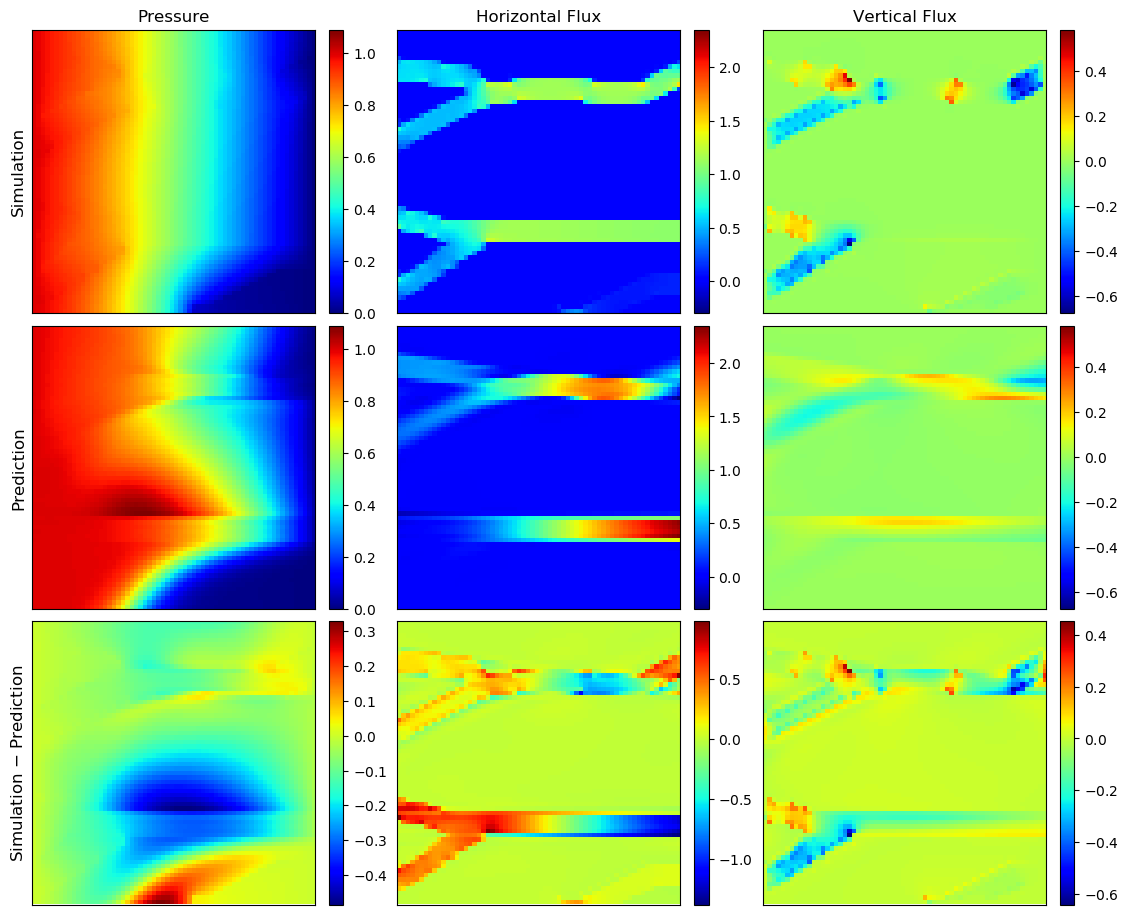}
        \caption{FC-NN, iteration $2000$.}
    \end{subfigure}
    \caption{Solving Darcy flow for one sample of the channelized field. The same network and training setup as in Fig.~\ref{fig:solve_k1024} are used. The FC-NN parameterization fails to converge.}
    \label{fig:solve_channel}
\end{figure}

\paragraph{Comparison of CNNs and FC-NNs to solve Darcy flow} 
We compare convolutional decoder networks and fully-connected networks presented in Section~\ref{sec:solve_det_pde} to solve the PDE system in Eq.~(\ref{eq:darcy}). The input permeability field is sampled from GRF KLE$1024$ over a $64 \times 64$ uniform grid. We optimize the CNN and the FC-NN with mixed residual loss using L-BFGS optimizer for $500$ and $2000$ iterations, respectively. The results are shown in Fig.~\ref{fig:solve_k1024}. The solution learned with CNN in iteration $250$ is even better than the solution learned with FC-NN in iteration $2000$, in terms of accuracy and retaining multiscale features of the flux fields. The same phenomenon is observed for input GRFs with other intrinsic dimensionalities. We further experiment on input sampled from the  channelized field, as shown in Fig.~\ref{fig:solve_channel}. For this  case, however, we observe that  the FC-NN fails to converge to a small enough error in contrast to the CNN.

The experiments on solving deterministic PDEs show that CNNs can capture the multiscale features of the solution much more effectively than the FC-NNs, as reflected by the resolved flux fields. This is mostly because of the difference in their parameterizations of a field solution and the ways to obtain spatial gradients. FC-NNs turn to generate images that look like light-paintings\footnote{https://distill.pub/2018/differentiable-parameterizations/\#section-xy2rgb}, but not rugged field. More broadly, this type of parameterization is intensively explored named compositional pattern producing networks~\cite{stanley2007compositional}.
CNNs can represent images with multiscale features quite efficiently as is evident in our experiments and the rapid advances in image generation applications. 
Due to the discretization of spatial gradients with Sobel filters, the error of the learned solution is mainly on the boundaries, and the checkerboard artifact becomes more severe in the pressure field as the flux fields becomes more rugged, as shown in Fig.~\ref{fig:solve_k4096} in~\ref{appendix:solve}.

\paragraph{Nonlinear flow in porous media}
\begin{figure}[h]
    \centering
    \begin{subfigure}[b]{0.48\textwidth}
        \centering
        \includegraphics[width=\textwidth]{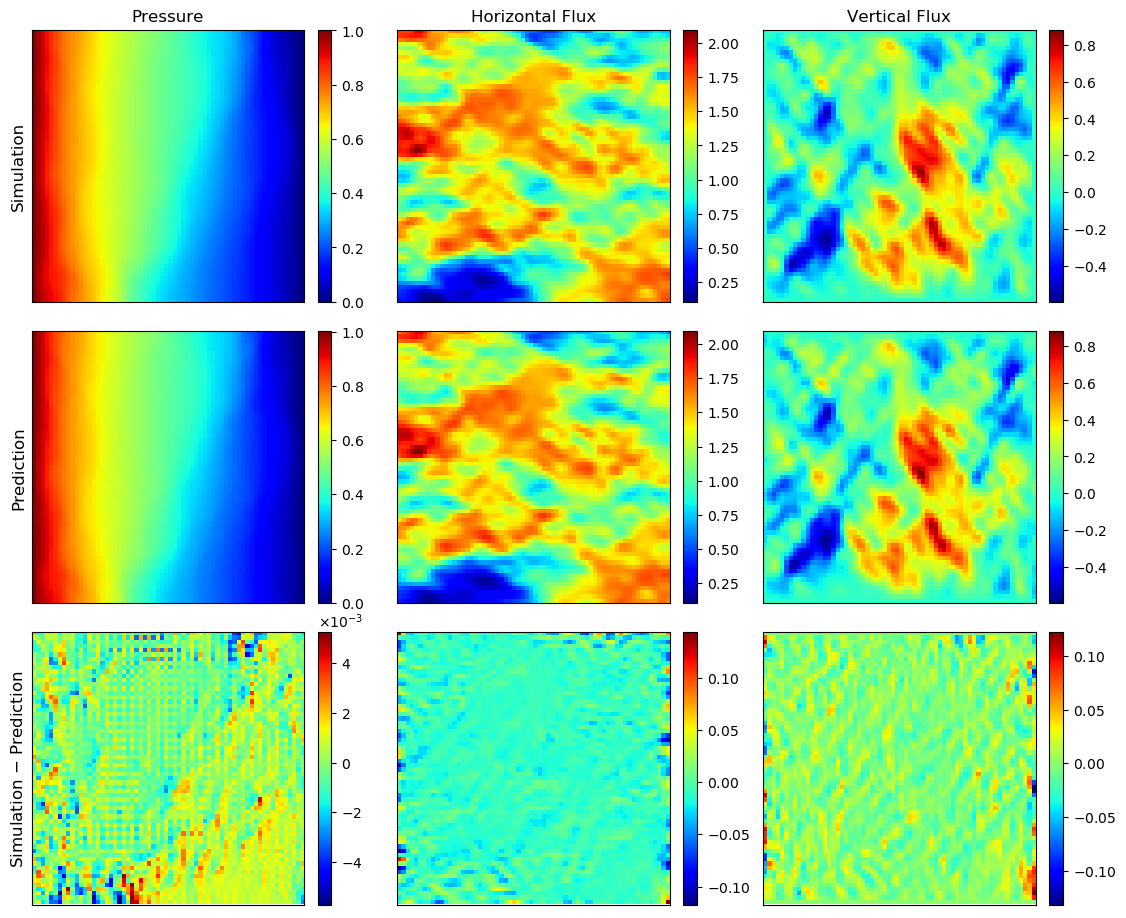}
        \caption{GRF KLE$1024$.}
    \end{subfigure}
    ~
    \begin{subfigure}[b]{0.48\textwidth}
        \centering
        \includegraphics[width=\textwidth]{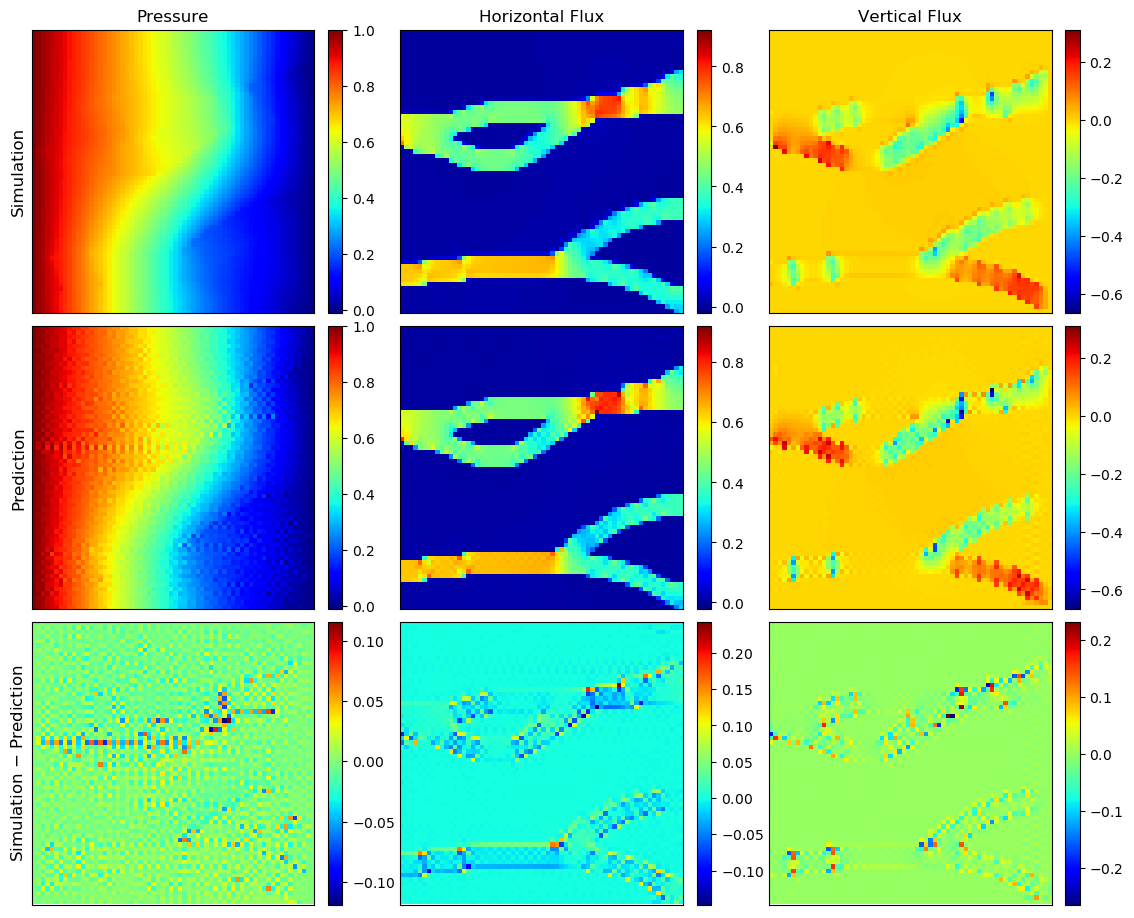}
        \caption{Channelized field.}
    \end{subfigure}
    \caption{Simulation (FEniCS) and learned solution (prediction) with CNN for the nonlinear flow for (a) GRF KLE1024 with $\alpha_1=0.1$ and $\alpha_2=0.1$, and (b) channelzied field with $\alpha_1=1.0$ and $\alpha_2=1.0$.}
    \label{fig:darcy_nonlinear_solve}
\end{figure}
Darcy's law $\bmtau = - K \nabla u$ is a well established linear constitutive relationship for flow through porous media when the Reynolds number $Re$ approaches zero. It has been shown both theoretically~\cite{mei1991effect} and experimentally~\cite{firdaouss1997nonlinear, rojas1998nonlinear} that the constitutive relation undergoes a cubic transitional regime at low $Re$, and then a quadratic Forchheimer~\cite{forchheimer1901wasserbewegung} when $Re\sim O(1)$. To show that our approach also works for nonlinear PDEs, we look at the nonlinear correction of Darcy's law as the following
\begin{equation}\label{eq:constitutive_nonlinear}
    - \nabla u = \frac{1}{K} \bmtau + \frac{\alpha_1}{K^{\frac{1}{2}}} \bmtau^2 + \alpha_2 \bmtau^3,
\end{equation}
where $\alpha_1, \alpha_2$ are usually obtained by fitting to experiment data.
We use CNNs to solve this nonlinear flow with the constitutive Eq.~(\ref{eq:constitutive_nonlinear}), the continuity equation $\nabla \cdot \bmtau = 0$ and the same boundary condition with the linear Darcy case. The reference solution is obtained with FEniCS (dual mixed formulation with Newton solver that converges in $5 \sim 6$ iterations with relative tolerance below $10^{-6}$). We experiment on input fields from GRF KLE$1024$ and the channelized field, with $\alpha_1=0.1$ and $\alpha_2=0.1$ in the first case, and $\alpha_1=1.0$ and $\alpha_2=1.0$ in the second case.
The convolutional decoder network is the same as in the previous section, and is trained with mixed residual loss. The results is shown in Fig.~\ref{fig:darcy_nonlinear_solve}.

For GRF KLE$1024$, the effect of the cubic constitutive relation is actually smoothing out the flux field in comparison to the linear case in Fig.~\ref{fig:solve_k1024} using the same input field. The nonlinearity of PDEs does not seem to increase the burden for the CNN training except for a few more steps of forward and backpropagation due to the nonlinear operations in the constitutive equation. This is a negligible cost w.r.t. the computations in the decoder network itself. However, note that solving nonlinear PDEs with the Newton solver requires $N$ iterations, thus increasing the computation by $N$ times.
For surrogate modeling, the mapping that the CNN learns from $K$ to $u$ is nonlinear even when the PDE to solve is linear.
We expect it will be easier to learn a surrogate in the nonlinear case due to the smoother output fields. We leave further investigation of surrogate modeling and uncertainty quantification for nonlinear stochastic PDEs for our future work.

\subsection{Deterministic Surrogate}\label{sec:exp_det_surrogate}

The experiments in solving deterministic PDEs lead us to choose CNNs over FC-NNs for surrogate modeling, with less training time and comparable accuracy, especially for high-dimensional input. We train both the physics-constrained surrogates and data-driven surrogates, and compare their accuracy and generalizability.

\begin{figure}[h]
    \centering
    \includegraphics[width=1\textwidth]{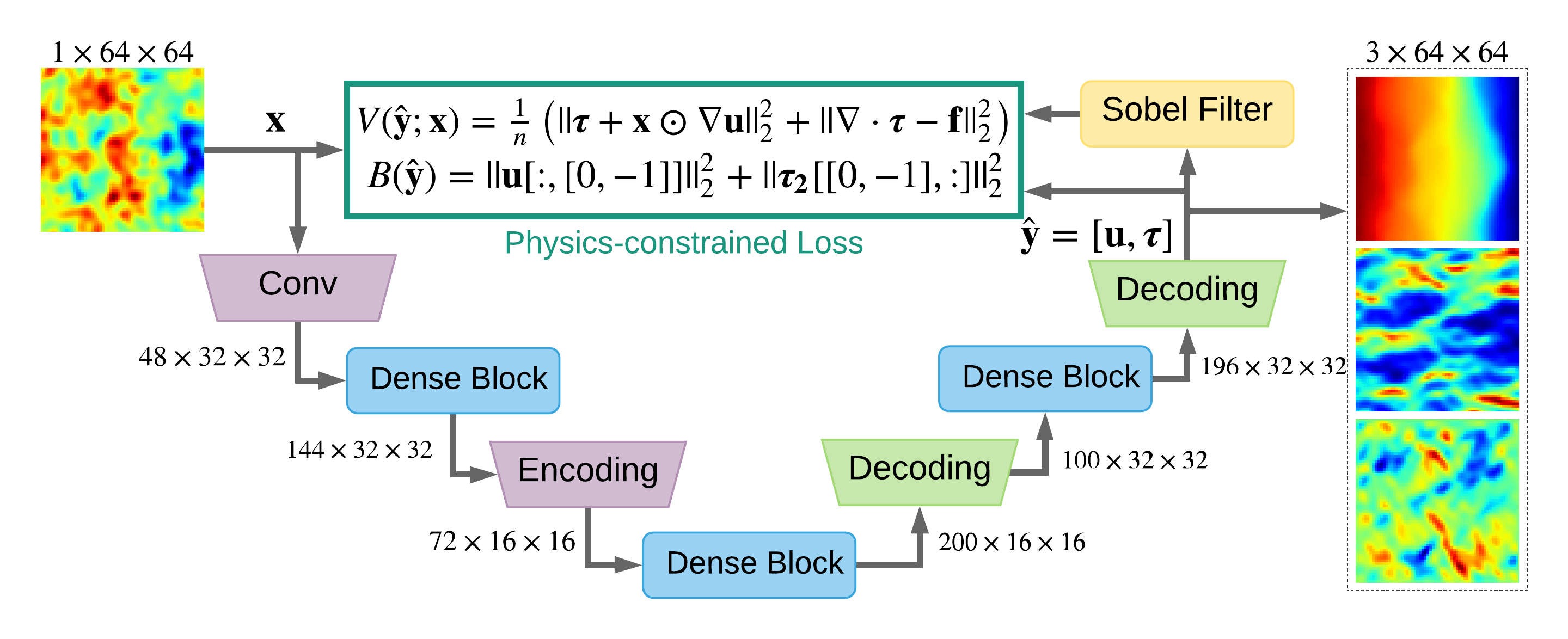}
    \caption{Dense convolutional encoder-decoder network as the deterministic surrogate. The model's input is the realization of a random field, the model's output is the prediction for each input field including 3 output fields, i.e. pressure and two flux fields. The model is trained with physics-constrained loss without target data.}
    \label{fig:codec}
\end{figure}
\paragraph{Network}
Dense convolutional encoder-decoder network~\cite{zhu2018bayesian} is used as the surrogate model, with one input channel $\bmx$ and three output channels $[\bmu, \bmtau_1, \bmtau_2]$, as shown in Fig.~\ref{fig:codec}.
The upsampling method in the decoding layers in the current implementation is nearest upsampling followed by convolution, different from transposed convolution used in the data-driven case. This is essential to avoid the checkerboard effect\footnote{https://distill.pub/2016/deconv-checkerboard/}, partially severed by Sobel filter besides the natural tendency of transposed convolution.
The resolution of the input fields is reduced by 4 times through the encoding path, from $64\times 64$ to $16 \times 16$, then increased to the size of the output fields, $64\times 64$. The number of layers in the three dense blocks are $6, 8, 6$, with growth rate $16$. There are $48$ initial feature maps after the first convolution layer.

\begin{figure}[h]
    \centering
    \begin{subfigure}[b]{0.48\textwidth}
        \centering
        \includegraphics[width=\textwidth]{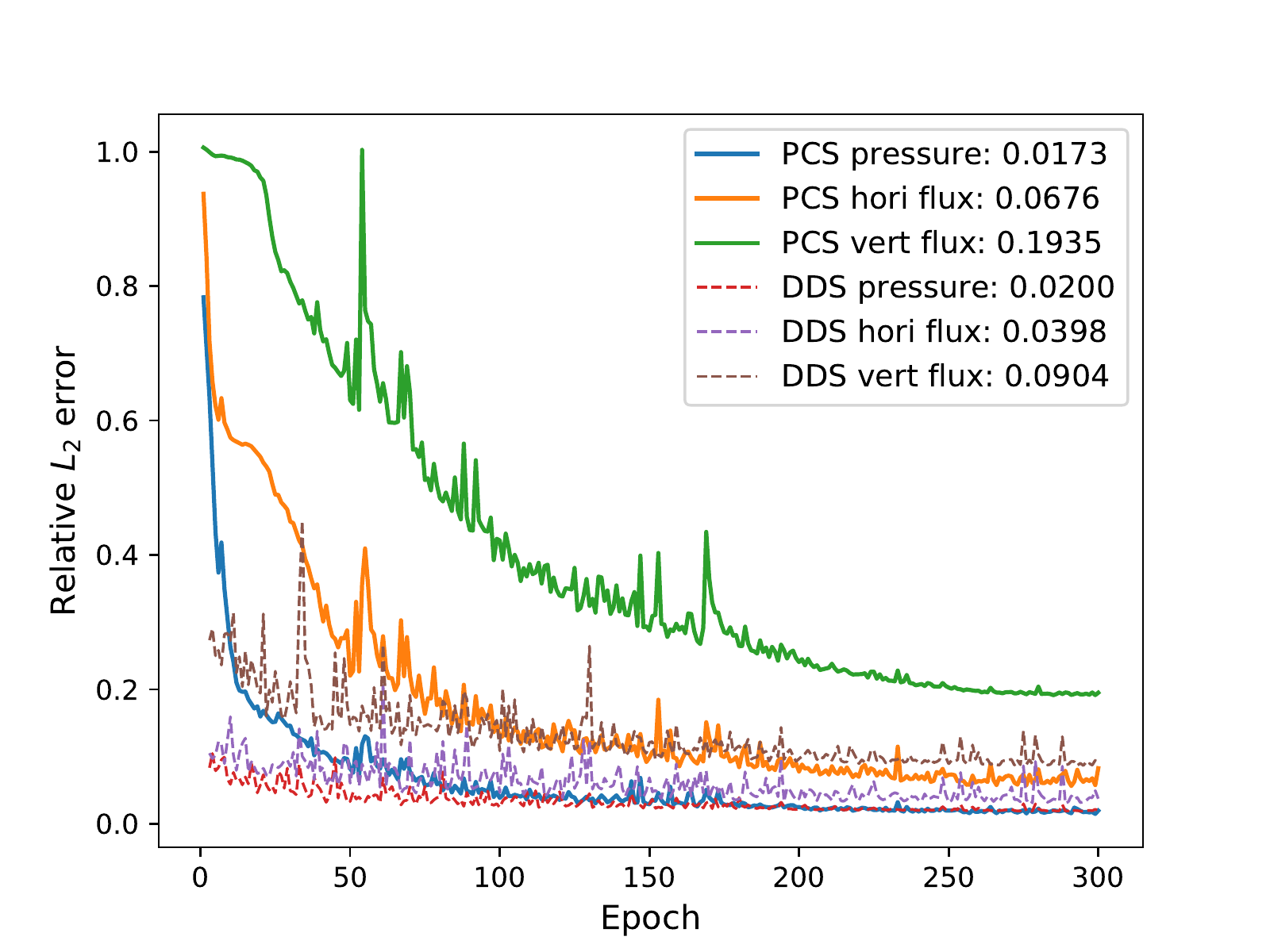}
        \caption{Test relative $L_2$ error.}
        \label{fig:test_nrmse_training_curve}
    \end{subfigure}
    ~
    \begin{subfigure}[b]{0.48\textwidth}
        \centering
        \includegraphics[width=\textwidth]{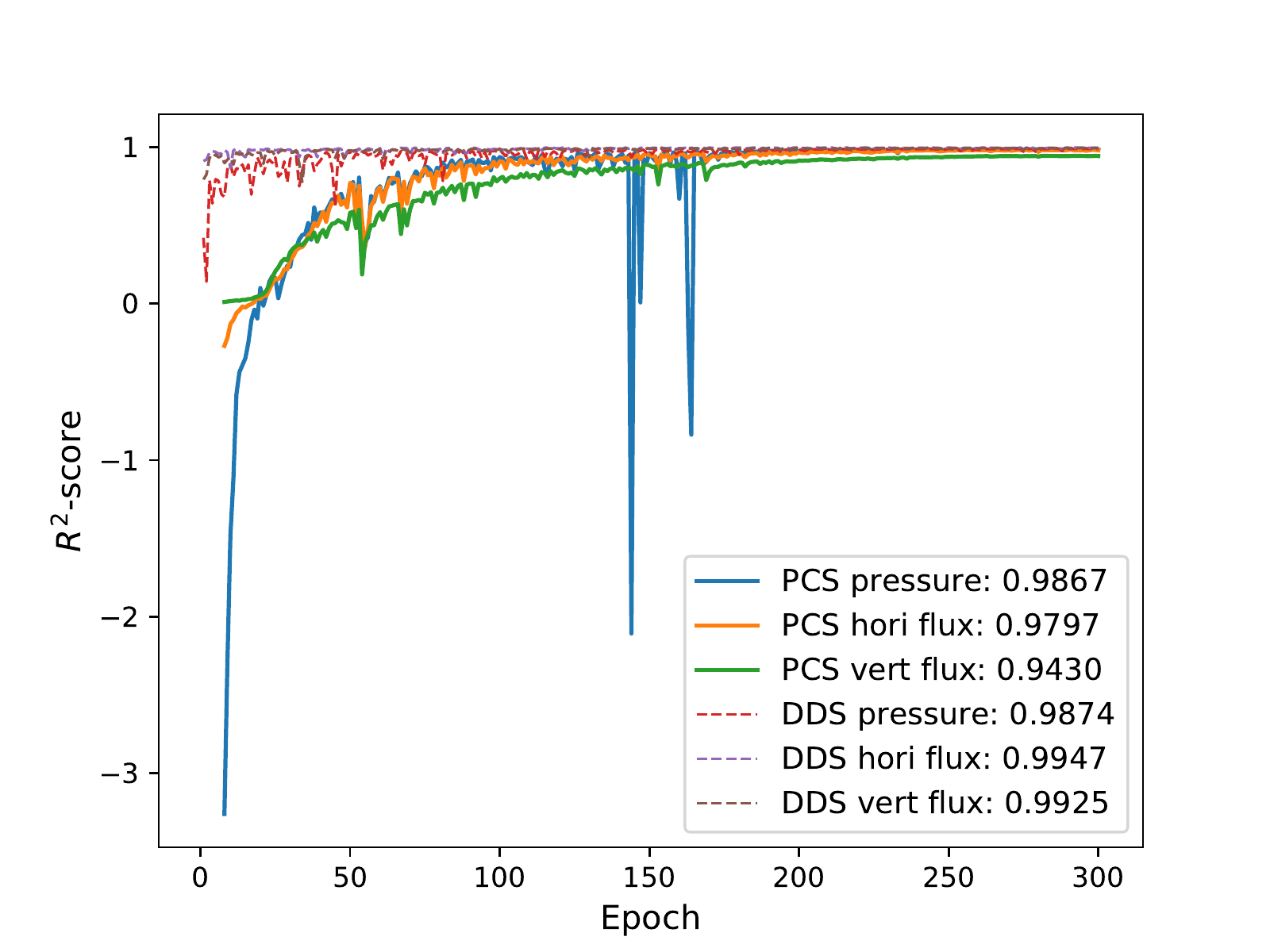}
        \caption{Test $R^2$ score.}
        \label{fig:test_r2_training_curve}
    \end{subfigure}
    \caption{Test relative $L_2$ error and $R^2$ score during training. The solid lines shows the error for the PCSs and the dashed lines for the DDSs. Both surrogates are trained on $8192$ samples of GRF KLE$512$ and tested on the same $512$ samples of GRF KLE$512$.}
    \label{fig:training_curve}
\end{figure}

\paragraph{Training}
We train the PCS with mixed residual loss as in Eq.~(\ref{eq:mixed_residual_impl}) with only input data, and compare it with the DDS with the same network architecture but trained with additional output data. The number of training data, mini-batch size and the category of test distributions vary in different experiments, but all with $T=512$ test data and employing the Adam~\cite{kingma2014adam} optimizer paired with one cycle policy\footnote{\url{https://github.com/fastai/fastai/blob/master/fastai/callbacks/one\_cycle.py}} (learning rate scheduler) where the maximum learning rate is $0.001$. The mini-batch size ranges from $8$ to $32$ depending on the number of training data. The weight coefficient for the boundary conditions is $\lambda = 10$. 
The evaluation metrics for prediction are relative $L_2$ error and $R^2$ score,
\begin{equation}
    \epsilon_j = \frac{1}{T} \sum_{i=1}^T \frac{\norm{\hat{\bmy}_j^\si - \bmy_j^\si}_2}{\norm{\bmy_j^\si}_2}, \qquad R^2_j = 1 - \frac{\sum_{i=1}^T \norm{\hat{\bmy}_j^\si - \bmy_j^\si}_2^2}{\sum_{i=1}^T \norm{\bmy_j^\si - \bar{\bmy}_j}_2^2},
\end{equation}
where $\hat{\bmy}_j^\si$ is the surrogate prediction of the $j$-th output channel/field ($j=1,2,3$ for pressure, horizontal flux and vertical flux field respectively), $\bmy_j^\si$ is the corresponding simulator output, $\bar{\bmy}_j=\frac{1}{T}\sum_{i=1}^T \bmy_j^\si$, $T$ is the total number of test inputs, $\norm{\cdot}_2$ is the $L_2$ norm. We mainly use relative $l_2$ error as evaluation metric.
The PCS is trained for $300$ epochs and the DDS is trained for $200$ epochs, since DDS is faster to converge than the PCS in general, as shown in Fig.~\ref{fig:training_curve}.

\paragraph{Prediction} 

\begin{figure}[h!]
    \centering
    \begin{subfigure}[b]{0.48\textwidth}
        \centering
        \includegraphics[width=\textwidth]{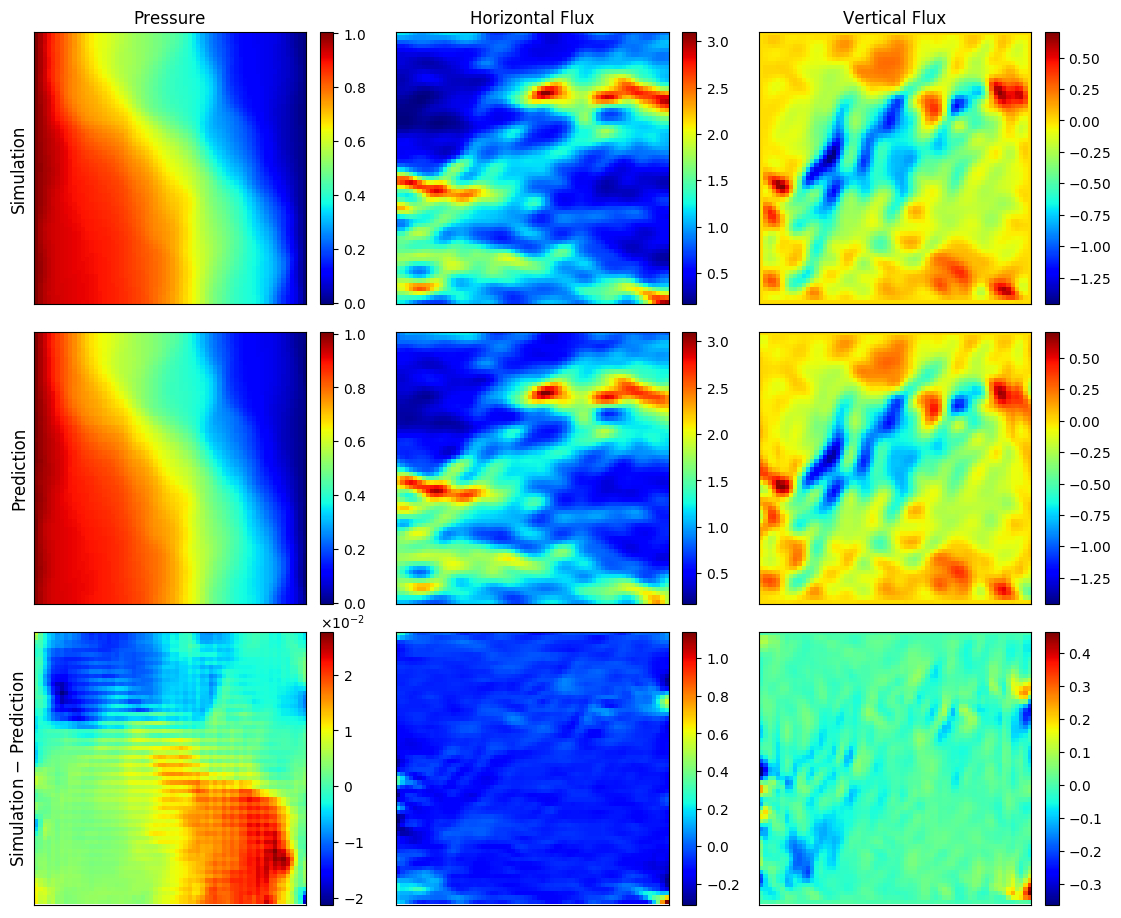}
        \caption{GRF KLE$512$, test $1$.}
    \end{subfigure}
    ~
    \begin{subfigure}[b]{0.48\textwidth}
        \centering
        \includegraphics[width=\textwidth]{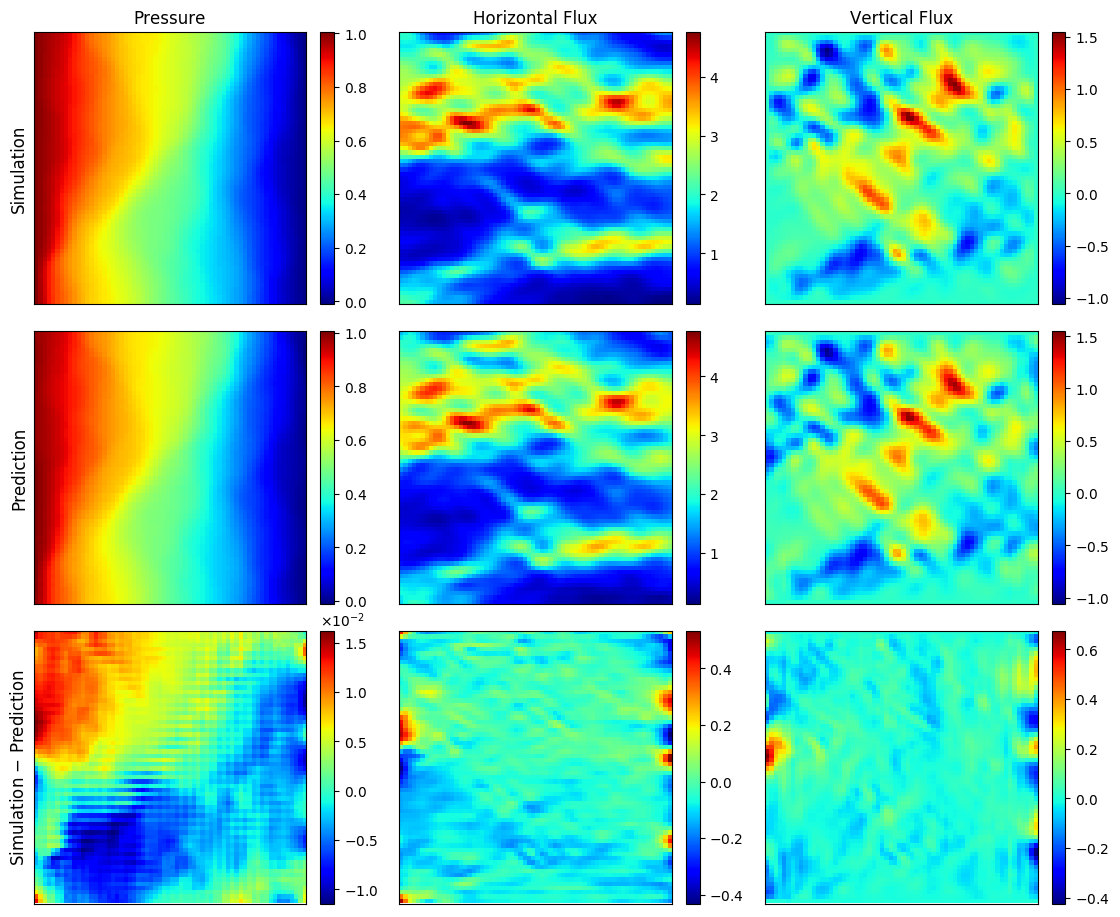}
        \caption{GRF KLE$512$, test $2$.}
        \label{fig:PCS_mixed_res_pred_examples_kle512_sample2}
    \end{subfigure}
    
    \begin{subfigure}[b]{0.48\textwidth}
        \centering
        \includegraphics[width=\textwidth]{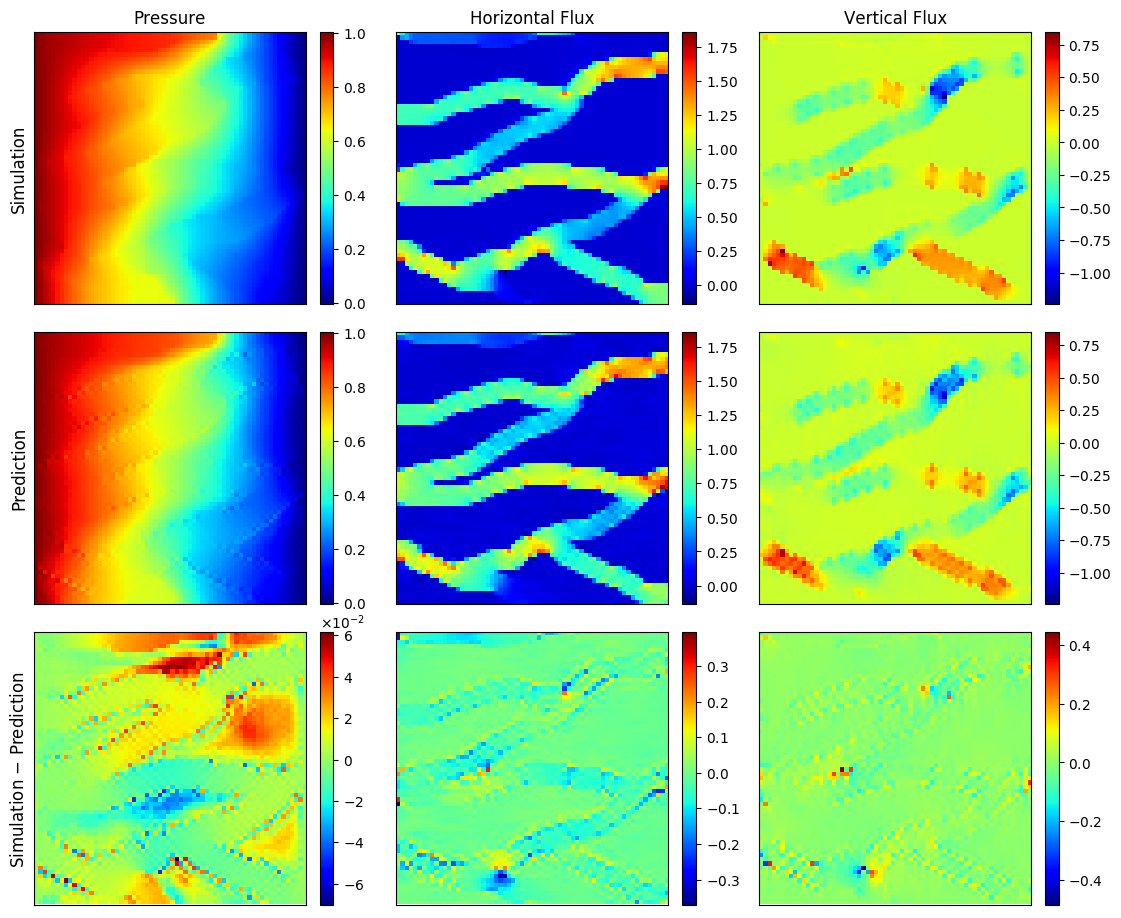}
        \caption{Channelized, test $1$.}
    \end{subfigure}
    ~
    \begin{subfigure}[b]{0.48\textwidth}
        \centering
        \includegraphics[width=\textwidth]{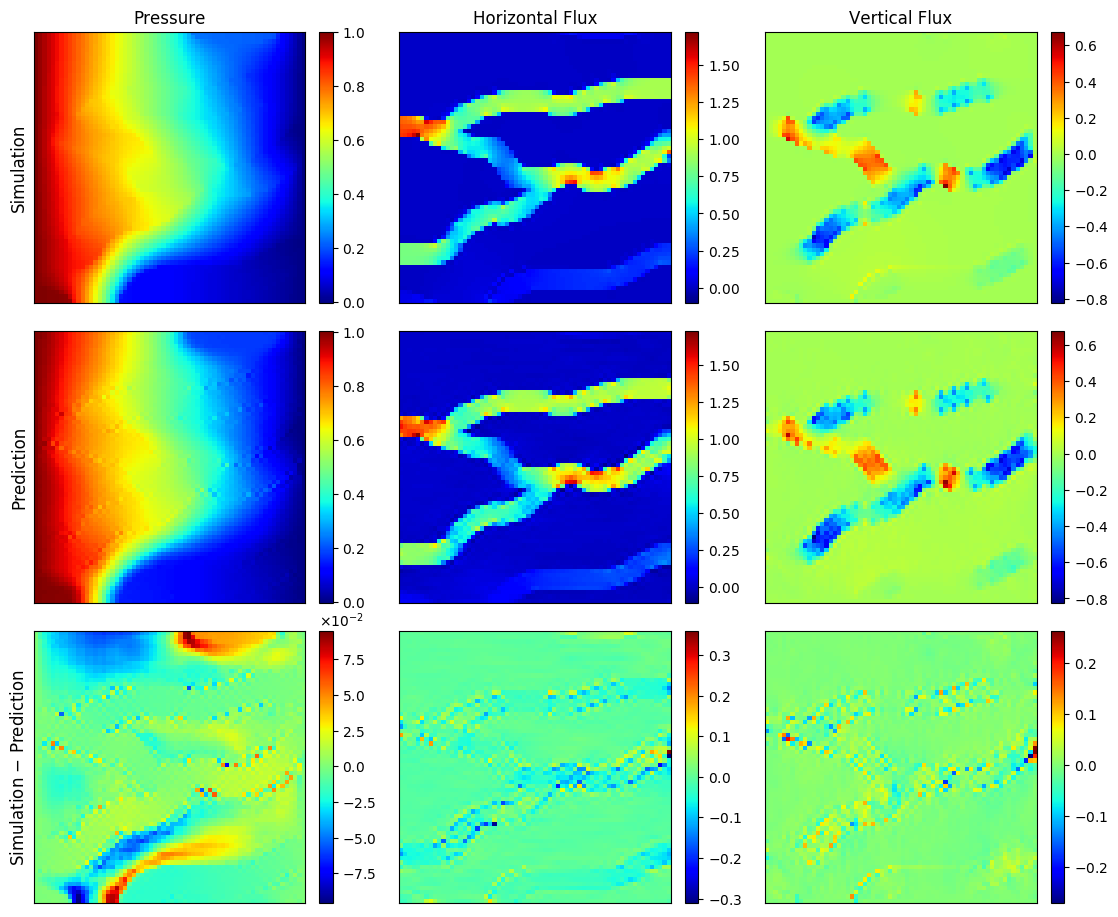}
        \caption{Channelized, test $2$.}
    \end{subfigure}
    
    \caption{Prediction examples of the PCS under the mixed residual loss. (a) and (b) are $2$ test results for the PCS trained with $8192$ samples of GRF KLE$512$; (c) and (d) are $2$ test results for the PCS trained with $4096$ samples of channelized fields.}
    \label{fig:PCS_mixed_res_pred_examples}
\end{figure}

To show that the physics-constrained approach to learn surrogate works well, we train the PCS on two datasets, i.e. GRF KLE$512$ ($8192$ samples) and channelized fields ($4096$ samples), respectively.
The prediction examples of the PCS for test GRFs and channelized fields are shown in Fig.~\ref{fig:PCS_mixed_res_pred_examples}.

We show the test relative $L_2$ error and $R^2$ score during training in Fig.~\ref{fig:training_curve}. Overall the PCS takes longer to converge than the DDS, which is reasonable since the PCS has to solve the PDE and learn the surrogate mapping at the same time. Compared with the DDS, the accuracy of the PCS' predictions of the pressure field are similar when trained with the same number of data, but the PCS' predictions of the  flux fields are worse. For the later case, the evaluation metric is dominated by the error on the boundary which is induced by the approximation of spatial derivatives. However, the predictions within the boundary are accurate, as shown in Fig.~\ref{fig:PCS_mixed_res_pred_examples}. Also the relative $L_2$ error is more sensitive than $R^2$ when the error is small, which can be seen by comparing Figs.~\ref{fig:test_nrmse_training_curve} and~\ref{fig:test_r2_training_curve}.

\remark{The quantitative results are mainly for the pressure field, not the flux fields even through we use the mixed formulation loss to train the model. Using the loss functions in Eqs.~(\ref{eq:loss_det_surr}) and~(\ref{eq:loss_det_surr_data}), we observe that the DDS focuses more on the flux fields than the pressure field, but the PCS has better predictability on the pressure field, which is often desirable. For the PCS trained with the mixed formulation, we can either output the pressure and flux fields directly, or re-compute the flux field with the predicted pressure field using the constitutive equation. The other reason for using the mixed residual loss over the primal variational loss is the better predictive accuracy of the pressure field.}

\begin{figure}[t]
    \centering
    \includegraphics[width=0.9\textwidth]{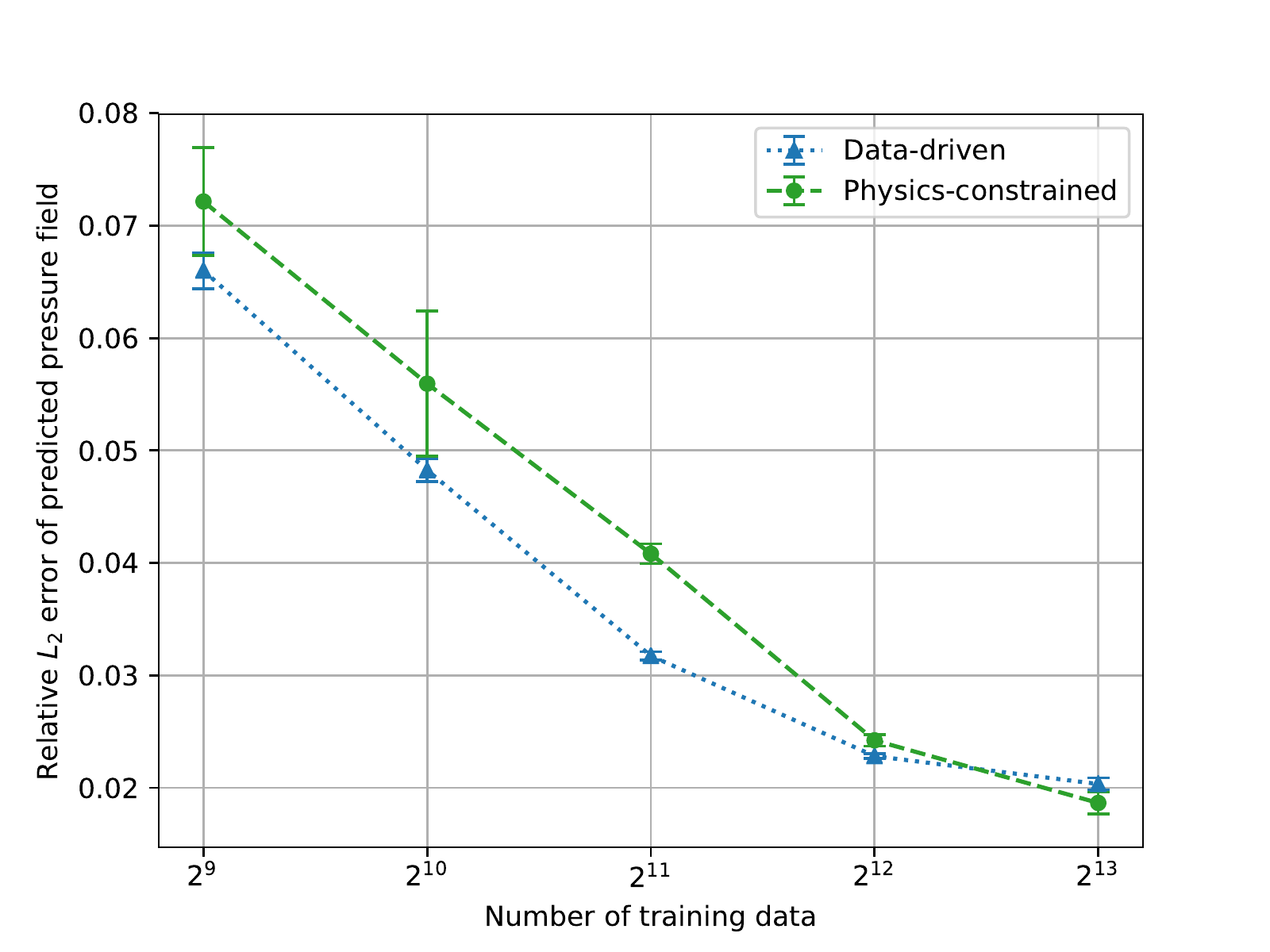}
    \caption{The relative $L_2$ error of the predicted pressure field of physics-constrained and data-driven surrogates trained with $512, 1024, 2048, 4096, 8192$ GRF KLE$512$ data, each with $5$ runs to obtain the error bars. The test set contains $512$ samples from GRF KLE$512$ as well. We emphasize that training the DDS requires an equal number of output data i.e.  solutions of the governing PDE. The reference to compute relative $L_2$ error is simulated with FEniCS.} 
    \label{fig:nrmse_pressure}
\end{figure}
\paragraph{Varying the number of training inputs} We train the PCS with different number of samples from GRF KLE$512$, and compare its predictive performance against the DDS in Fig.~\ref{fig:nrmse_pressure}. From the figure, the relative $L_2$ error decreases as the PCS is trained with more input data. While this is not surprising, it shows the convergence behavior of physics-constraint learning approach. Moreover, the PCS achieves similar relative $L_2$ error of predicted pressure field with the DDS when there are enough training input samples, and even lower when the number of training input samples is $8192$. 

The common requirement for data-driven modeling of physical systems is data efficiency, since we need \textit{expensive} simulated output data to supervise the training. Taking~\cite{zhu2018bayesian} for an example, the number of training data is often less than $1024$. The comparison here is not really appropriate. The DDS does not require physics while the PCS does not require output data.
Overall, Fig.~\ref{fig:nrmse_pressure} suggests that with physical knowledge, we can achieve \textit{comparable} predictive performance with the state-of-the-art DDS \textit{without} any simulation output (but only samples from the random input).

\begin{figure}[t]
    \centering
    \includegraphics[width=0.9\textwidth]{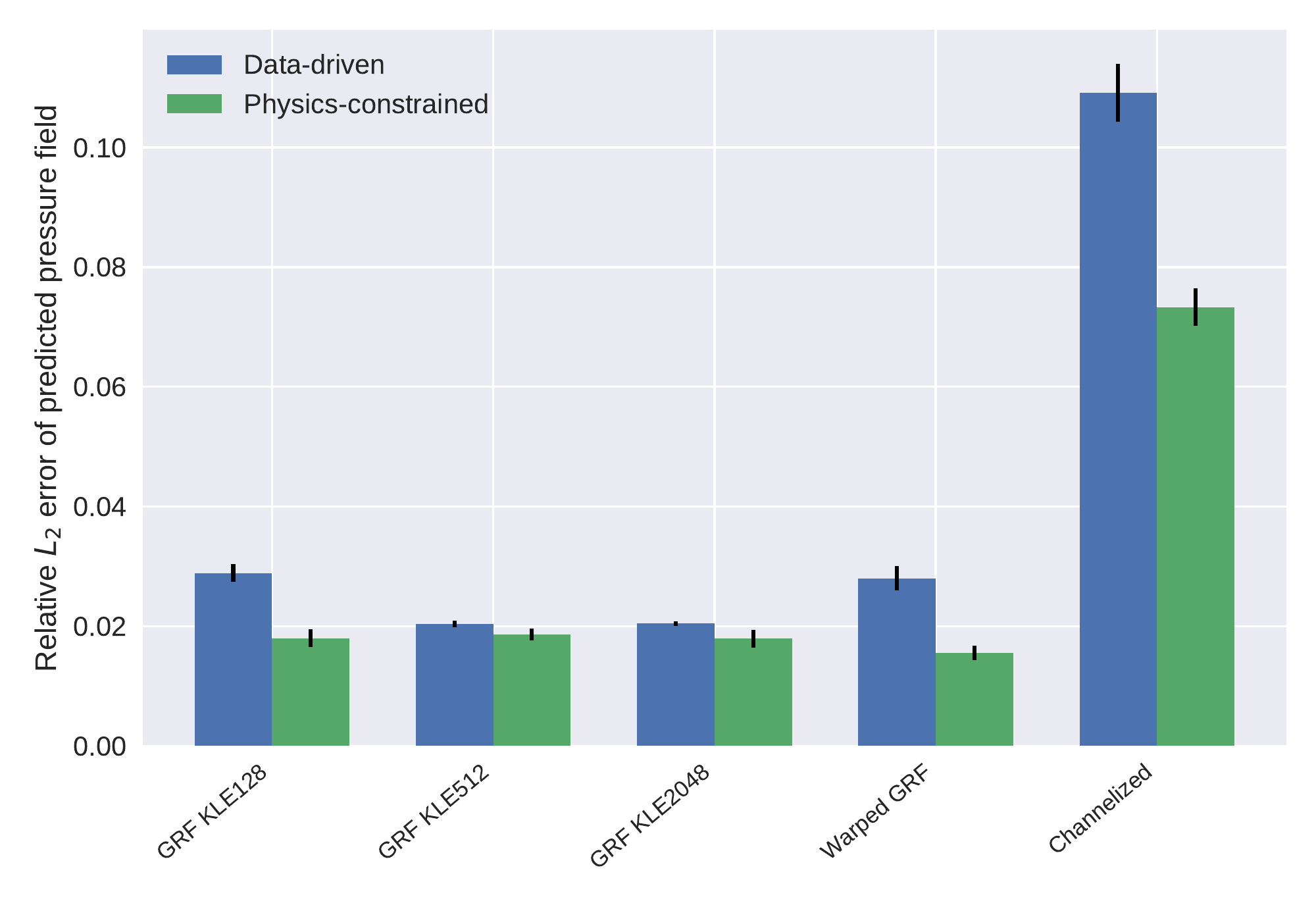}
    \caption{Generalization to new input distributions, which are GRF KLE$128$, KLE$512$ (interpolation), KLE$2048$, warped GRF, and channelized fields. The surrogates are trained with $8192$ samples from GRF KLE$512$. Each test set contains $512$ samples.}
    \label{fig:generalization_nrmse_pressure_ntrain8192}
\end{figure}
\paragraph{Generalization} Apart from computational time, the PCS can `generalize' to \textit{any} input by directly solving the governing equations, i.e. minimizing the loss function in Eq.~(\ref{eq:loss_det_surr}) over this particular input, as shown to work properly in Section~\ref{sec:solve_det_pde}.
Thus generalization here evaluates how accurate the model's prediction is when we need to predict fast, e.g. pass the input through the surrogate, or fine-tuning the surrogate for few steps.

Figure~\ref{fig:nrmse_pressure} shows the surrogates' interpolation performance for the test input from the same distribution as the training input, i.e. GRF KLE$512$. 
Here, we further examine the surrogates' \textit{extrapolation} to out-of-distribution input. We select two other GRFs with different KLE terms, in particular we take KLE$128$ which is smoother than KLE$512$ and KLE$2048$ that leads to higher-variability than KLE$512$. The third test input is warped GRF which is two layers of Gaussian processes. The fourth test input is the channelized field. The samples from those test distributions are shown in Fig.~\ref{fig:generalization_test_input_samples}. 

\begin{figure}
    \centering
    \begin{subfigure}[b]{0.48\textwidth}
        \centering
        \includegraphics[width=\textwidth]{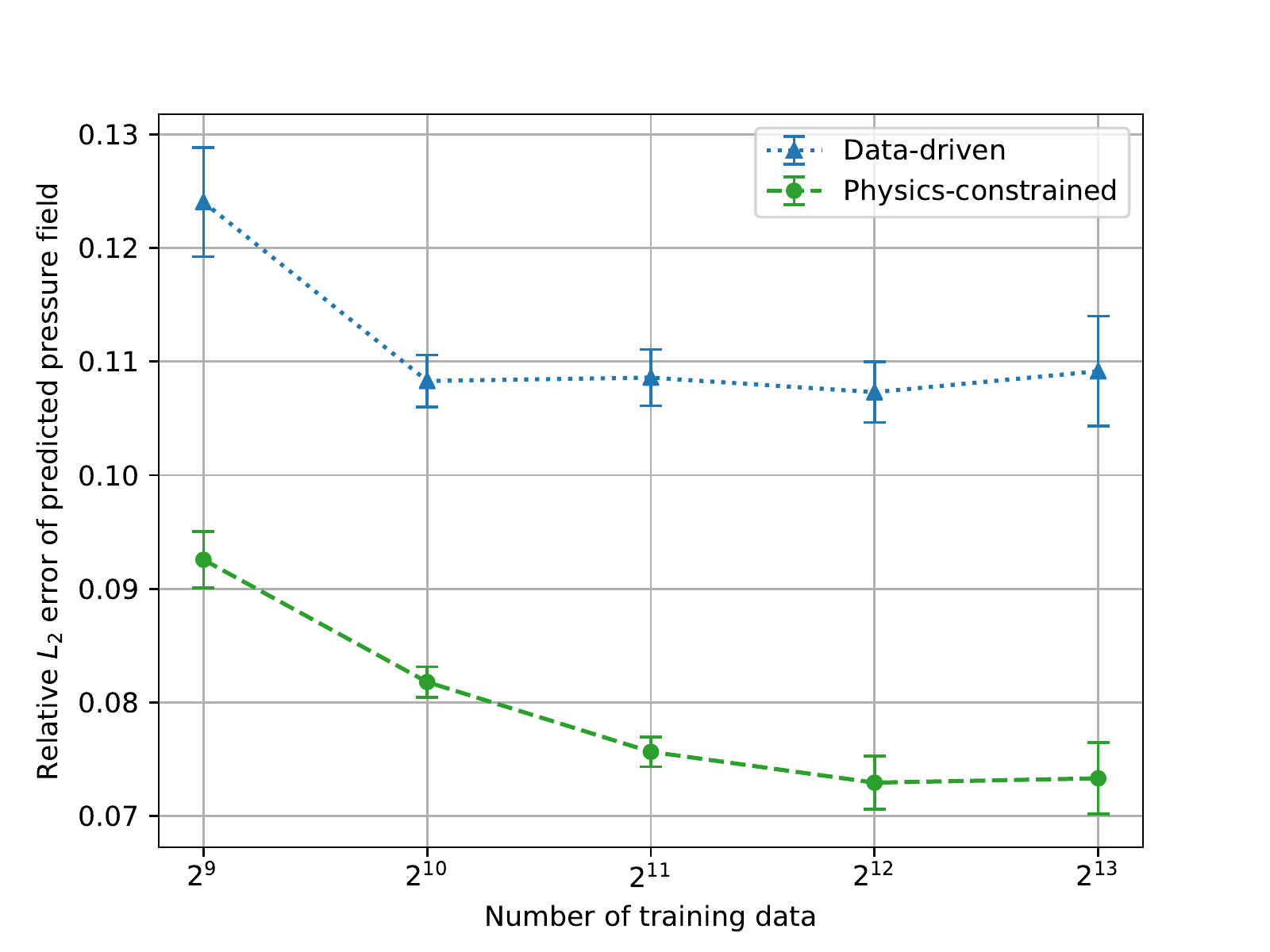}
        \caption{}
        \label{fig:channel_nrmse_pressure}
    \end{subfigure}
    ~
    \begin{subfigure}[b]{0.48\textwidth}
        \centering
        \includegraphics[width=\textwidth]{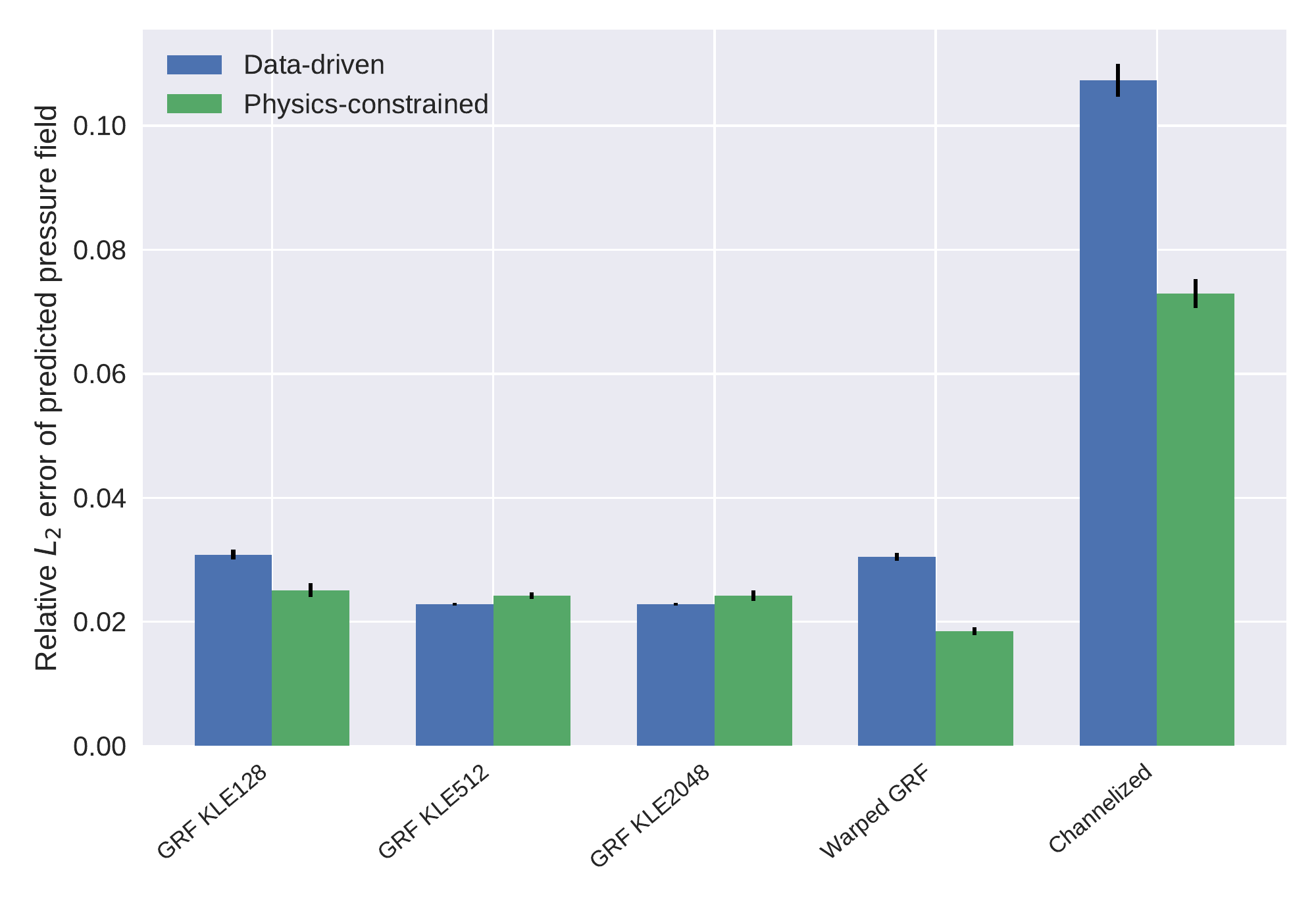}
        \caption{}
        \label{fig:generalization_nrmse_pressure_ntrain4096}
    \end{subfigure}
    \caption{(a) The relative $L_2$ error of predicted pressure field with PCSs and DDSs trained with $512, 1024, 2048, 4096, 8192$ GRF KLE512 data (the same surrogates as Fig.~\ref{fig:nrmse_pressure}), each with $5$ runs. The test set contains $512$ samples from channelized field, with completely different distribution from the training GRF. (b) Generalization across new test input distributions for surrogates trained with $4096$ samples from GRF KLE$512$.}
\end{figure}

We take the surrogates trained on GRF KLE$512$ as in the previous experiment, and test them on the four new input distributions. The relative $L_2$ error of predicted pressure field is shown in Fig.~\ref{fig:generalization_nrmse_pressure_ntrain8192} for the surrogates trained with $8192$ data. The figure shows both PCSs and DDSs generalize well to other test GRF input, including the warped one, but less so when it comes to the channelized field, which is completely different from the training input. 
Notably, the PCS has better generalization than the DDS when tested on warped GRF and channelized fields, which are further away from the training input distribution than the other two GRFs. This is highlighted in Fig.~\ref{fig:channel_nrmse_pressure}. This holds as well for surrogates trained on $512, 1024, 2048, 4096$ samples. Figure~\ref{fig:generalization_nrmse_pressure_ntrain4096} shows the generalization performance when the training sample size is $4096$.

\subsection{Probabilistic Surrogate}\label{sec:prob_surrogate}
This section presents the experiments on using the conditional Glow model shown in Fig.~\ref{fig:glow_msc} as the probabilistic surrogate. We are interested in how the conditional Glow captures predictive uncertainty, uncertainty calibration and its generalization performance to unseen test input. We choose to work on $32\times 32$ discretization instead of $64\times 64$ with the input GRF KLE$100$ because of the large model size of the current implementation of the model.

\paragraph{Network} In our experiment, we use $L=3$ levels, each of which contains $F=6$ steps of flow. Both the dense blocks and coupling networks \fbox{$\bms$} and \fbox{$\bmt$} in affine coupling layers use DenseNet~\cite{huang2017densely} as the building block. The number of dense layers within each dense block in the encoder is $3, 4, 4$ (from the input to the latent direction). The coupling networks $\texttt{CouplingNN}$ as in Table~\ref{tab:affine_coupling} for scaling and shift have $3$ dense layers, followed by a $3\times3$ convolution layer with zero initialization to reduce the number of output features to be the same as its input features. The model has $1,535,549$ parameters, including $179$ convolution layers. For other hyperparameters of the model, please refer to our open-source code.

\paragraph{Training} The model is trained with $4096$ input samples from GRF KLE$100$ over $32 \times 32$ grid for $400$ epochs with mini-batch size $32$. No output data is needed for training.  We use the Adam optimizer with initial learning rate $0.0015$, and one-cycle learning rate scheduler. The weight for boundary conditions $\lambda$ is $50$. The inverse temperature $\beta$ is prefixed to certain values. Training the model with the above setting on a single NVIDIA GeForce GTX $1080$ Ti GPU card takes about $3$ hours.

\begin{figure}
    \centering
    \begin{subfigure}[b]{0.5\textwidth}
        \centering
        \includegraphics[width=\textwidth]{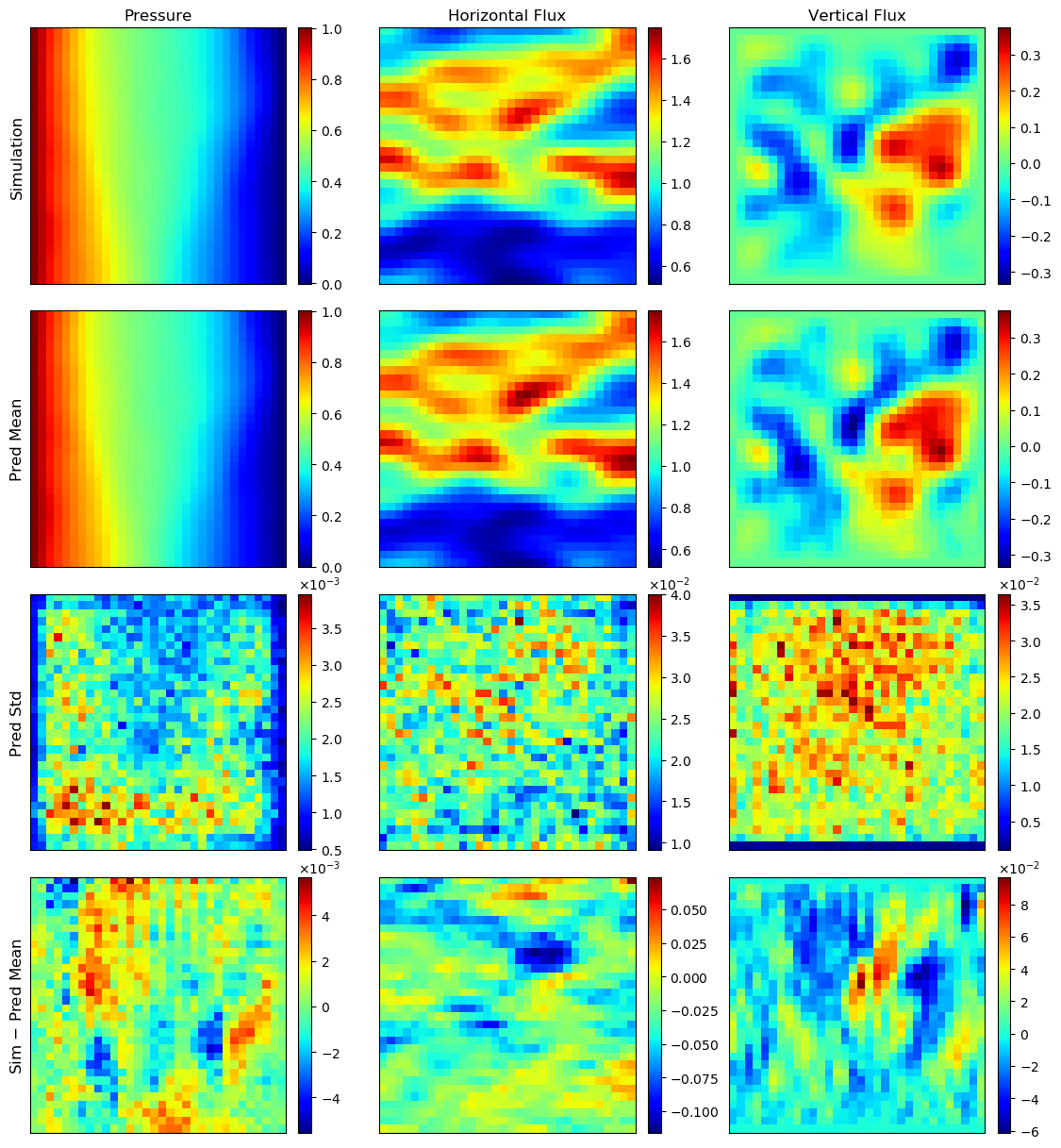}
        \caption{Predictive mean and variance.}
        \label{fig:cflow_pred_at_x}
    \end{subfigure}
    ~
    \begin{subfigure}[b]{0.45\textwidth}
        \centering
        \includegraphics[width=\textwidth]{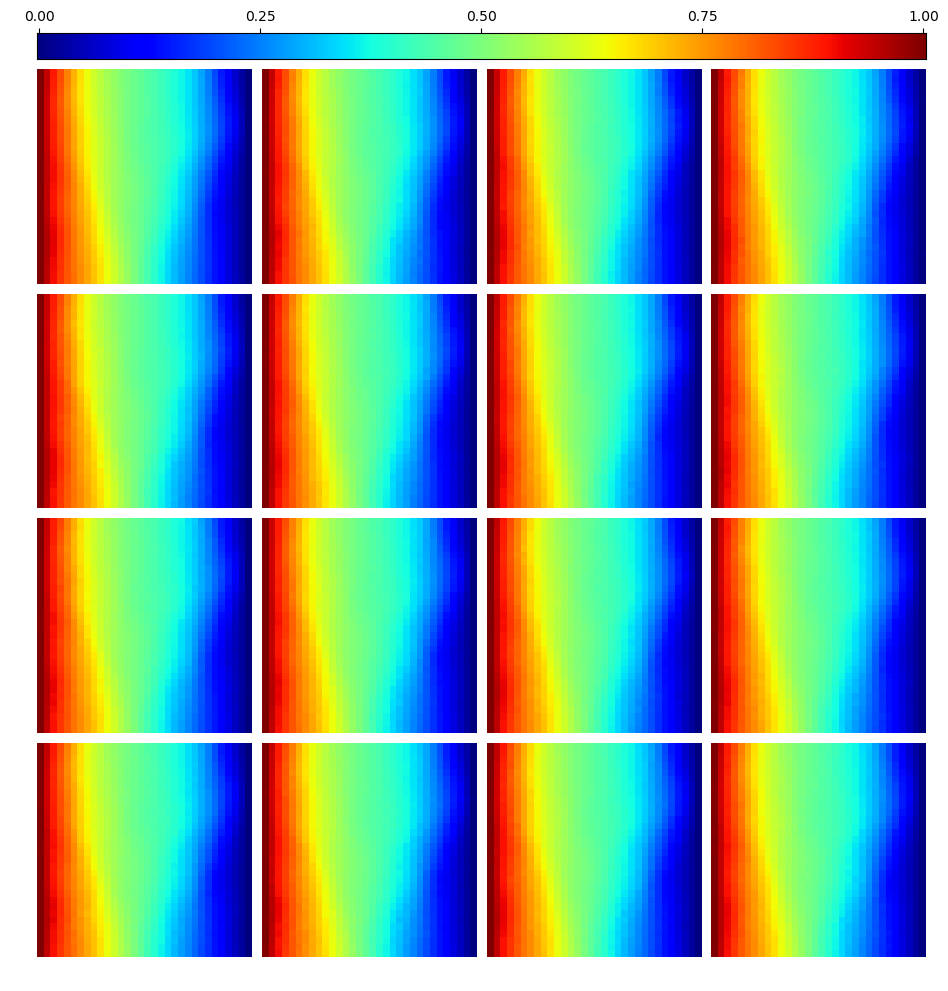}
        \caption{Samples of predicted pressure.}
        \label{fig:cflow_pred_at_x_samples_pressure}
    \end{subfigure}
    
    \begin{subfigure}[b]{0.47\textwidth}
        \centering
        \includegraphics[width=\textwidth]{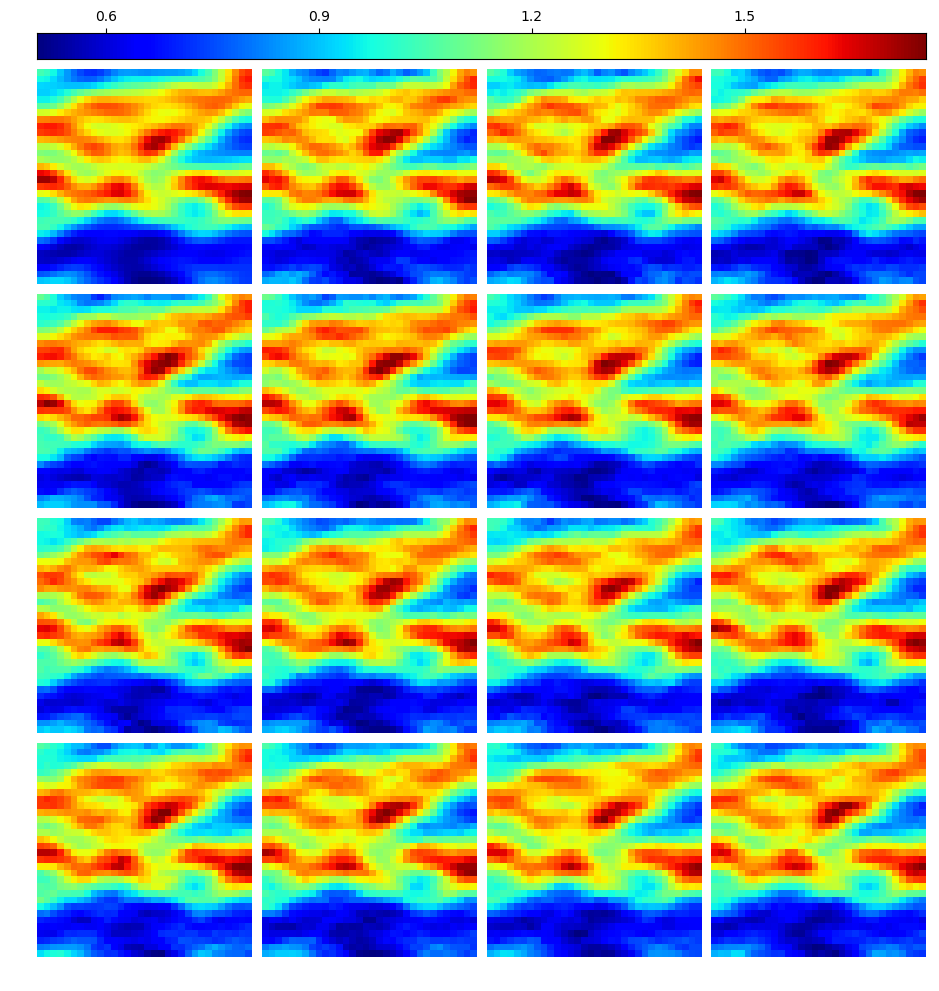}
        \caption{Samples of predicted horizontal flux.}
        \label{fig:cflow_pred_at_x_samples_hor_flux}
    \end{subfigure}
    ~
    \begin{subfigure}[b]{0.47\textwidth}
        \centering
        \includegraphics[width=\textwidth]{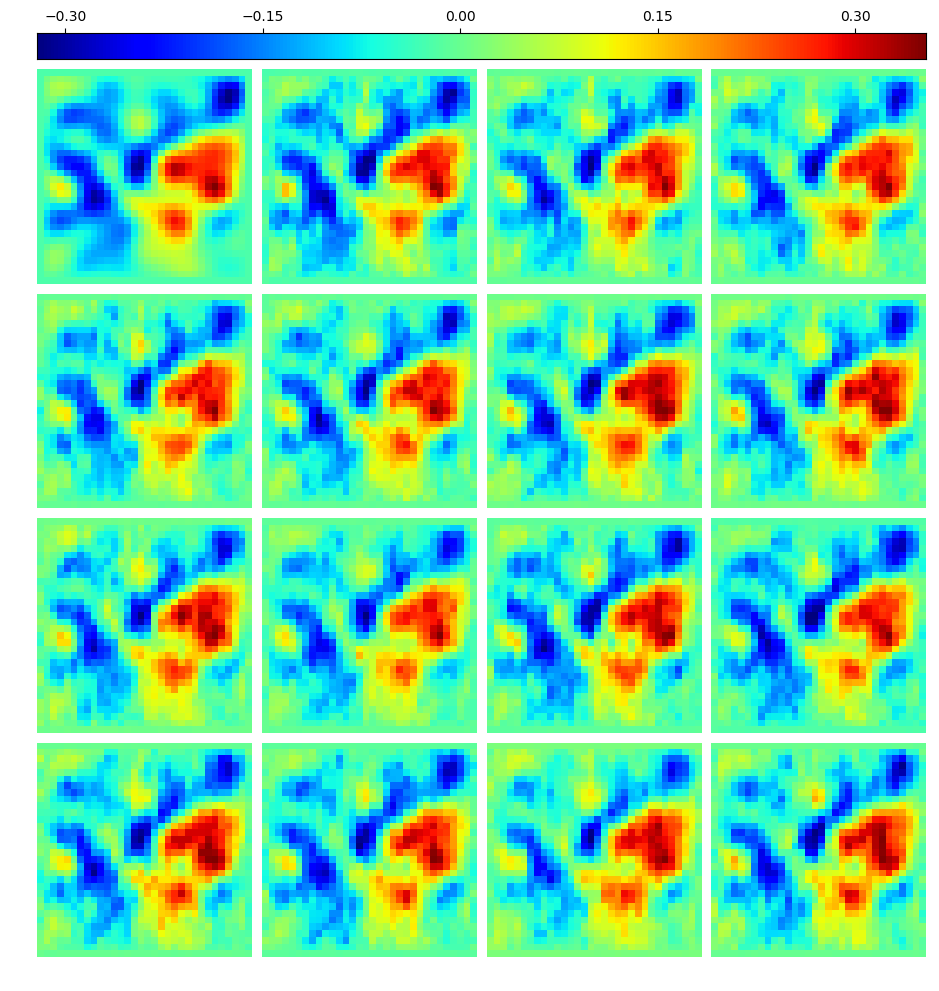}
        \caption{Samples of predicted vertical flux.}
        \label{fig:cflow_pred_at_x_samples_ver_flux}
    \end{subfigure}
    \caption{Prediction of the multiscale conditional Glow ($\beta=150$) for a test input which is sampled from GRF KLE$100$ over $32\times 32$ grid. (a) The predictive mean ($2$nd row) and one standard derivation ($3$rd row) are obtained with 20 output samples. The first row show three simulated output fields, the $4$-th row shows the error between the reference and the predictive mean. In (b), (c), (d), the top left corner shows the simulated output, the rest $15$ are samples from the conditional predictive density $p_\bmtheta(\bmy | \bmx)$. The relative $L_2$ error for the predicted pressure field is $0.0038$ when tested on 512 samples from GRF KLE$100$.}
    \label{fig:cflow_kle100_grid32}
\end{figure}
\paragraph{Predictive distribution}  Fig.~\ref{fig:cflow_kle100_grid32} shows the prediction for a test input from GRF KLE100, where in Fig.~\ref{fig:cflow_pred_at_x} the predictive mean and variance are estimated pixel-wise with $20$ samples from the conditional density by sampling $20$ realizations of noise $\{\bmepsilon_l^\si\}_{l=1, i=1}^{L, 20}$ as in Algorithm~\ref{algo:cglow}. 
The test relative $L_2$ error for the pressure field (comparing predictive mean against simulated output) achieves $0.0038$, which is comparable to the relative $L_2$ error of the deterministic surrogate ($0.0035$). The predictive variance of the pressure  and vertical flux fields reflect correctly the boundary conditions, which are close to zero on the left-right boundaries and top-bottom boundaries, respectively. We  also draw $15$ samples from the predictive distribution for each output field, which are shown in Figs.~\ref{fig:cflow_pred_at_x_samples_pressure},~\ref{fig:cflow_pred_at_x_samples_hor_flux},~\ref{fig:cflow_pred_at_x_samples_ver_flux}. The predictive output samples are still diverse despite the predictive mean being highly accurate. Mode collapse is a well-known problem for conditional GANs~\cite{anonymous2019diversity-sensitive, zhu2017toward} and VAEs~\cite{anonymous2019lagging}, which seems not much of a concern for flow-based generative models as demonstrated with the diversity of samples.

\begin{figure}
    \centering
    \begin{subfigure}[b]{0.48\textwidth}
        \centering
        \includegraphics[width=\textwidth]{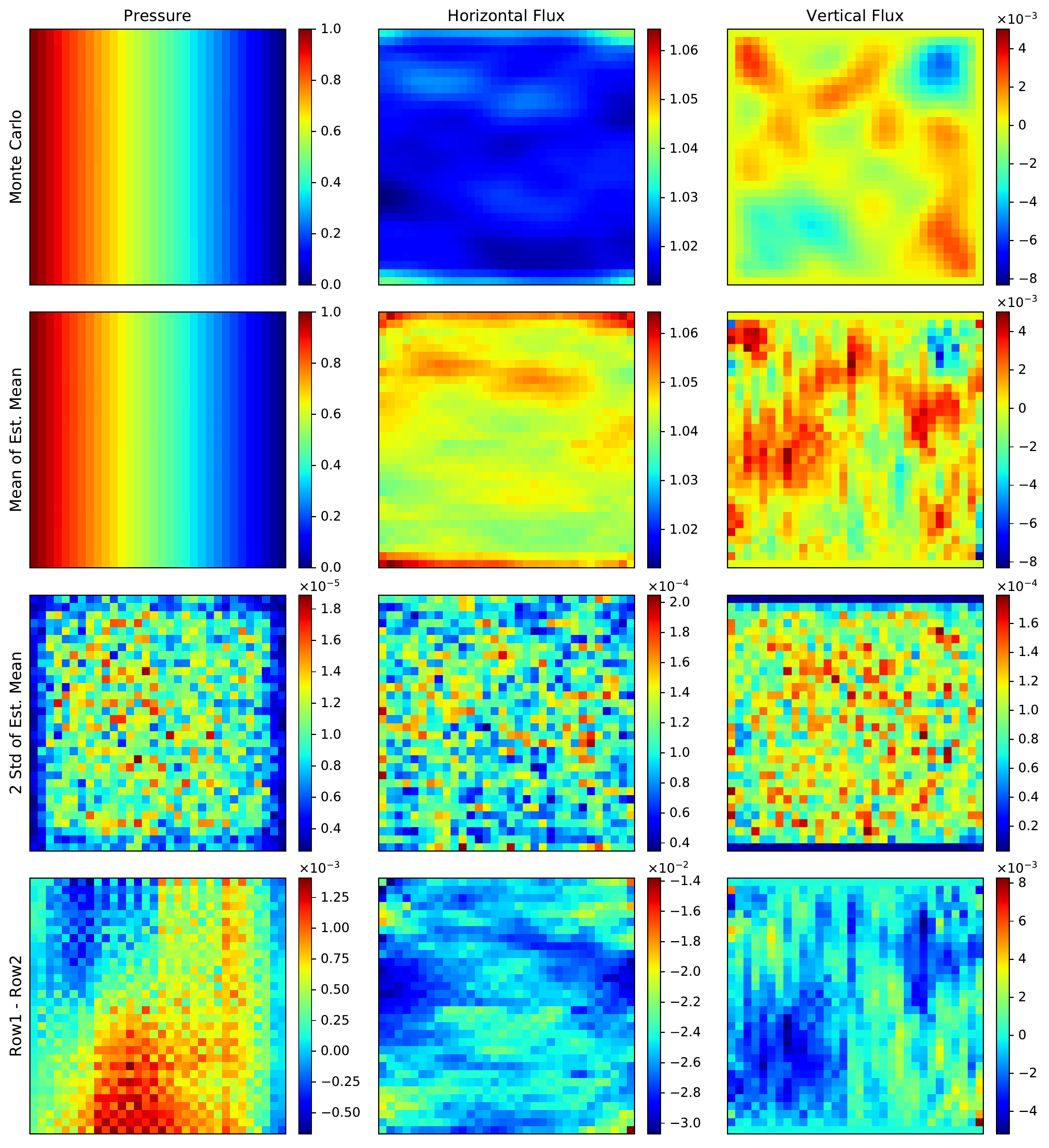}
        \caption{Estimate of output mean.}
        \label{fig:cglow_up_mean}
    \end{subfigure}
    ~
    \begin{subfigure}[b]{0.48\textwidth}
        \centering
        \includegraphics[width=\textwidth]{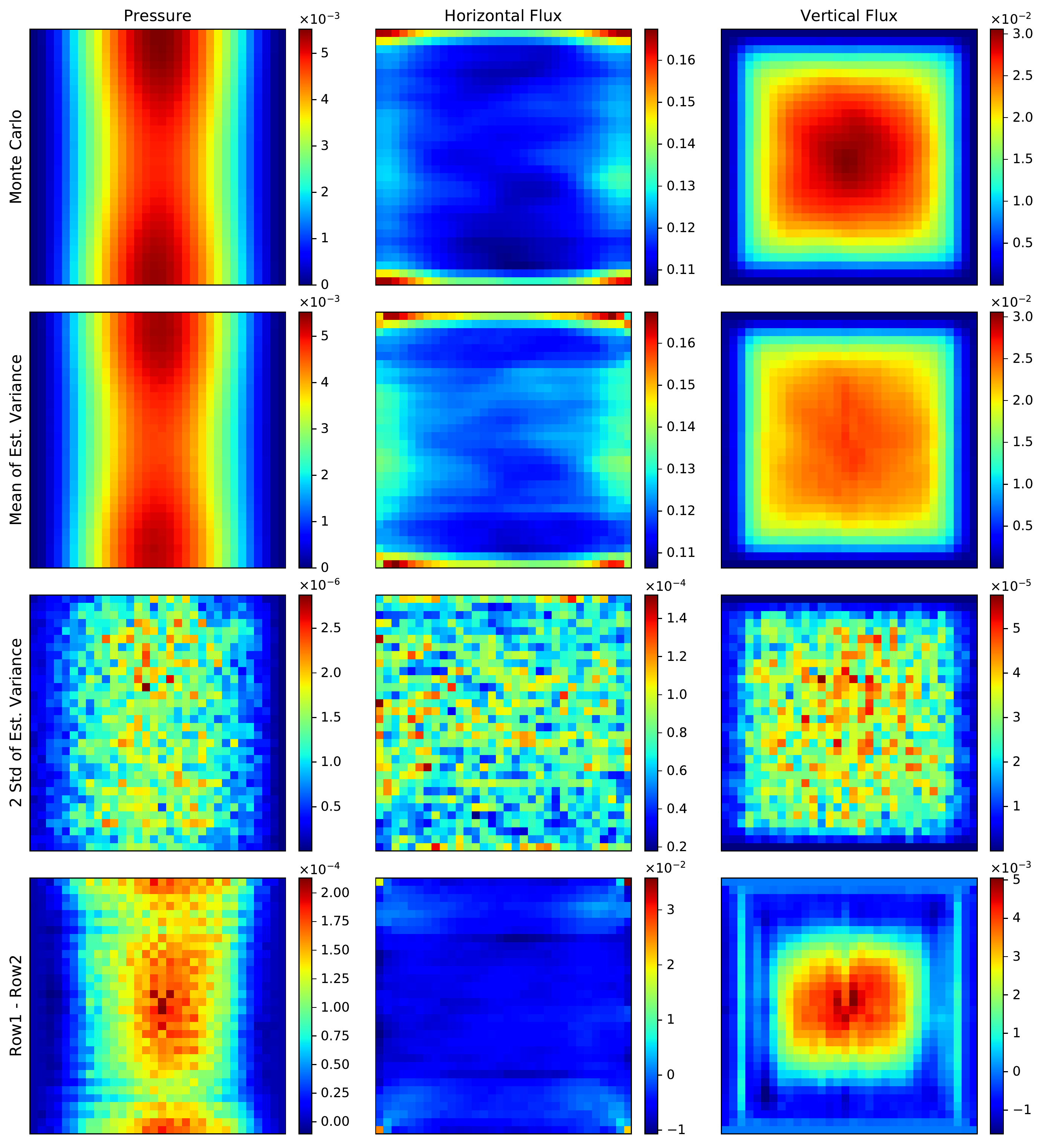}
        \caption{Estimate of output variance.}
    \end{subfigure}
    \caption{Uncertainty propagation with multiscale conditional Glow, $\beta=150$. (a) The first row shows the sample mean of $10,000$ simulated output, the second and third rows show the sample mean and two standard deviation of the estimate mean of $10,000$ predicted output with the probabilistic surrogate, and the fourth row shows the error between the first two rows. (b) The results for output variance.}
    \label{fig:cflow_up}
\end{figure}
\paragraph{Uncertainty propagation}
We use the trained conditional Glow as a surrogate to quickly predict the output for $10,000$ input samples from GRF KLE100, then compute the mean and variance of the estimated output mean, and output variance, then compare against the Monte Carlo estimate using the corresponding $10,000$ simulated output. We generate 20 samples for each input with the trained surrogate, then estimate the mean and variance of the output with the law of total expectation and the law of total variance. By repeating this process for 10 times, we obtain 10 estimates of the mean and variance for the output. Then the sample mean and variance of the 10 estimate means and estimate variances can be computed, which are shown in the second and third row of Fig.~\ref{fig:cflow_up}. The statistics of the surrogate output matches that of the simulation output very well, especially for the output variance which are typically underestimated when using surrogates. Note that there is only small error (around $3 \%$ relative error) between the estimated mean of the horizontal flux field despite the noticeable difference in color as in Fig.~\ref{fig:cglow_up_mean}.

\begin{figure}[h]
    \centering
    \begin{subfigure}[b]{0.95\textwidth}
        \centering
        \includegraphics[width=\textwidth]{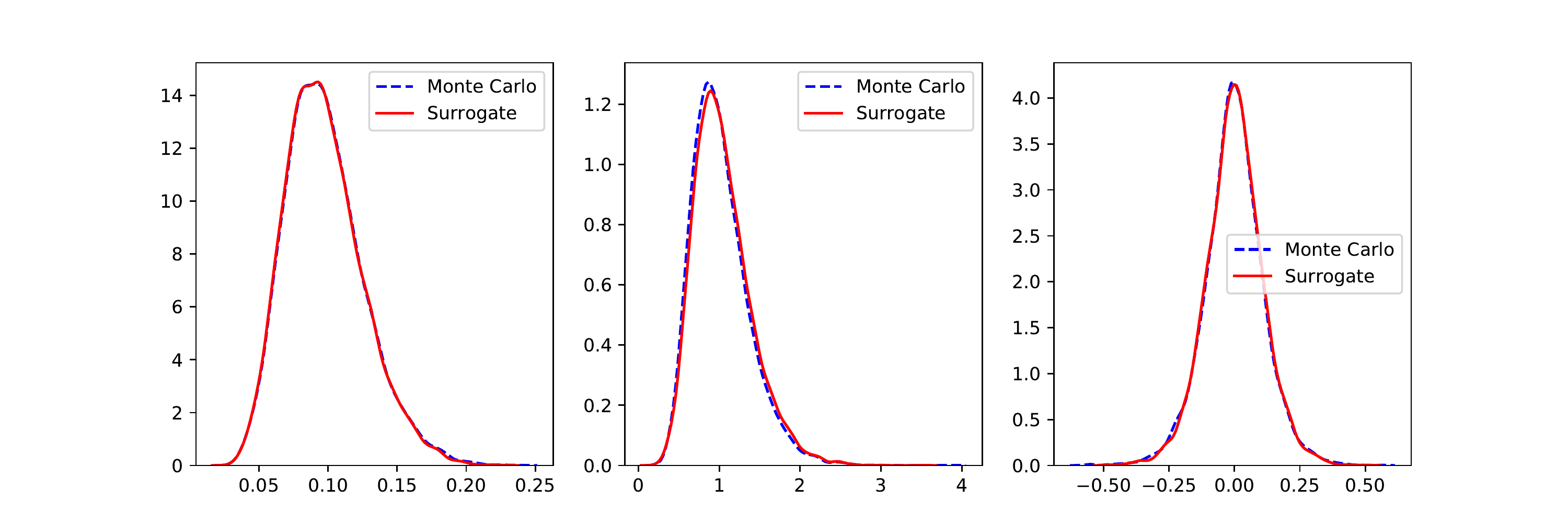}
        \caption{Distribution estimation at location $(0.2875, 0.8875)$.}
    \end{subfigure}
    
    \begin{subfigure}[b]{0.95\textwidth}
        \centering
        \includegraphics[width=\textwidth]{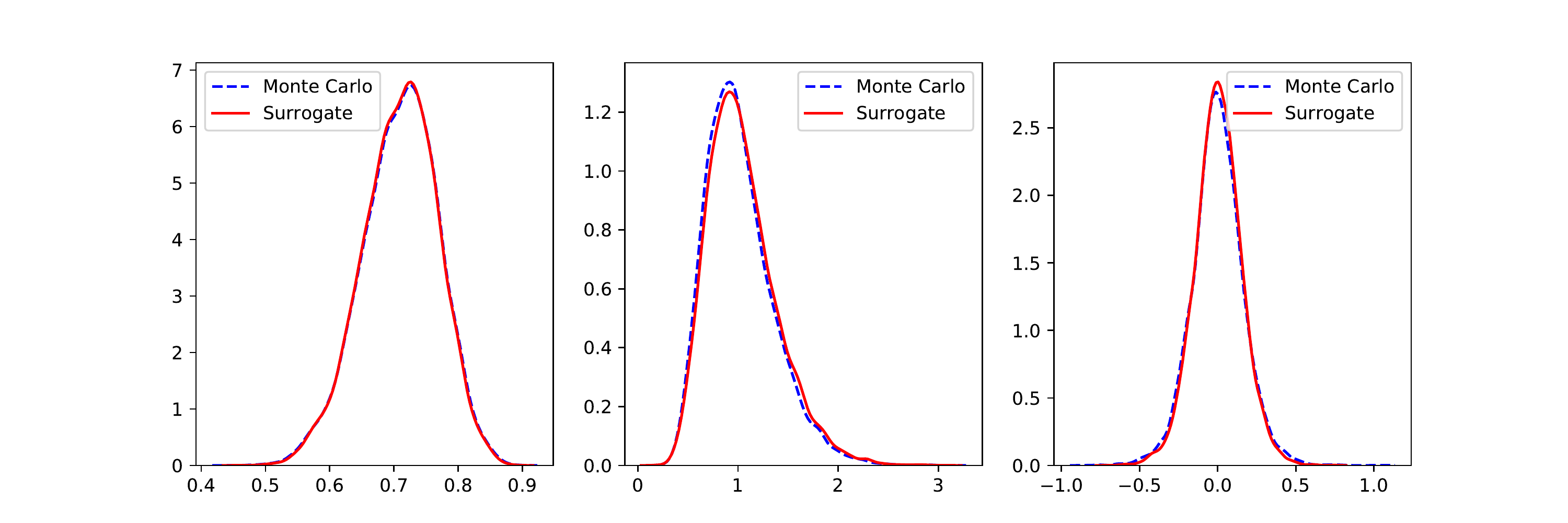}
        \caption{Distribution estimation at location $(0.4625, 0.2875)$.}
    \end{subfigure}
    \caption{Distribution estimate with conditional Glow, $\beta=150$. From left to right shows the density estimate of pressure, horizontal flux and vertical flux at certain locations of the domain $[0, 1]^2$. }
    \label{fig:cflow_dist_est}
\end{figure}

\paragraph{Distribution estimate} We show in Fig.~\ref{fig:cflow_dist_est} the kernel density estimation for the values of three output fields at random locations in the domain using the $10,000$ output samples from simulation and the ones propagated with the trained conditional Glow. 

\begin{figure}
    \centering
    \centering
    \begin{subfigure}[b]{0.7\textwidth}
        \centering
        \includegraphics[width=\textwidth]{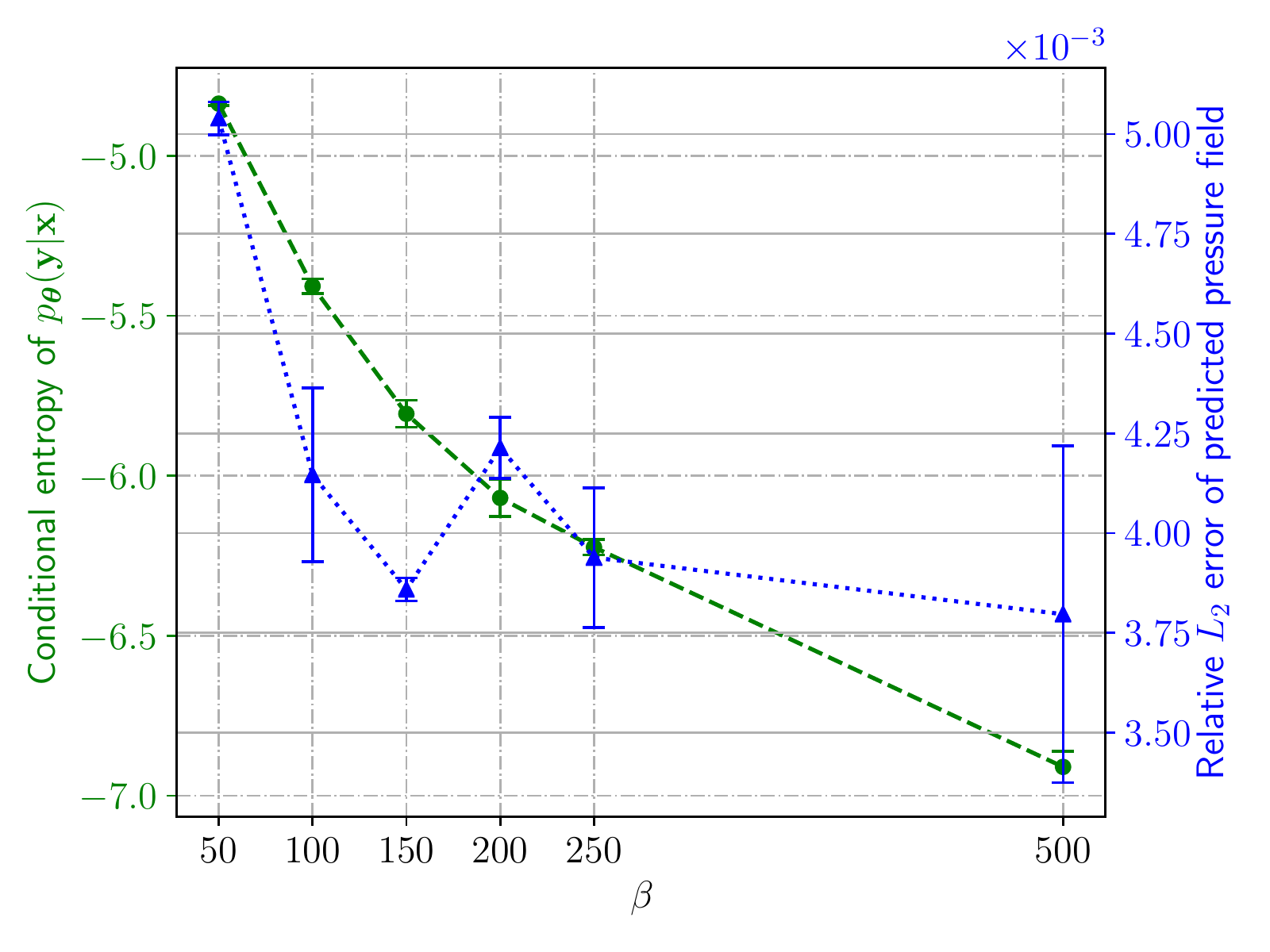}
    \caption{}
    \label{fig:cglow_entropy_nrmse_vs_beta}
    \end{subfigure}
    
    \begin{subfigure}[b]{0.8\textwidth}
        \centering
        \includegraphics[width=\textwidth]{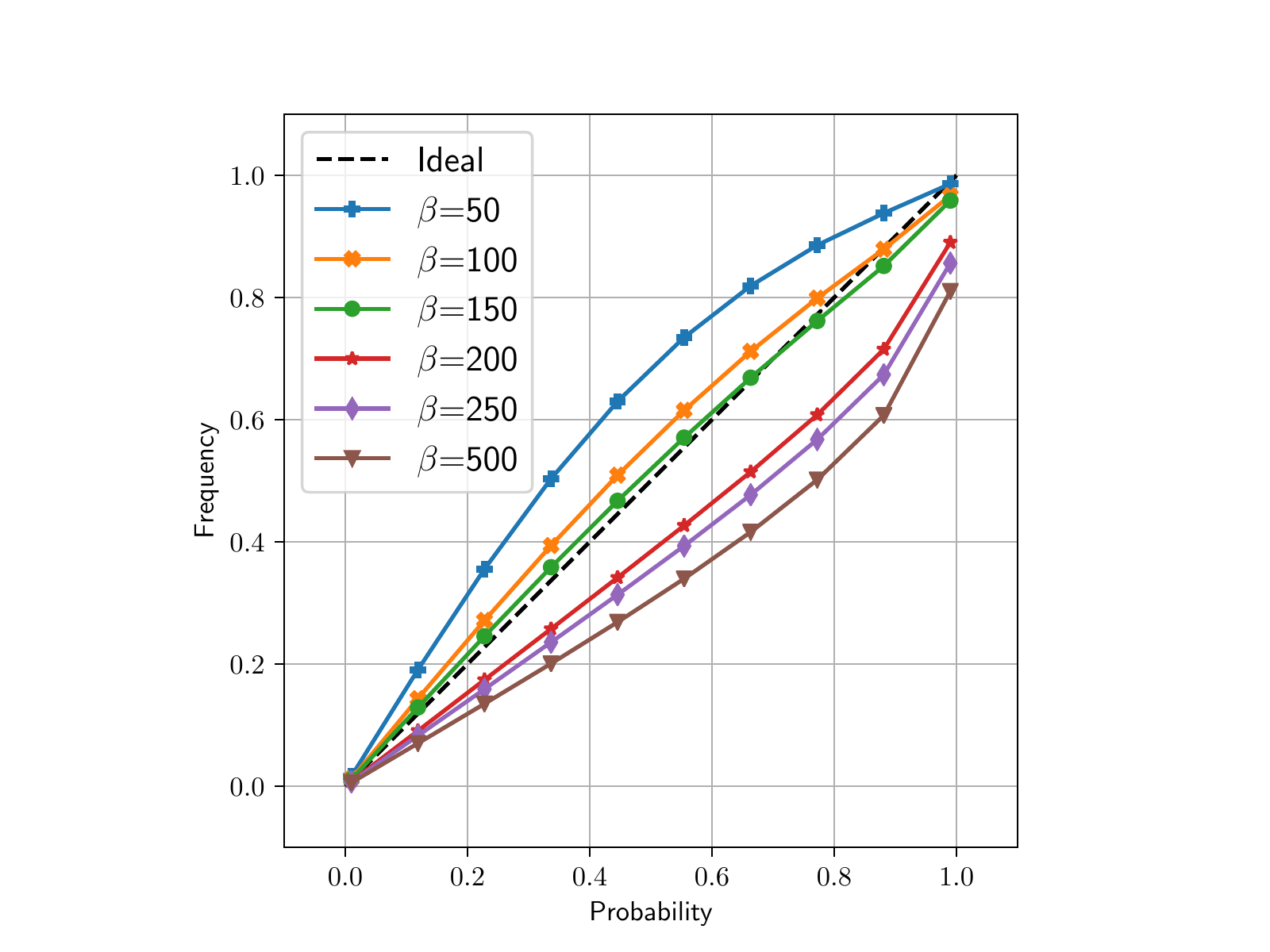}
    \caption{}
    \label{fig:cglow_reliability}
    \end{subfigure}
    \caption{(a) Conditional entropy of $p_\bmtheta(\bmy|\bmx)$ and relative $L_2$ error of predicted pressure field w.r.t. $\beta$. Conditional entropy is evaluated in bits per dimension.
    The surrogate is tested on $512$ input samples from GRF KLE$100$. The error bar is obtained with $3$ independent runs. (b) Reliability diagram of predicted pressure field with conditional Glow trained with different $\beta$, which is evaluated with $10,000$ input-output data pairs. The closer the diagram is to the diagonal, the better the probabilistic surrogate is calibrated.}
    \label{fig:cglow_uncertainty}
\end{figure}
\paragraph{Uncertainty calibration by tuning $\beta$} 
Given the PDEs and boundary conditions, the prediction of the surrogate can be evaluated directly with the loss $L(\bmy, \bmx)$, without requiring the reference solution (e.g. simulation output). However, this loss cannot be readily translated to the uncertainty of the solution, e.g. the upper and lower bound of the solution at every grid point in the domain. The probabilistic surrogate trained under the reverse KL divergence can provide the uncertainty estimate, but may be at the expense of the accuracy of the mean prediction. The precision parameter $\beta$ controls the overall variance of the reference density, which is reflected from the conditional entropy of the model density $p_\bmtheta(\bmy | \bmx)$ in Fig.~\ref{fig:cglow_entropy_nrmse_vs_beta}. The influence of $\beta$ on the accuracy and the entropy of the model can be seen from the two competing terms in the reverse KL divergence as well. Larger $\beta$ puts more penalty of the PDE loss term $L(\bmy, \bmx)$ and less on the negative conditional entropy, thus the predictions become more accurate but less diverse, and to some extent, the probabilistic surrogate becomes over confident, as shown in Fig.~\ref{fig:cglow_reliability} when $\beta=250$. On the other hand, when $\beta$ is too small, the probabilistic surrogate is prudent (large uncertainty estimate) and less accurate about the solution, e.g. the case of $\beta=50$. From the figure, the model trained under $\beta=150$ is well-calibrated (its reliability diagram is close to the diagonal dashed line) and achieves high accuracy at the same time. 

\begin{figure}
    \centering
    \begin{subfigure}[b]{0.48\textwidth}
        \centering
        \includegraphics[width=\textwidth]{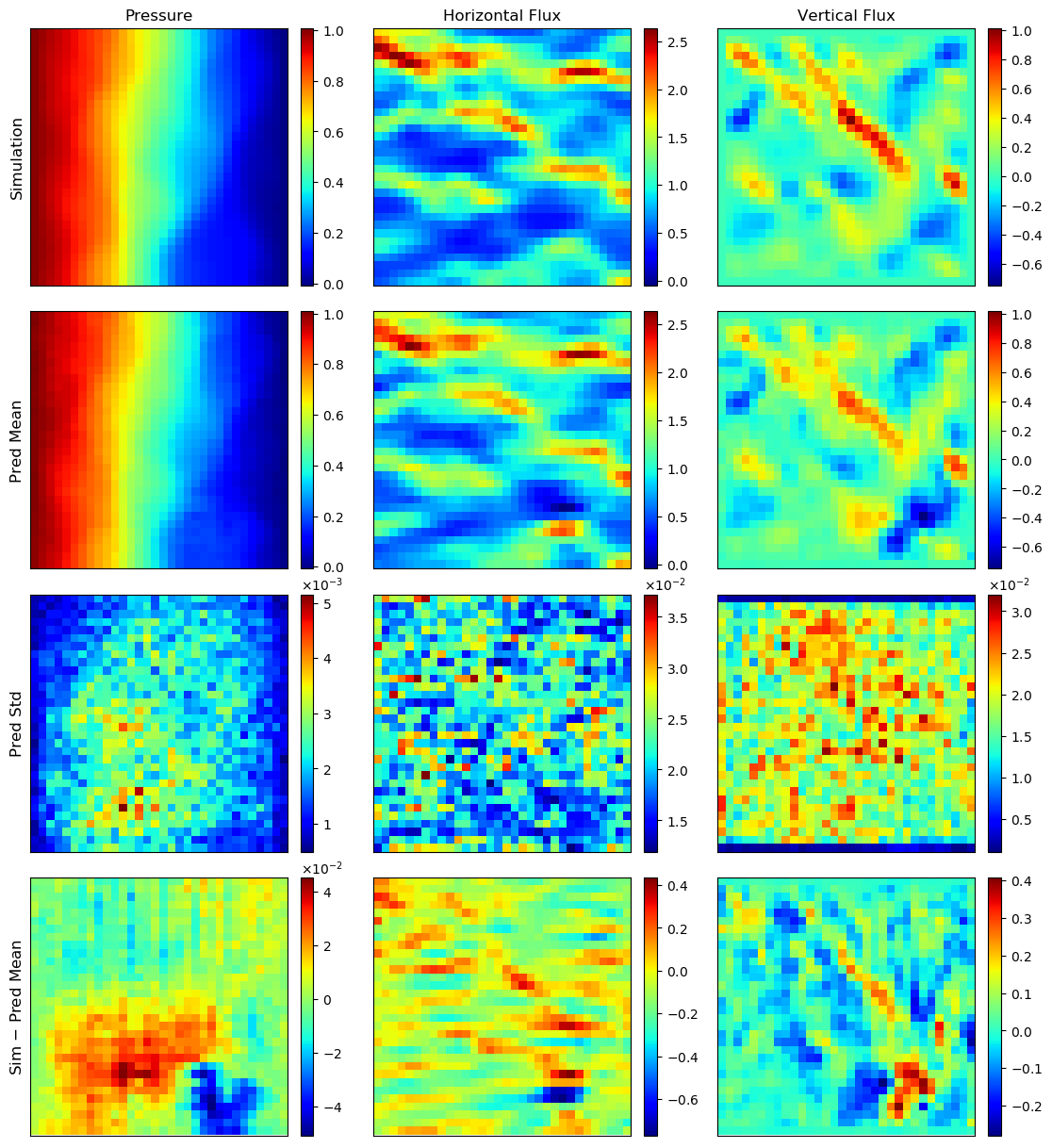}
        \caption{GRF KLE$256$.}
    \end{subfigure}
    ~
    \begin{subfigure}[b]{0.48\textwidth}
        \centering
        \includegraphics[width=\textwidth]{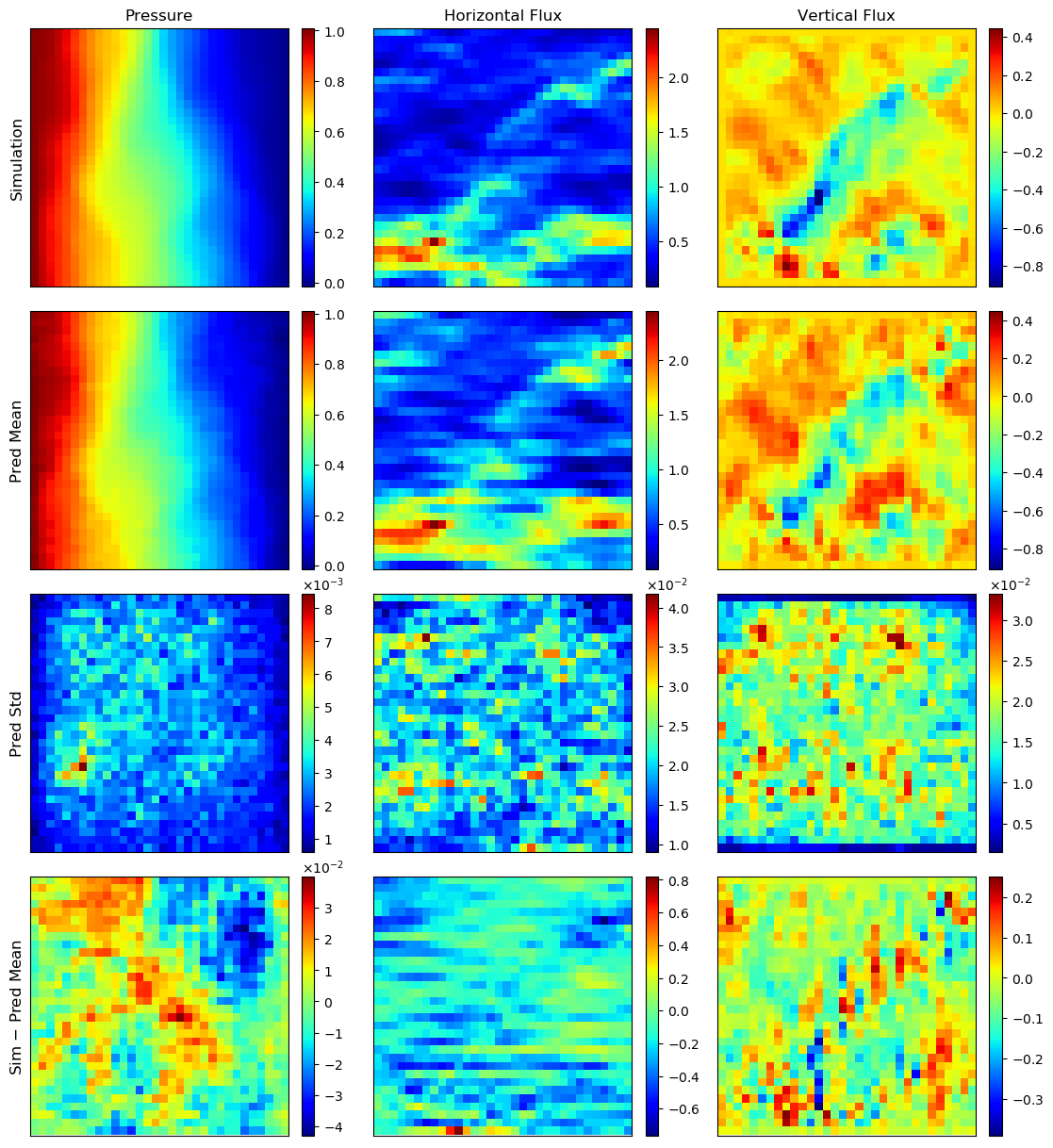}
        \caption{GRF KLE$512$.}
    \end{subfigure}
    
    \begin{subfigure}[b]{0.48\textwidth}
        \centering
        \includegraphics[width=\textwidth]{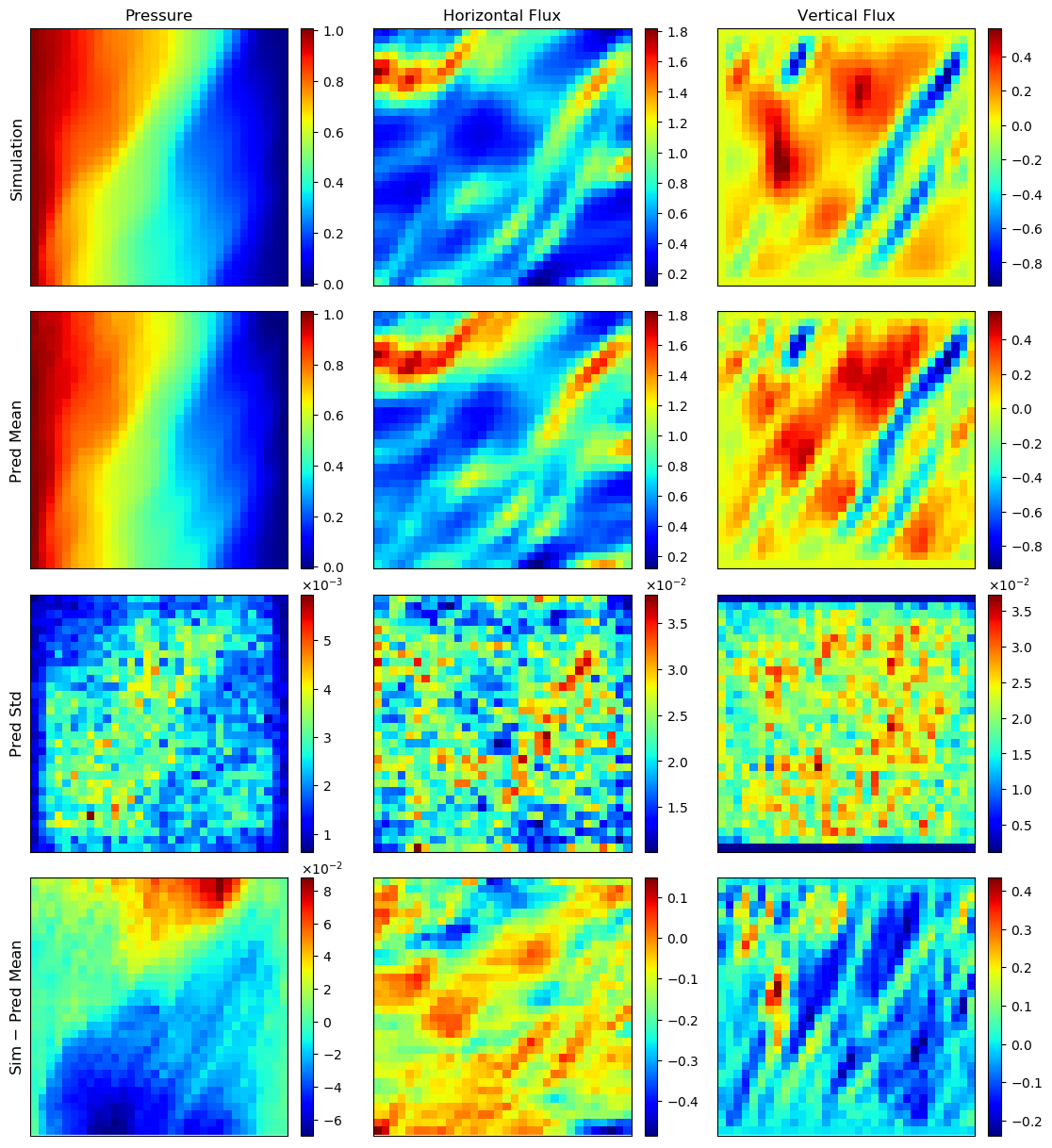}
        \caption{Warped GRF.}
    \end{subfigure}
    ~
    \begin{subfigure}[b]{0.48\textwidth}
        \centering
        \includegraphics[width=\textwidth]{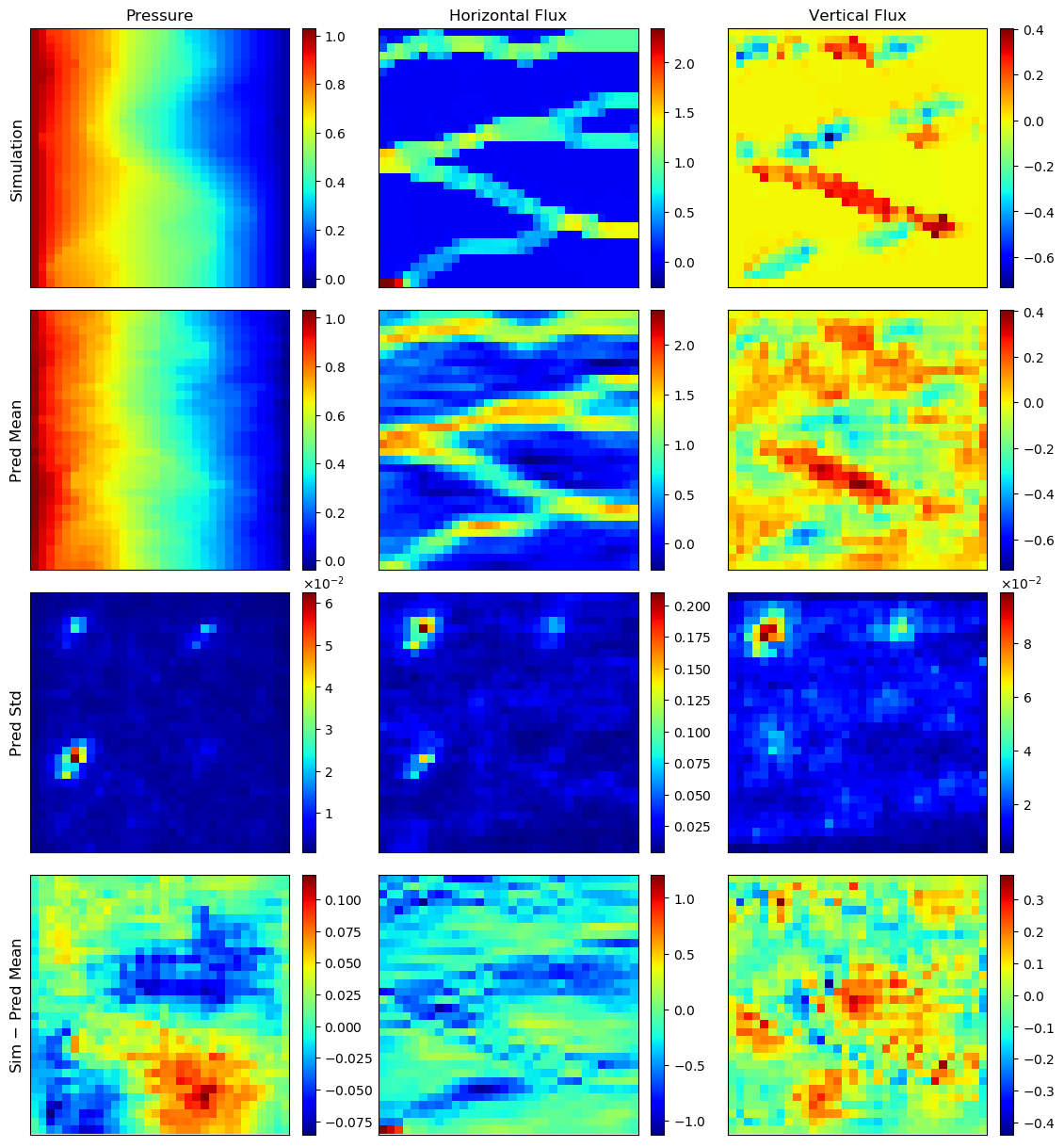}
        \caption{Channelized field.}
    \end{subfigure}
    \caption{Generalization of conditional Glow to out-of-distribution input. The model is trained on GRF KLE$100$, and tested on (a) GRF KLE$256$, (b) GRF KLE$512$, (c) Warped GRF, (d) channelized field. Note that the results are cherry-picked.}
    \label{fig:cglow_generalization}
\end{figure}

\paragraph{Generalization}
We test the generalization of conditional Glow on input distributions different from the training input (GRF KLE$100$), including GRF KLE$256$, GRF KLE$512$, warped GRF, and channelized fields, as in Fig.~\ref{fig:cglow_generalization}. However, we could not observe larger uncertainty when the test input is far away from the training input. The error between the predictive mean and simulation is in general one magnitude larger than the uncertainty. Thus the current surrogate cannot express what it does not known which in practice is a highly desirable outcome.

\section{Conclusions}
\label{sec:Conclusions}
This paper has offered a foray in physics-aware machine learning for surrogate modeling and uncertainty quantification, with emphasis on the solution of PDEs. The most significant contribution of the proposed framework and simultaneously the biggest difference with other efforts along these lines, is that \textit{no labeled data} are needed i.e. one does not need to solve governing PDEs for the training inputs. This is accomplished by incorporating appropriately the governing equations into the loss/likelihood functions. We have demonstrated that convolutional encoder-decoder network-based surrogate models can achieve high predictive accuracy for high-dimensional stochastic input fields. Furthermore, the generalization performance of the physics-constrained surrogates proposed is consistently better than data-driven surrogates for out-of-distribution test inputs. The probabilistic surrogate built on the flow-based conditional generative model and trained by employing the reverse KL-divergence loss, is able to capture predictive uncertainty as demonstrated in several uncertainty propagation and calibration tasks.

Many important unresolved tasks have been identified that will be addressed in forthcoming works. They include 
(a) Extension of this work to surrogate modeling for dynamical systems, (b) Improving generalization on out-of-distribution input, e.g. fine-tuning the trained surrogate on test input~\cite{devlin2018bert, radford2018improving}, learned gradient update~\cite{adler2017solving, hammernik2018learning}, meta-learning on a distribution of regression tasks~\cite{finn2017model}, etc., (c) Combining physics-aware and data-driven approaches when only limited simulation data and partially known physics are available~\cite{yang2018physics}, (d) Scale the flow-based conditional generative models to higher dimensions~\cite{2018arXiv181001367G}, (e) More reliable probabilistic models, e.g. being able to express what the model does not know~\cite{nalisnick2018deep, choi2018generative} by showing larger predictive uncertainty when tested on out-of-distribution input, (f) Exploring ways to increase the expressiveness of FC-NNs to better capture the multiscale features of PDE solutions, e.g. by evolving network architectures~\cite{stanley2002evolving} and (g) Exploring the solution landscape with the conditional generative surrogates~\cite{farrell2015deflation}.

\section*{Acknowledgements}
The authors acknowledge support  from the Defense Advanced Research Projects Agency (DARPA) under the Physics of Artificial Intelligence (PAI) program (contract HR$00111890034$). 
Additional computing resources were provided by the University of Notre Dame's Center for Research Computing (CRC). 

\appendix
\section{Sobel filter to estimate spatial gradients}\label{appendix:sobel_filter}
Sobel filter is used to estimate horizontal and vertical spatial gradients by applying one convolution with the following $3 \times 3$ kernels, respectively:
\[
\mc H = 
\begin{bmatrix}
    1 & 0 & -1 \\
    2 & 0 & -2 \\
    1 & 0 & -1
\end{bmatrix}, \quad
\mc V =
\begin{bmatrix}
    1 & 2 & 1 \\
    0 & 0 & 0 \\
    -1 & -2 & -1
\end{bmatrix}.
\]
Intuitively it is a smoothed finite difference method.
The convolution operation goes natural with CNN representation of solution fields and is highly efficient. Sobel filter is way more efficient than using automatic differentiation to obtain spatial gradients in the FC-NN parameterization, with the compromise of reduced accuracy, especially for locations close to the boundaries. 

To improve the accuracy of gradient estimate on the boundary, we use the following correction.
For 2D image matrix $\bmI$ of size $H\times W$,  Sobel kernel $\mc H$, and correction matrix $M_{\mc H}$ of size $W\times W$,
\[
M_{\mc H} = 
\begin{bmatrix}
    4  & 0 & 0 &        &  & 0 \\
    -1 & 1 & 0 &        &  &   \\
    0  & 0 & 1 &        &  &  \\
       &   &   & \cdots & 0 & 0  \\
       &   &   &        & 1 & -1 \\
     0  &   &   &       0 & 0 & 4  \\
\end{bmatrix},
\]
the horizontal gradient is estimated as $(\bmI \star \mc H) M_{\mc H}$, where $\star$ is convolution with replicate padding on the boundary. This is effectively using forward finite differences on the left boundary and backward finite differences on the right boundary. The vertical gradient estimate is corrected similarly. We found that this correction reduces the error of the learned solution by several times. However, there are still errors in four corners, which can be further improved by more refined correction.

\section{Solving PDEs with FC-NNs and CNNs}\label{appendix:solve}
\subsection{Network architecture}
The FC-NN used in the experiments in Section~\ref{sec:solve_det_pde} has $8$ hidden layers and $512$ nodes per hidden layer, with the input and output dimensions being $2$ and $3$, respectively. The nonlinear activation is \texttt{Tanh}. The total number of parameters is $1,841,155$. We increased the number of nodes in the hidden layer from $20$ to $512$ to overfit the solution. We considered the collocation points to be at random locations in the domain, and increased their number. However, none of these modifications lead to improvement of the learned solution.

The convolutional decoder network uses two dense blocks with $8$ and $6$ dense layers respectively to transform the latent $\bmz$ of size $1\times 16 \times 16$ to the output $\bmy$ of size $3 \times 64 \times 64$. The decoding layers use nearest upsampling followed by one $3\times 3$ convolution. The network has $514,278$ parameters and $20$ convolution layers.

We train FC-NNs and CNNs with mixed residual loss using L-BFGS optimizer (with history size $50$ and maximum iteration $20$), learning rate $0.5$. The weight for boundary loss is $\lambda = 10$.

\subsection{Supplementary Numerical Experiments}
\begin{figure}[h!]
    \centering
    \begin{subfigure}[b]{0.48\textwidth}
        \centering
        \includegraphics[width=\textwidth]{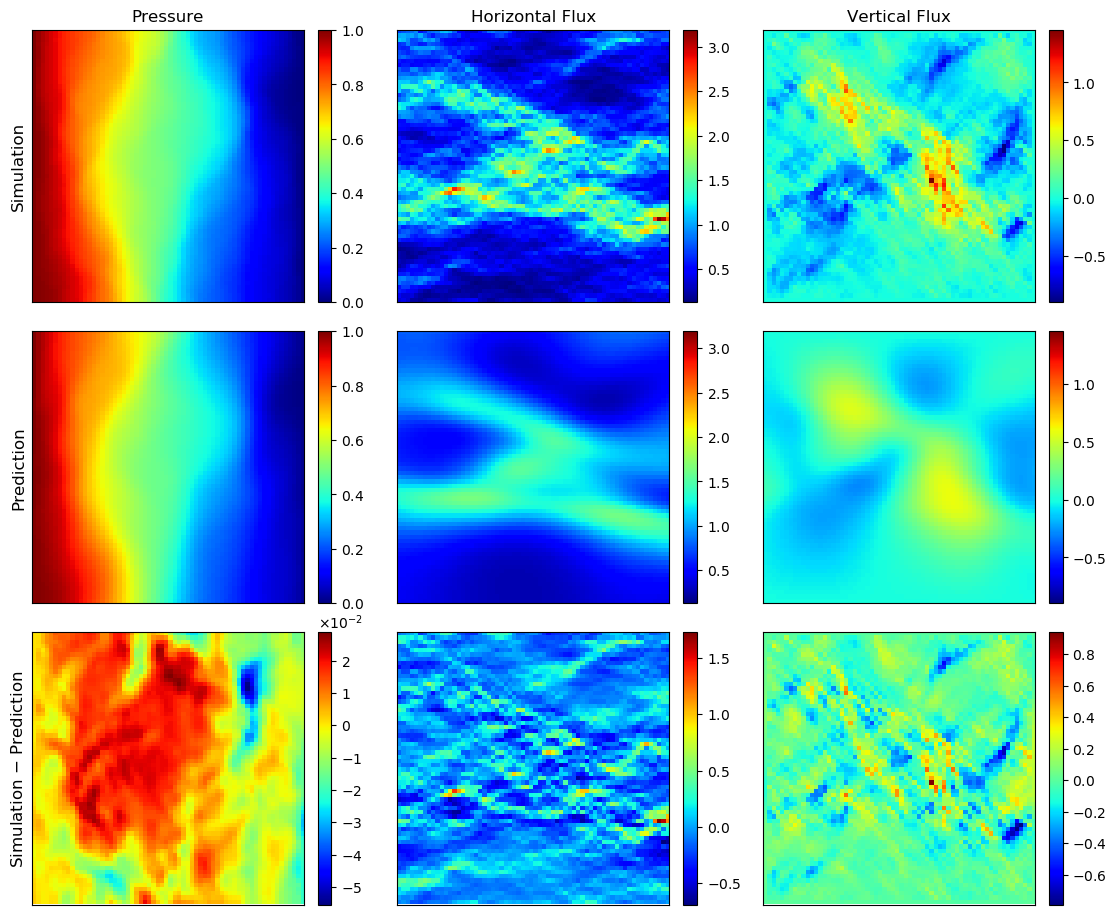}
        \caption{FC-NN, iteration $500$.}
    \end{subfigure}
    ~
    \begin{subfigure}[b]{0.48\textwidth}
        \centering
        \includegraphics[width=\textwidth]{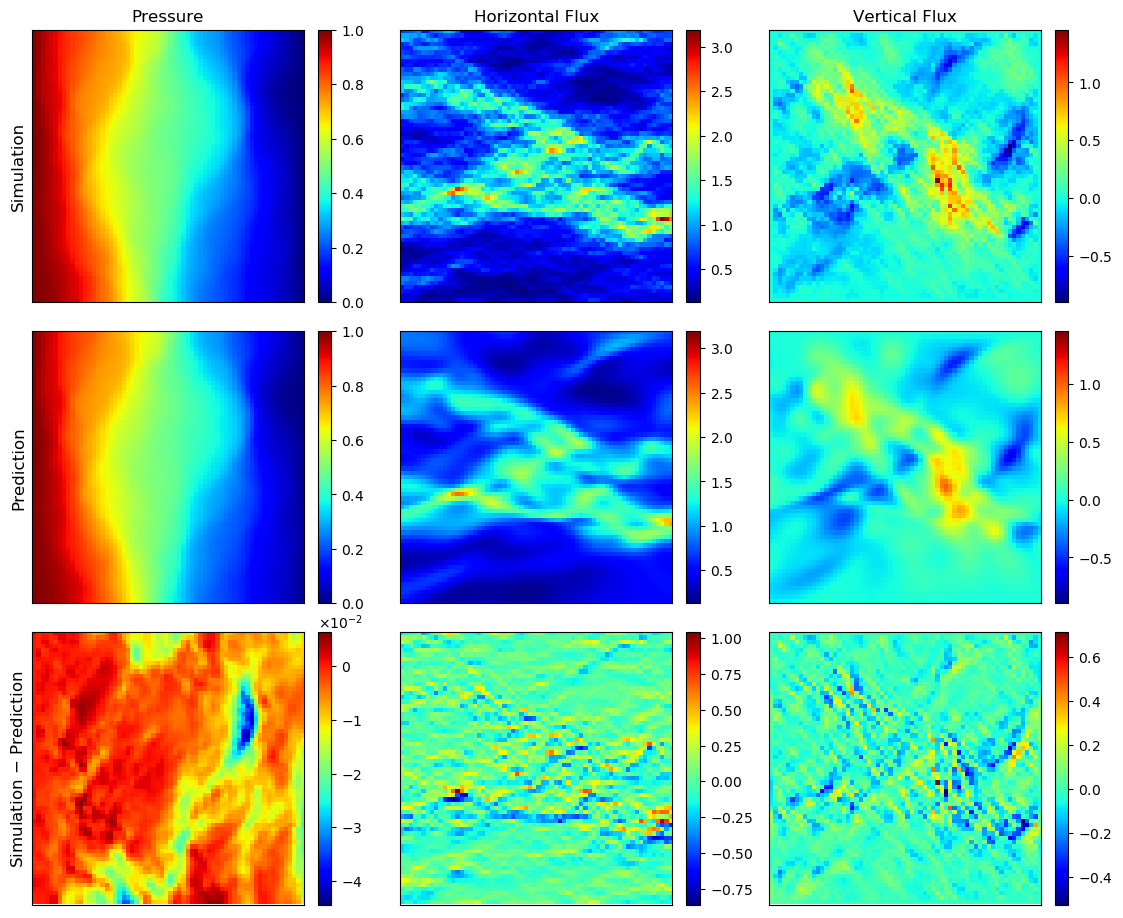}
        \caption{FC-NN, iteration $2000$.}
    \end{subfigure}
    
    \begin{subfigure}[b]{0.48\textwidth}
        \centering
        \includegraphics[width=\textwidth]{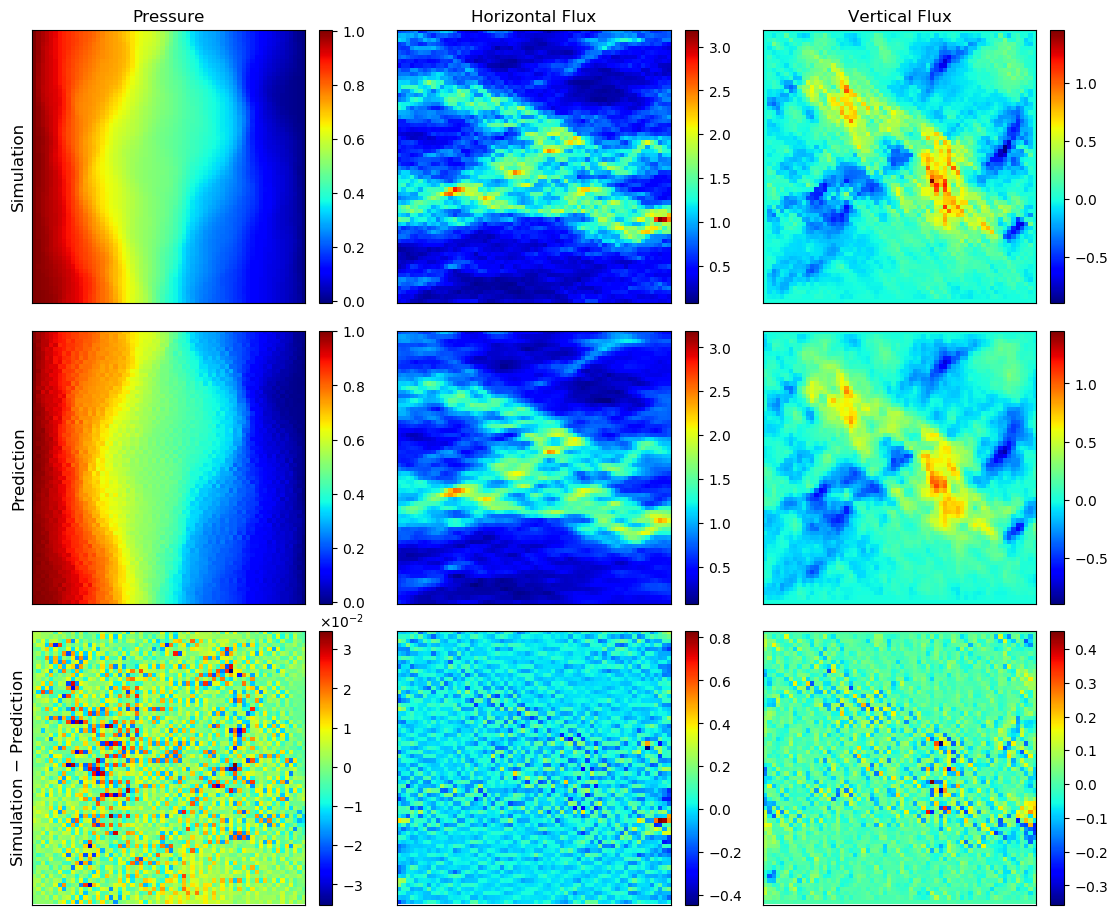}
        \caption{CNN, iteration $50$.}
    \end{subfigure}
    ~
    \begin{subfigure}[b]{0.48\textwidth}
        \centering
        \includegraphics[width=\textwidth]{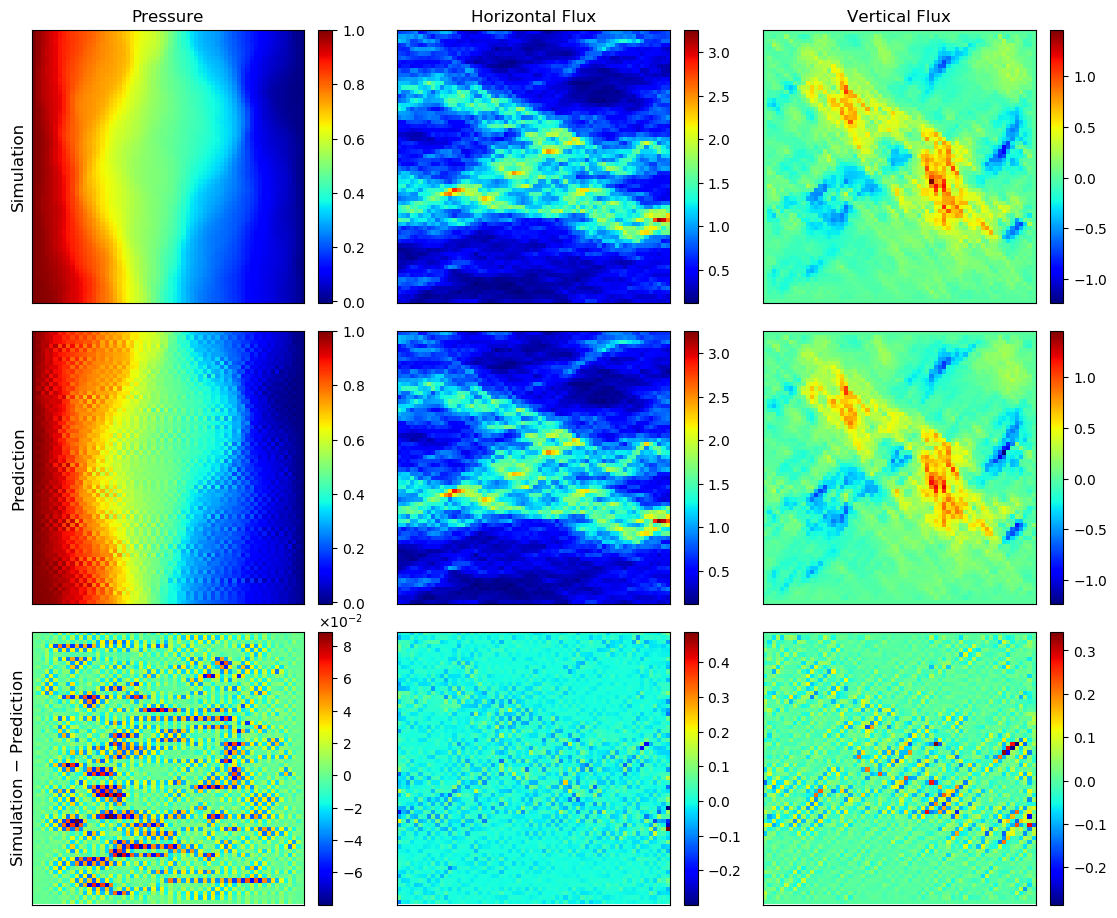}
        \caption{CNN, iteration $500$.}
    \end{subfigure}

    \caption{Solving Darcy flow for one sample from GRF KLE4096 under mixed residual loss.}
    \label{fig:solve_k4096}
\end{figure}

We also show learned solutions for GRF KLE$4096$ in Fig.~\ref{fig:solve_k4096}. Again the CNN can capture the flux field much faster and better than FC-NN, but in this case the pressure field begins to show severe checkerboard artifact that the largest error being larger than that of the pressure solution of FC-NN.

\section{Details on conditional Glow and supplementary results}\label{appendix:cglow}
\subsection{Details on the network structure}
In Fig.~\ref{fig:glow_msc_sub}, the encoder network includes a cascade of $L$ \texttt{Dense Block}s (that maintain the feature map size) and $L-1$ \texttt{Trans Down} layers (that typically half the feature map size, e.g. from $32\times 32$ to $16\times 16$). The features extracted after each dense block are treated as input features $\{\bmxi_l\}_{l=1}^L$. For details of \texttt{Dense Block}s and \texttt{Trans Down} layers (encoding layer), please refer to Section~$2.2$ in~\cite{zhu2018bayesian}.

In Fig.~\ref{fig:glow_msc_sub}, the \texttt{Squeeze} operator rearranges the features of size $C \times H \times W$ into $4C \times \frac{1}{2} H \times \frac{1}{2}W$ if the squeeze factor is $2$. The \texttt{Split} operator splits out half of the features/channels as latent variable $\bmz_l$, which is diagonal Gaussian parameterized by the other half features $\bmh_l$ with a $3 \times 3$ convolution, stride 1 and zero initialization.
In Fig.~\ref{fig:flow_step}, one step of flow contains an activation normalization layer (\texttt{ActNorm}), an invertable $1\times1$ convolution layer and an affine coupling layer. \texttt{ActNorm} performs affine transformation of the activation with a scale and bias parameter per-channel. 
The $1\times 1$ convolution layer with equal number of input and output channels is a learnable permutation operation to mix the two parts of the flow features before passing them  to the affine coupling layer.

We show the detailed computation of the forward and reverse paths of the affine coupling layer (Fig.~\ref{fig:affine_coupling}) in Table~\ref{tab:affine_coupling}. The nonlinear transform \texttt{CouplingNN} includes $3$ dense layers, followed by a $3\times3$ convolution layer with zero initialization, whose output channels split into two parts, i.e. $(\hat{\bms}, \bmt)$.
\begin{table}[]
    \caption{Forward (from $\bmy'$ to $\bmz'$) and reverse paths of affine coupling layer with condition of input features $\bmxi_l$ as in Figure~\ref{fig:affine_coupling}.}
    \label{tab:affine_coupling}
    \centering
    \begin{tabularx}{\textwidth}{ X | X }
        Forward & Reverse
        \\\hline
        \begin{tabular}[t]{@{}l@{}}
            $\bmy_1', \bmy_2' = \texttt{split}(\bmy')$\\
            $\hat{\bmy}_1 = \texttt{concat}(\bmy_1', \bmxi_l)$\\
            $(\hat{\bms}, \bmt) = \texttt{CouplingNN}(\hat{\bmy}_1)$\\
            $\bms = \texttt{sigmoid}(\hat{\bms} + 2)$\\
            $\bmz_2' = \bms \odot \bmy_2' + \bmt$\\
            $\bmz_1' = \bmy_1'$\\
            $\bmz' = \texttt{concat}(\bmz_1', \bmz_2')$
        \end{tabular}
        &
        \begin{tabular}[t]{@{}l@{}}
            $\bmz_1', \bmz_2' = \texttt{split}(\bmz')$\\
            $\hat{\bmz}_1 = \texttt{concat}(\bmz_1', \bmxi_l)$\\
            $(\hat{\bms}, \bmt) = \texttt{CouplingNN}(\hat{\bmz}_1)$\\
            $\bms = \texttt{sigmoid}(\hat{\bms} + 2)$\\
            $\bmy_2' = (\bmz_2' - \bmt) / \bms$\\
            $\bmy_1' = \bmz_1'$\\
            $\bmy' = \texttt{concat}(\bmy_1', \bmy_2')$
        \end{tabular}
    \end{tabularx}
\end{table}

\subsection{Results on higher input dimension}
\begin{figure}
    \centering
    \begin{subfigure}[b]{0.47\textwidth}
        \centering
        \includegraphics[width=\textwidth]{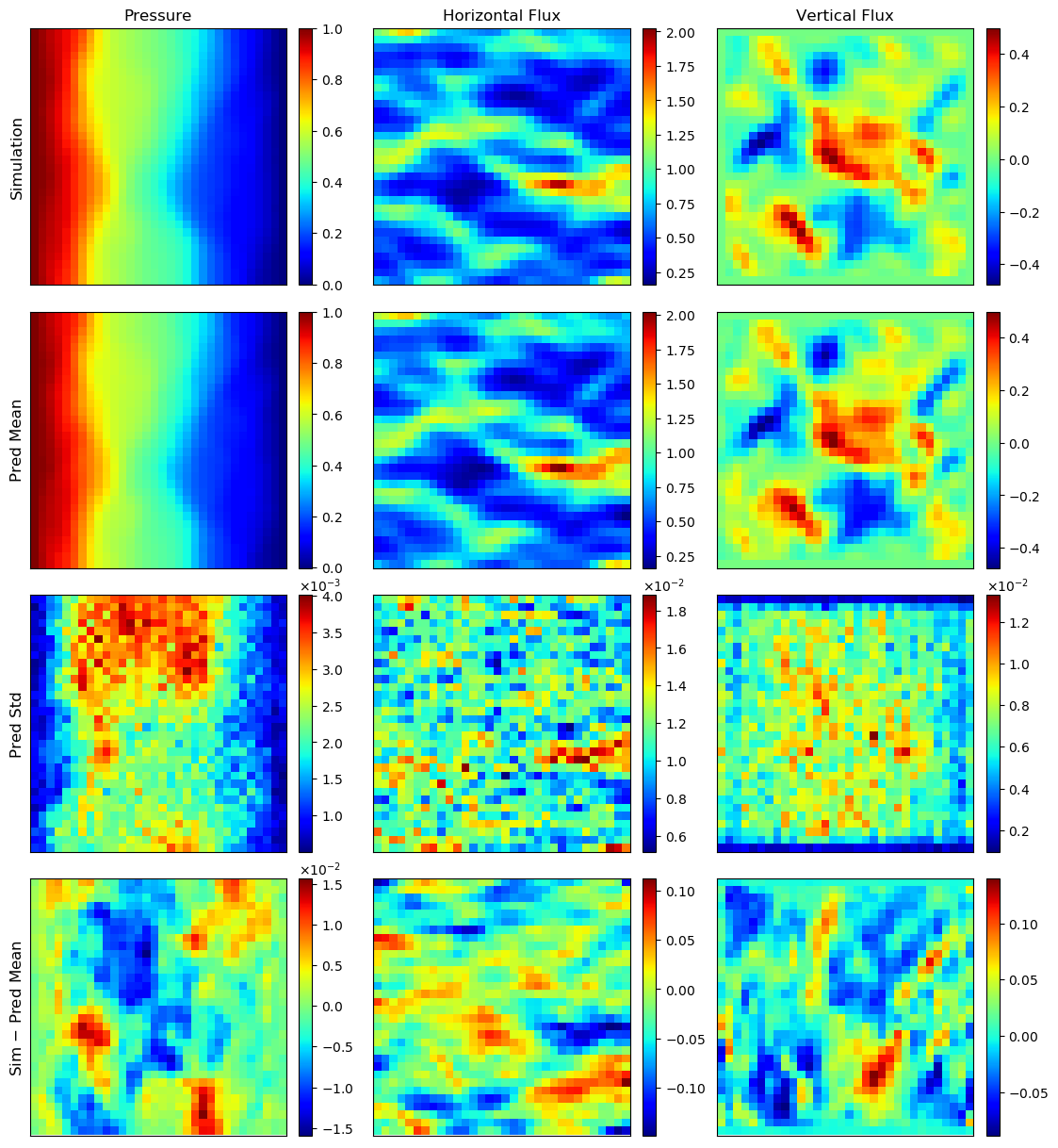}
        \caption{Test realization $1$.}
    \end{subfigure}
    ~
    \begin{subfigure}[b]{0.47\textwidth}
        \centering
        \includegraphics[width=\textwidth]{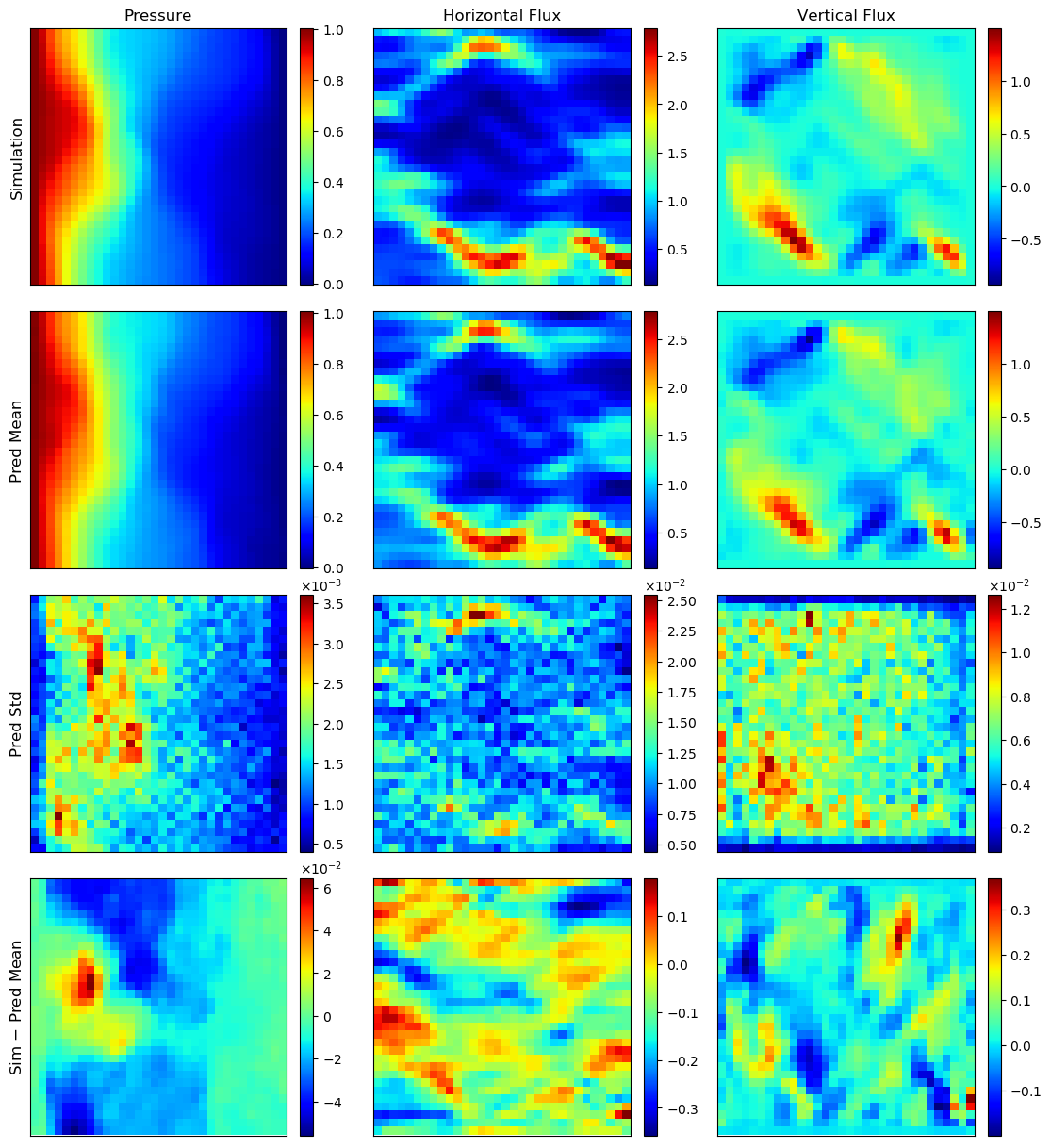}
        \caption{Test realization $2$.}
    \end{subfigure}
      
    \caption{Prediction of the multiscale conditional Glow ($\beta=200$) for two test inputs in (a) and (b) which are sampled from GRF KLE$256$ over $32\times 32$ grid. The predictive mean ($2$nd row) and one standard derivation ($3$rd row) are obtained with 20 output samples. The first row shows three simulated output fields, and the fourth row shows the error between the simulation and predictive mean. The relative $L_2$ error for the predicted pressure field is $0.019875$, evaluated on 512 test samples from GRF KLE256.}
    \label{fig:cglow_kle256_grid32}
\end{figure}

We also trained the conditional Glow with $4096$ samples from GRF KLE$256$ over $32\times 32$ grid. The prediction results for two test inputs are shown in Fig.~\ref{fig:cglow_kle256_grid32}.

\bibliographystyle{elsarticle-num}
\bibliography{refs}

\end{document}